\documentclass[preprint,12pt]{emulateapj}
\usepackage{epsfig}
\usepackage{natbib}
\usepackage{amssymb}
\usepackage{amsbsy}
\usepackage{natbib}
\usepackage{url}
\usepackage[mathcal]{euscript}
\pdfoutput=1
\bibpunct{(}{)}{,}{a}{}{,}

\newcommand{\vsinis}{$v \sin i$'s}
\newcommand{\vsini}{$v \sin i$}

\begin{document}
 
\def\simlt{\vcenter{\hbox{$<$}\offinterlineskip\hbox{$\sim$}}}
\def\simgt{\vcenter{\hbox{$>$}\offinterlineskip\hbox{$\sim$}}}
\def\etal{et al.\ }
\def\kms{km s$^{-1}$}

\title{CSI~2264: Characterizing Young Stars in NGC~2264 with Stochastically
  Varying Light Curves \footnotemark[*]}
\footnotetext[*]{Based on data from the {\em Spitzer} and {\em CoRoT}
missions, as well as the Canada France Hawaii Telescope (CFHT) MegaCam
CCD, and  the European Southern Observatory Very Large Telescope,
Paranal Chile, under program 088.C-0239. The {\em CoRoT} space mission
was developed and is  operated by the French space agency CNES, with
particpiation of ESA's RSSD and Science Programmes, Austria, Belgium,
Brazil, Germany, and Spain.  MegaCam is a joint project of CFHT and
CEA/DAPNIA, which is operated by the National Research Council (NRC)
of Canada, the Institute National des  Sciences de l'Univers of the
Centre National de la Recherche Scientifique of France, and the
University of Hawaii.}
\author{John Stauffer\altaffilmark{1}, Ann Marie Cody\altaffilmark{2}, 
Luisa Rebull\altaffilmark{1}, Lynne A. Hillenbrand\altaffilmark{3},  
Neal J. Turner\altaffilmark{4}, John Carpenter\altaffilmark{3},
Sean Carey\altaffilmark{1}, 
Susan Terebey\altaffilmark{5}, Mar\'ia Morales-Calder\'on\altaffilmark{6},  
Silvia H. P. Alencar\altaffilmark{7}, 
Pauline McGinnis\altaffilmark{7}, 
Alana Sousa\altaffilmark{7}, 
Jerome Bouvier\altaffilmark{8}, 
Laura Venuti\altaffilmark{8}, 
Lee Hartmann\altaffilmark{9}, Nuria Calvet\altaffilmark{9}, 
Giusi Micela\altaffilmark{10}, Ettore Flaccomio\altaffilmark{10}, 
Inseok Song\altaffilmark{11}, Rob Gutermuth\altaffilmark{12},  
David Barrado\altaffilmark{6}, 
Frederick J. Vrba\altaffilmark{13}, Kevin Covey\altaffilmark{14}, 
William Herbst\altaffilmark{15}, 
Edward Gillen\altaffilmark{16}, 
Marcelo Medeiros Guimar\~{a}es\altaffilmark{17},
Herve Bouy\altaffilmark{6}, 
Fabio Favata\altaffilmark{18} }
\altaffiltext{1}{Spitzer Science Center, California Institute of
Technology, Pasadena, CA 91125, USA}
\altaffiltext{2}{NASA Ames Research Center, Kepler Science Office, Mountain
View, CA 94035}
\altaffiltext{3}{Astronomy Department, California Institute of
Technology, Pasadena, CA 91125, USA}
\altaffiltext{4}{Jet Propulsion Laboratory, California Institute
of Technology, Pasadena, CA 91109, USA}
\altaffiltext{5}{Department of Physics and Astronomy, 5151 State University
Drive, California State  University at Los Angeles, Los Angeles, CA 90032}
\altaffiltext{6}{Centro de Astrobiolog\'ia, Dpto. de
Astrof\'isica, INTA-CSIC, PO BOX 78, E-28691, ESAC Campus, Villanueva de
la Ca\~nada, Madrid, Spain}
\altaffiltext{7}{Departamento de F\'{\i}sica -- ICEx -- UFMG, 
Av. Ant\^onio Carlos, 6627, 30270-901, Belo Horizonte, MG, Brazil}
\altaffiltext{8}{Universit\'e de Grenoble, Institut de Plan\'etologie 
et d'Astrophysique de Grenoble (IPAG), F-38000 Grenoble, France;
CNRS, IPAG, F-38000 Grenoble, France}
\altaffiltext{9}{Department of Astronomy, University of Michigan, 
500 Church Street, Ann Arbor, MI 48105, USA}
\altaffiltext{10}{INAF - Osservatorio Astronomico di Palermo, Piazza 
del Parlamento 1, 90134, Palermo, Italy}
\altaffiltext{11}{Department of Physics and Astronomy, The University 
of Georgia, Athens, GA 30602-2451, USA}
\altaffiltext{12}{Department of Astronomy, University of Massachusetts,
Amherst, MA 01003, USA}
\altaffiltext{13}{U.S. Naval Observatory, Flagstaff Station, 10391 
West Naval Observatory Road, Flagstaff, AZ 86001, USA}
\altaffiltext{14}{Department of Physics and Astronomy (MS-9164), Western
Washington Univ., 516 High St., Bellingham, WA 98225, USA}
\altaffiltext{15}{Astronomy Department, Wesleyan University,
Middletown, CT 06459, USA}
\altaffiltext{16}{Department of Physics, University of Oxford, Keble Road, 
Oxford, OX1 3RH, UK}
\altaffiltext{17}{Departamento de F\'isica - UFS - Rod. Marechal Rondon,
49100-000, S\~{a}o Cristov\~{a}o, SE, Brazil}
\altaffiltext{18}{European Space Agency, 8-10 rue Mario Nikis, 
F-75738 Paris Cedex 15, France}
\email{stauffer@ipac.caltech.edu}

\begin{abstract}

We provide {\em CoRoT} and {\em Spitzer} light curves, as well as
broad-band multi-wavelength photometry and high resolution, multi- and
single-epoch spectroscopy for seventeen classical T~Tauri stars in
NGC~2264 whose {\em CoRoT} light curves exemplify the ``stochastic"
light curve class as defined in Cody et al.\ (2014).   

The most probable physical mechanism to explain the optical
variability within this light curve class is time-dependent mass
accretion onto the stellar photosphere, producing transient hot
spots.   As evidence in favor of this hypothesis, multi-epoch high
resolution spectra for a subset of these stars shows that their
veiling levels also vary in time and that this veiling variability is
consistent in both amplitude and timescale with the optical light
curve morphology.  Furthermore, the veiling variability is
well-correlated with the strength of the  \ion{He}{1} 6678 \AA\
emission line, a feature predicted by models to arise in accretion
shocks on or near the stellar photosphere.   Stars with accretion
burst light curve morphology (Stauffer et al.\ 2014) are also
attributed to variable mass accretion.   Both the stochastic and
accretion burst light curves can be explained by a simple model of
randomly occurring flux bursts, with the stochastic light curve class
having a higher frequency of lower amplitude events.

Based on their UV excesses, veiling, and mean H$\alpha$ equivalent
widths, members of the stochastic light curve class have only moderate
time-averaged mass accretion rates.  The most common feature of their
H$\alpha$ profiles is for them to exhibit blue-shifted absorption
features, most likely originating in a disk wind.  The lack of
periodic signatures in the light curves suggests that little of the
variability is due to long-lived hot spots rotating into or out of our line of
sight; instead, the primary driver of the observed photometric
variability is likely to be instabilities in the inner disk  that lead
to variable mass accretion.

\end{abstract}


\section{Introduction}

The published literature on the photometric variability of T~Tauri stars
now extends back in time more than 150 years (Schmidt 1861, 1866), predating
by nearly 100 years the general acceptance that T associations are sites
of recent/current star formation (Salpeter 1954, Ambartsumian 1954), and
therefore that the photometric variability of T~Tauri stars is likely
a signpost of youth (Walker 1956).  Haro \& Herbig (1955) and Walker (1956)
discovered that many of the photometrically variable T~Tauri stars also
had UV excesses and abnormally weak photospheric absorption lines (now
usually described as spectral ``veiling"); the former authors concluded
that the UV excesses might originate from a small hot spot on or near the
stellar photosphere.  However, it was not until the 1970s that the
modern paradigm for young stellar objects (YSOs) began to develop, 
first by linking the photospheric hotspots to accretion from a disk
(Walker 1972; Wolf, Appenzeller \& Bertout 1977), and eventually 
by having the accretion
stream be channeled by magnetic field lines connecting the star and
inner disk (Ghosh \& Lamb 1978; K\"onigl 1991)

Based on photometry derived from visual observations or photographic plates,
Joy (1945) made photometric variability one of the defining
characteristics of the T~Tauri class of stars.  Only the largest amplitude
variables could be identified with those techniques.
The modern era in the study of the photometric variability of T~Tauri stars
began in the 1980s in a series of papers by Herbst and his students
(Herbst et al.\ 1982, 1987; Holtzman, Herbst \& Booth 1986) and 
by Vrba and his collaborators
(Rydgren \& Vrba 1983; Vrba et al.\ 1986, 1989).  They showed that 
very young stars often have spotted photospheres, allowing their rotation
periods to be derived (which showed that many of them
are rapidly rotating relative to the Sun, as had long been known
from spectroscopy, e.g., Walker 1956).  In many cases, the data were
compatible with cold spots as found on the Sun; however, in some cases --
particularly when photometry over a broad wavelength range was available --
the data were best fit by spots hotter than the stellar photosphere,
confirming the speculation of Haro \& Herbig (1955).  Other
T~Tauri stars instead showed apparently chaotic (or at least aperiodic)
variability.   Herbst et al.\ (1994) summarized the existing data on T
Tauri variability, and proposed sorting the light curves into several
physically motivated classes:
\begin{itemize}
\item Type I - light curves resulting from stable, long-lived cold spots that
 are not distributed in an axisymmetric way over the stellar photosphere.
 The resultant light curves are periodic, and persist with little change
 often for months to years;

\item Type II - light curves dominated by relatively short lived hot spots.
 Often these do not show obvious periodicity, indicating that the hot
 spots evolve significantly on timescales less than the stellar rotation
 period;

\item Type IIp - hot spot light curves where the spot lifetime is instead
 comparable or larger than the stellar rotation period, resulting in
 more or less periodic variations -- but which generally evolve rapidly;

\item Type III - young stars with large amplitude flux variations in their
  light curves which are not accompanied by significant variations in the
  veiling inferred from spectroscopy.  The physical interpretation of this
  class was uncertain, though the leading candidate was extinction by
  clumps of dust (most plausibly connected to structures in the inner
  circumstellar disk).
\end{itemize}

In the past two decades, time series photometry in star forming regions has
been used primarily to determine rotation periods in order to study the
initial angular momentum distribution for low mass stars (Herbst et al.\ 2000;
Rebull 2001; Lamm et al.\ 2005;
Moraux et al.\ 2013).  Those
programs have been quite successful, with rotation periods now known for
more than a thousand low mass stars with ages less than a few Myr.  Some programs have
continued to attempt to characterize YSO variability not due to spots (e.g.
Carpenter, Hillenbrand and Skrutskie 2001), including 
a series of campaigns aimed at a particularly intriguing
YSO -- AA~Tau (Bouvier et al.\ 1999; Terquem \& Papaloizou 2000;
Bouvier et al.\ 2003).  Based on intensive ground-based
time series photometry and spectroscopy over several years, the latter set of papers
provided evidence that the light curve for AA~Tau is dominated by
variable extinction caused by a warped inner disk periodically intersecting
our line of sight.   More recently, using three weeks of continuous,
high signal-to-noise ratio (S/N) photometry
from the CoRoT satellite, Alencar et al.\ (2010) showed that many of the 
classical T~Tauri stars (CTTS)
in the star-forming region NGC~2264 have light curves similar to AA~Tau,
indicating that variable extinction from structures in the circumstellar
disk passing through our line of sight is common.

\begin{figure*}
\begin{center}
\epsscale{0.80}
\plottwo{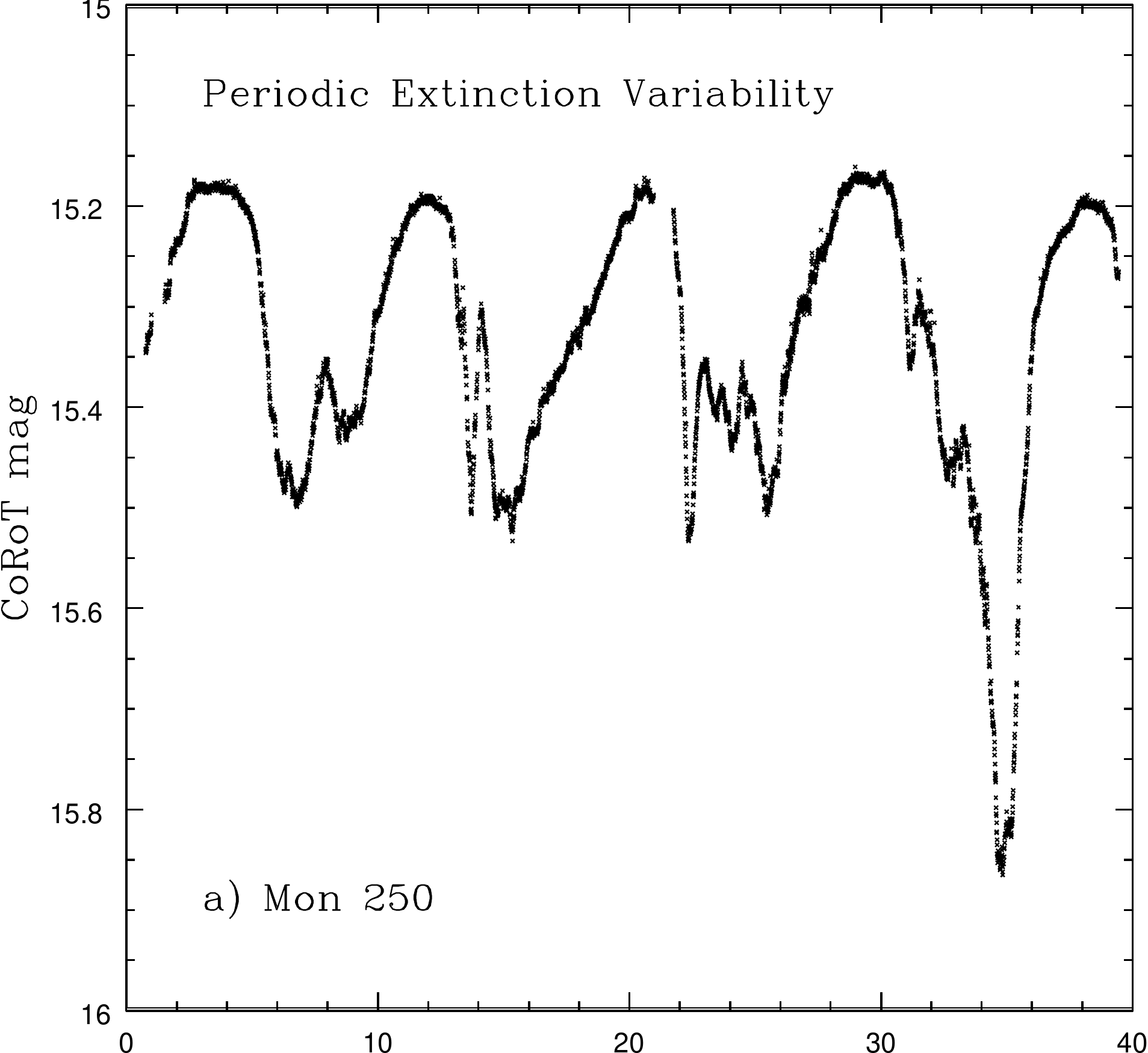}{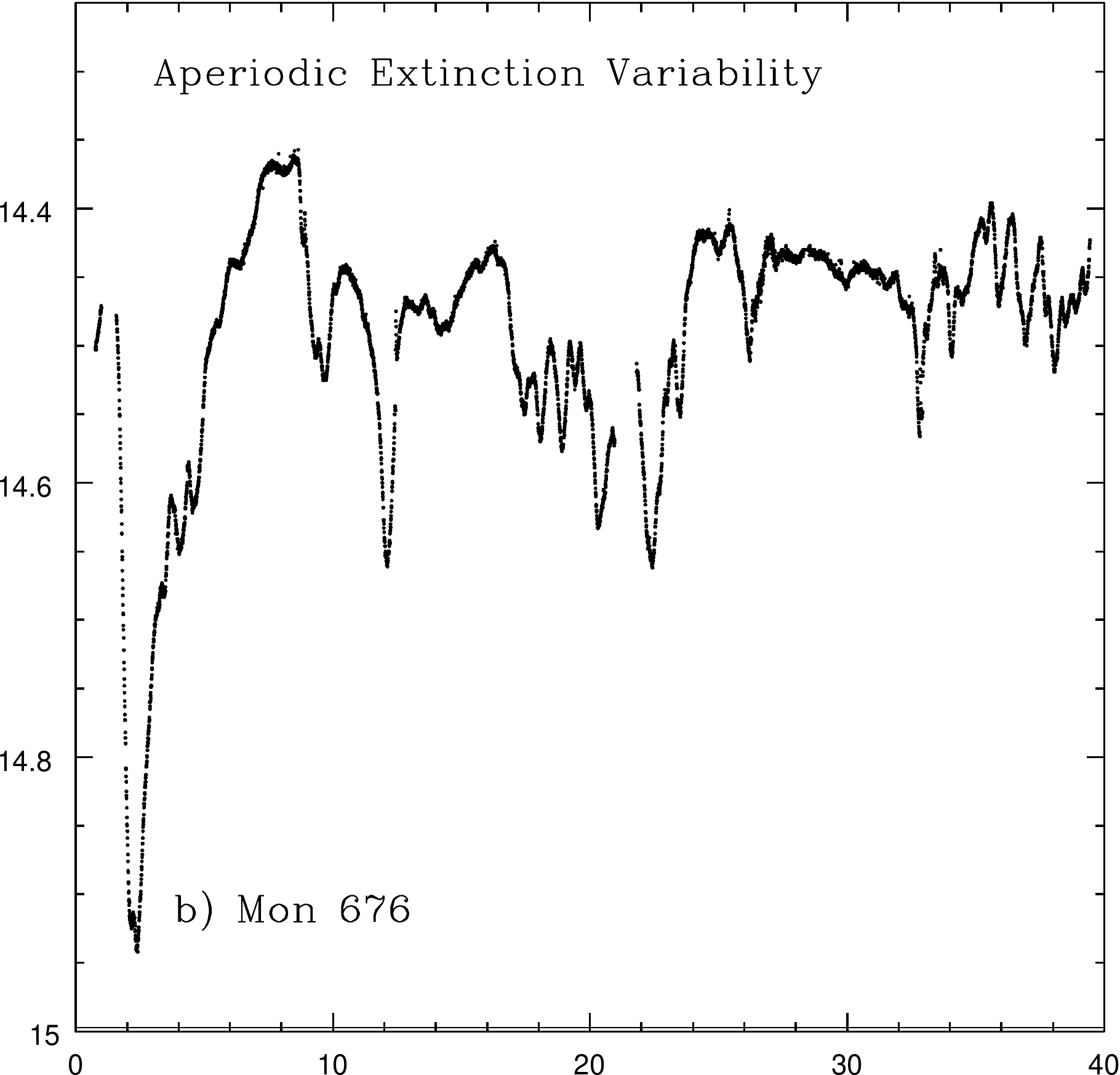}
\vspace{-0.cm}
\plottwo{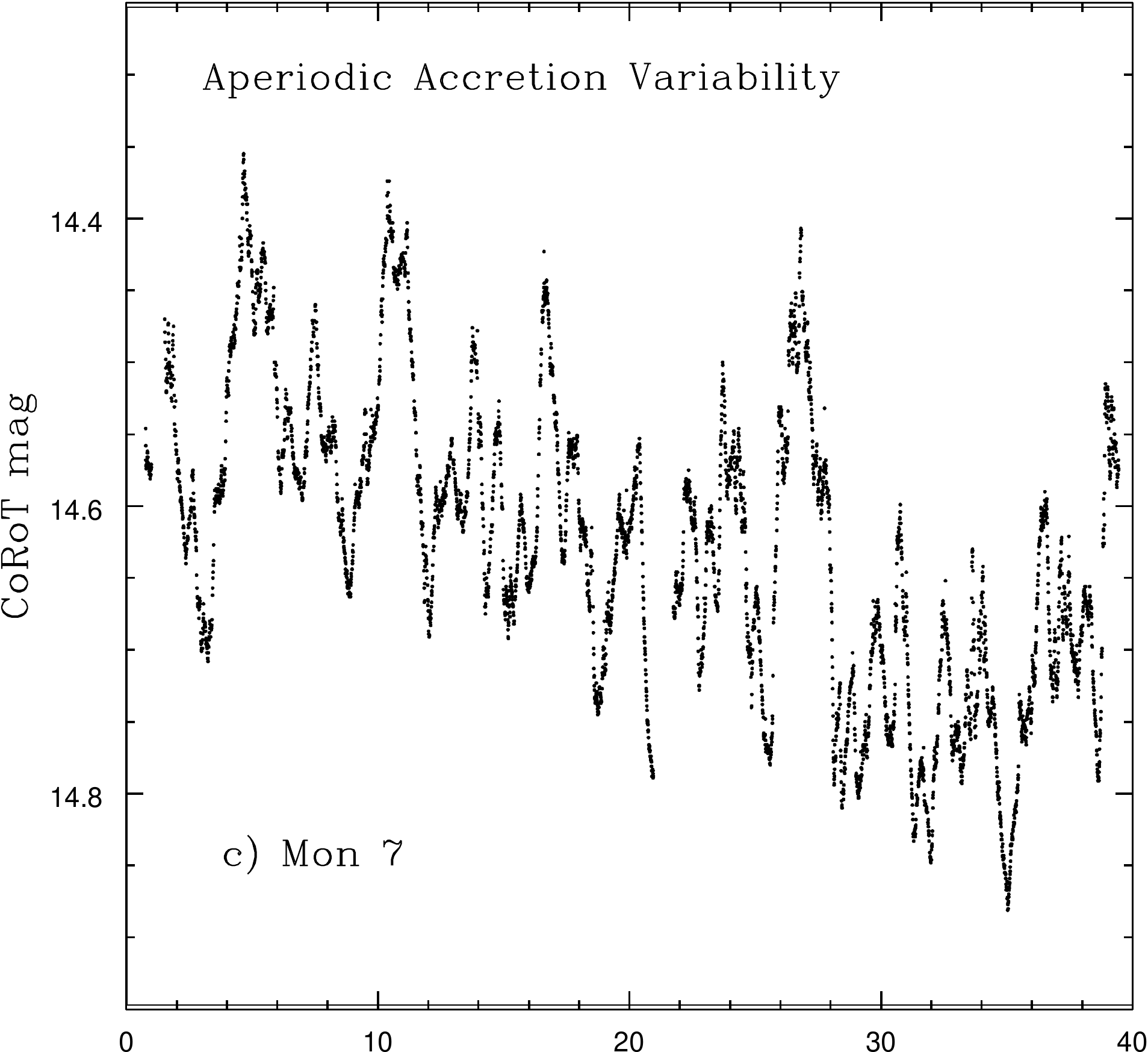}{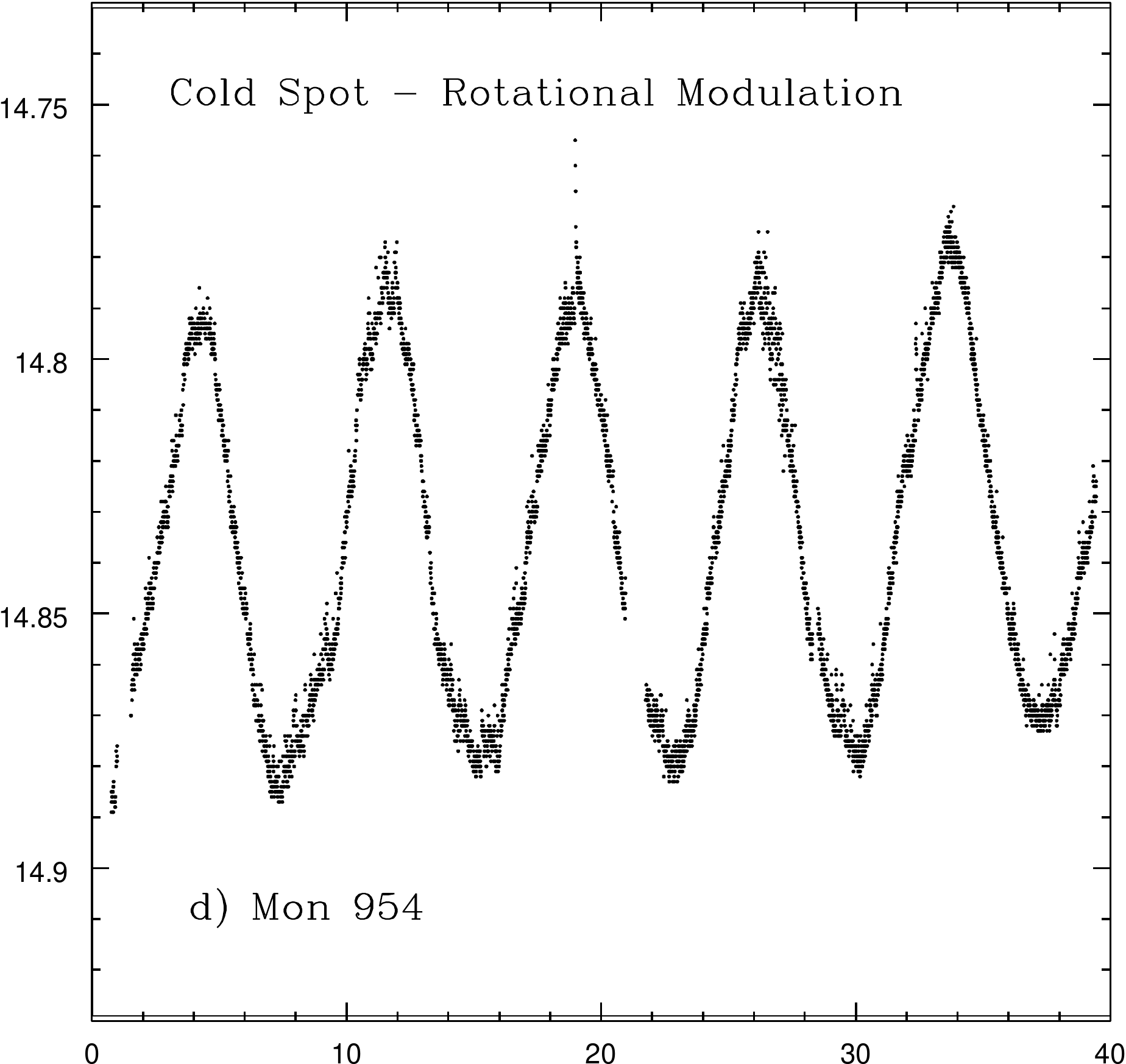}
\vspace{-0.cm}
\plottwo{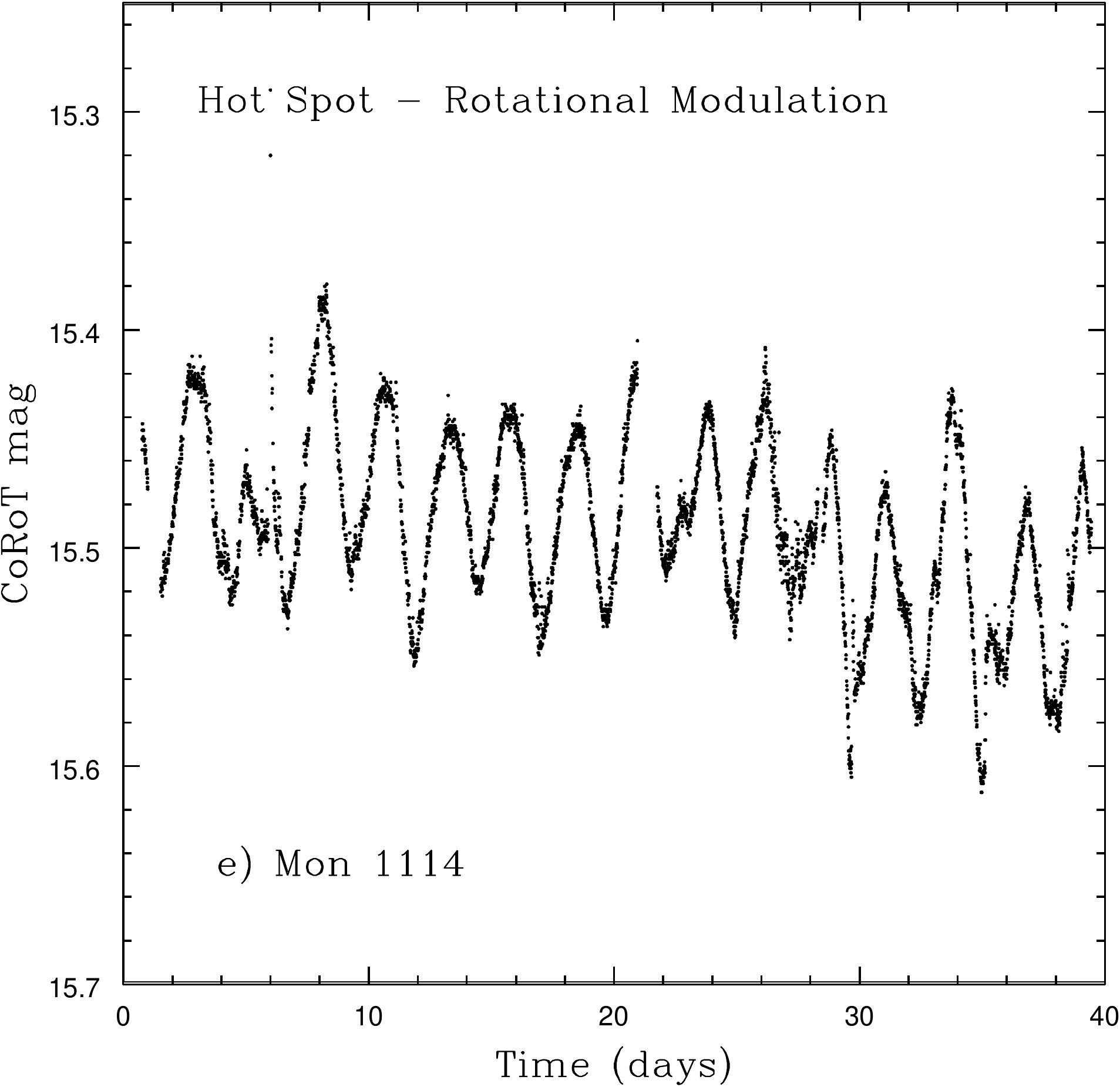}{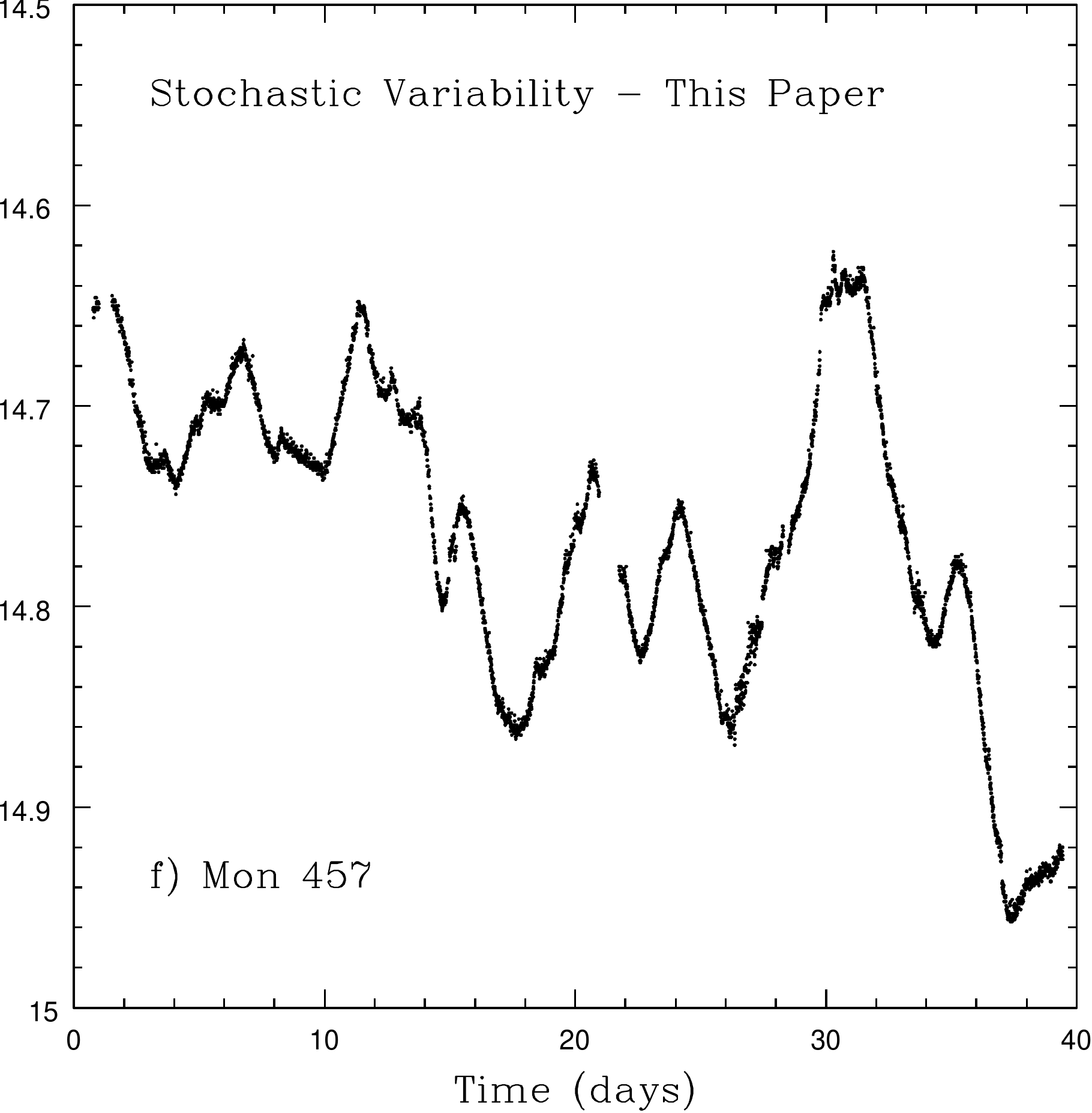}
\end{center}
\caption{Light curves for five CTTS members of NGC~2264 whose {\em CoRoT}
light curves illustrate morphologies to which we have ascribed physical
mechanisms including variable extinction (panels a and b), 
unstable accretion or accretion burst (panel c),
cold spots (panel d), and hot spots (panel e), and a sixth CTTS that
shows a typical
stochastic light curve (panel f).
The variable extinction exemplars include an AA~Tau analog (Mon-250) and a
star with aperiodic flux dips possibly due to MRI events briefly raising
dust structures above the quiescent disk photosphere (Mon-676).
The Mon-1114 light curve also includes two  isolated
short-duration flux dips near days 30 and 35 of the type which we associate with
extinction due to dust entrained in accretion streams (Stauffer et al.\ 2015).
\label{fig:sixctts}}
\end{figure*}

This paper is based on a number of observational efforts organized around
a Spitzer Space Telescope Exploration Science program entitled ``A
Coordinated, Synoptic Investigation of Star-Formation in NGC~2264"
(aka CSI~2264).  CSI~2264
combines simultaneous or contemporaneous time series observations
of the entire NGC~2264 star-forming region spanning a wide range in
wavelength and including both photometry and spectroscopy.  Cody et al.\
(2014) provide an overview of the program and propose a new, quantitative
light curve classification scheme for YSOs based on indices that measure
the degree of periodicity and the degree of symmetry of the light curves.
Stauffer et al.\ (2014) identified and characterized the members of one
of these classes -- CTTS 
whose light curves are dominated by short-duration flux bursts.
McGinnis et al.\ (2015) and Stauffer et al.\ (2015) similarly identify and
characterize CTTS whose light curves are dominated by variable extinction
events.  Venuti et al.\ (2015a) discuss all of the NGC~2264 stars whose
CoRoT light curves are periodic or quasi-periodic.  Figure~\ref{fig:sixctts} illustrates
stars whose light curves are dominated by variable extinction, accretion
bursts, cold or hot spots, and a prototype of the stochastic light curve
class, the topic of this paper.

The last remaining, major light curve class identified by Cody et al.\ (2014)
for which our group has not yet published a detailed characterization paper
are the stars designated as having stochastic light curves. 
Light curves in this class show variability with no definite
periodicity and also show symmetric shapes (that is, they do not appear to
have a continuum level punctuated by dips -- as for the variable extinction stars --
nor a continuum level punctuated by discrete flux excesses -- as for the accretion burst
stars).  The goal of this paper is to illustrate in more detail the 
characteristics of the stars with stochastic optical light curves, and to
propose a physical mechanism for their variability.  In \S 2, we give a
brief listing of the observational data obtained as part of the CSI~2264
program.  In \S 3, we identify a set of seventeen CTTS whose light curves we
consider to exemplify the stochastic class, we provide additional empirical
data for these stars, and we compare their properties to other CTTS members
of NGC~2264.   We believe the most plausible physical mechanism to explain
the optical variability of the stars with stochastic light curves to be
variable accretion onto the star's photosphere, and we provide our arguments
for that conclusion in \S 4.   Additional characterization of the stochastic
light curve stars is provided in \S 5.  In \S 6, we summarize the properties
of the stochastic light curve class members and discuss connections to
physical models and how future observations could improve our understanding
of these stars.

\section{Observational Data }

A detailed description of the data we collected in our
2008 and 2011 observing campaigns for the NGC~2264 star-forming region
is provided in Cody et al.\ (2014).\footnote{The {\em CoRoT} and {\em
Spitzer} light curves for all probable NGC~2264 members, as well as
our broad band photometry for these stars, are available at
\url{http://irsa.ipac.caltech.edu/data/SPITZER/CSI2264}.  As of June 2015,
calibrated versions of
all VLT/FLAMES spectra obtained prior to December 2014 - including those taken for both
CSI~2264 and ESO-Gaia - are available 
at \url{http://archive.eso.org/wdb/wdb/adp/phase3\_spectral}.}     
The
subset of the data sources that we use here is based primarily on
the synoptic data we obtained with {\em CoRoT}, {\em Spitzer}, VLT/FLAMES,
CFHT and the USNO 1.0m telescopes, which we summarize in 
Table~\ref{tab:synoptic_data}. 

\begin{deluxetable*}{lccccc}
\tabletypesize{\scriptsize}
\tablecolumns{12}
\tablewidth{0pt}
\tablecaption{Synoptic Data Used in This Paper\label{tab:synoptic_data}}
\tablehead{
\colhead{Telescope}  & \colhead{Data Type} &
\colhead{Wavelength(s)} & \colhead{Resolution} &
\colhead{Date Range} & \colhead{Cadence}  \\
\colhead{} & \colhead{} & \colhead{} & \colhead{} &
\colhead{(MJD)}  & \colhead{}
 }
\startdata
{\em CoRoT} & Optical Photometry & 6000 \AA\ & 2 & 54534-54556 & 32 sec or 512 sec \\
{\em CoRoT} & Optical Photometry & 6000 \AA\ & 2 & 55897-55936 & 32 sec or 512 sec \\
{\em Spitzer} & IR Photometry  & 3.6$\mu$m and 4.5 $\mu$m & 4 & 55900-55930 & 2 hours \\
USNO 1.0m &  Optical Photometry &  7000 \AA\ & 8 &  55890-55990 & few dozen per night \\
CFHT  &  Optical Photometry &  ugri & 10 & 55971-55986 & $\sim$3 times per night \\
VLT/FLAMES & Spectroscopy & 6450-6750 \AA\  & 17000 & 55899-55981 & $\sim$20 spectra in 90 nights \\
VLT/FLAMES & Spectroscopy & 6450-6750 \AA\  & 17000 & 56648-56660 & $\sim$1 spectrum per night \\
\enddata
\end{deluxetable*}

In addition to the synoptic data, we also use several single epoch data
sources in order to help characterize the overall properties of the NGC~2264
members.   These data sets include:
\begin{itemize}
\item Single epoch $ugri$ imaging of the entire cluster,  with standard
Sloan Digital Sky Survey (SDSS) filters,  obtained using Megacam on
the Canada-France-Hawaii Telescope (CFHT)  during two observing runs
in 2010 and 2012.  These data are presented in detail in Venuti et
al.\ (2014).  We also make use of single-epoch UBVRI data from
Sung et al.\ (2008).
\item {\em Spitzer} IRAC 3.6, 4.5, 5.8 and 8.0 $\mu$m and MIPS 24 $\mu$m
single epoch imaging photometry of the entire cluster obtained early
in the cryogenic mission, and reported in Sung et al.\ (2009) and
Teixeira et al.\ (2012).
\item Single epoch FLAMES
spectra for many of the NGC~2264 members, originally obtained as part of the
ESO-Gaia program (Randich et al.\ 2013). We downloaded the raw spectra
from the ESO-Gaia archive for these stars, and produced our own calibrated spectra.
\end{itemize}

We have used the VLT/FLAMES spectra to analyse H$\alpha$ profiles, spectral
veiling, projected rotational velocities and emission line FWHM as diagnostics
of accretion and the geometry of the CTTS disks.  Details of this analysis
are generally discussed in the Appendix.

\section{The YSOs in NGC~2264 With Stochastic Light Curves}

As noted in \S 1 and in Cody et al.\ (2014), the stochastic
light curve class includes stars whose {\em CoRoT} 
photometry shows little or no periodic signature and for which
the light curve has no significant preference for upward
excursions or downward excursions (they appear symmetric
when reflected about their median level).  Quantitatively,
this corresponds to Q $>$ 0.61 and -0.25 $<$ M $<$ 0.25
using the Cody et al.\ (2014) metrics, where the Q metric is
a measure of the light curve periodicity (or stochasticity) and the
M metric quantifies the degree of symmetry of the light curve relative
to its median level.  Light curves for prototypes of the stochastic
class are shown in Figure 18 of Cody et al. (2014).  
Those light curves
resemble model light curves produced by damped random walk
processes (Findeisen, Cody \& Hillenbrand 2015), which is the reason
for adopting ``stochastic" as the descriptive name for this class.

\begin{figure*}
\includegraphics[width=6.0in]{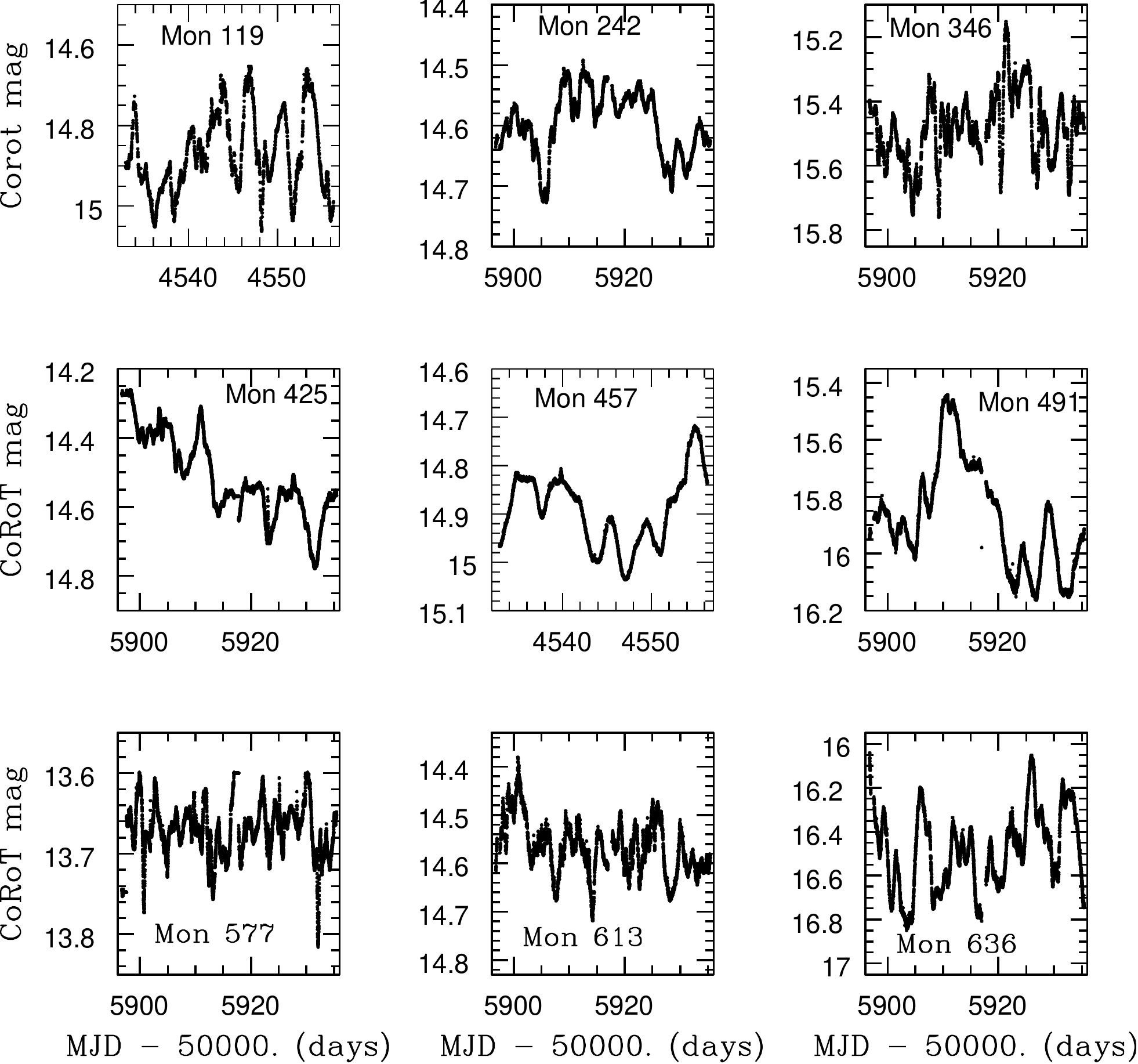}
\caption{{\em CoRoT} light curves for nine of the NGC~2264 CTTS which
we consider as having stochastic light curves.  Note that the full
amplitudes of the variability for these stars is generally in the
10\% to 30\% range, which is small compared to the stars with
variable-extinction dominated light curves (McGinnis et al.\ 2015).
\label{fig:stochastic_stars.set1}}
\end{figure*}

\begin{figure*}
\includegraphics[width=6.0in]{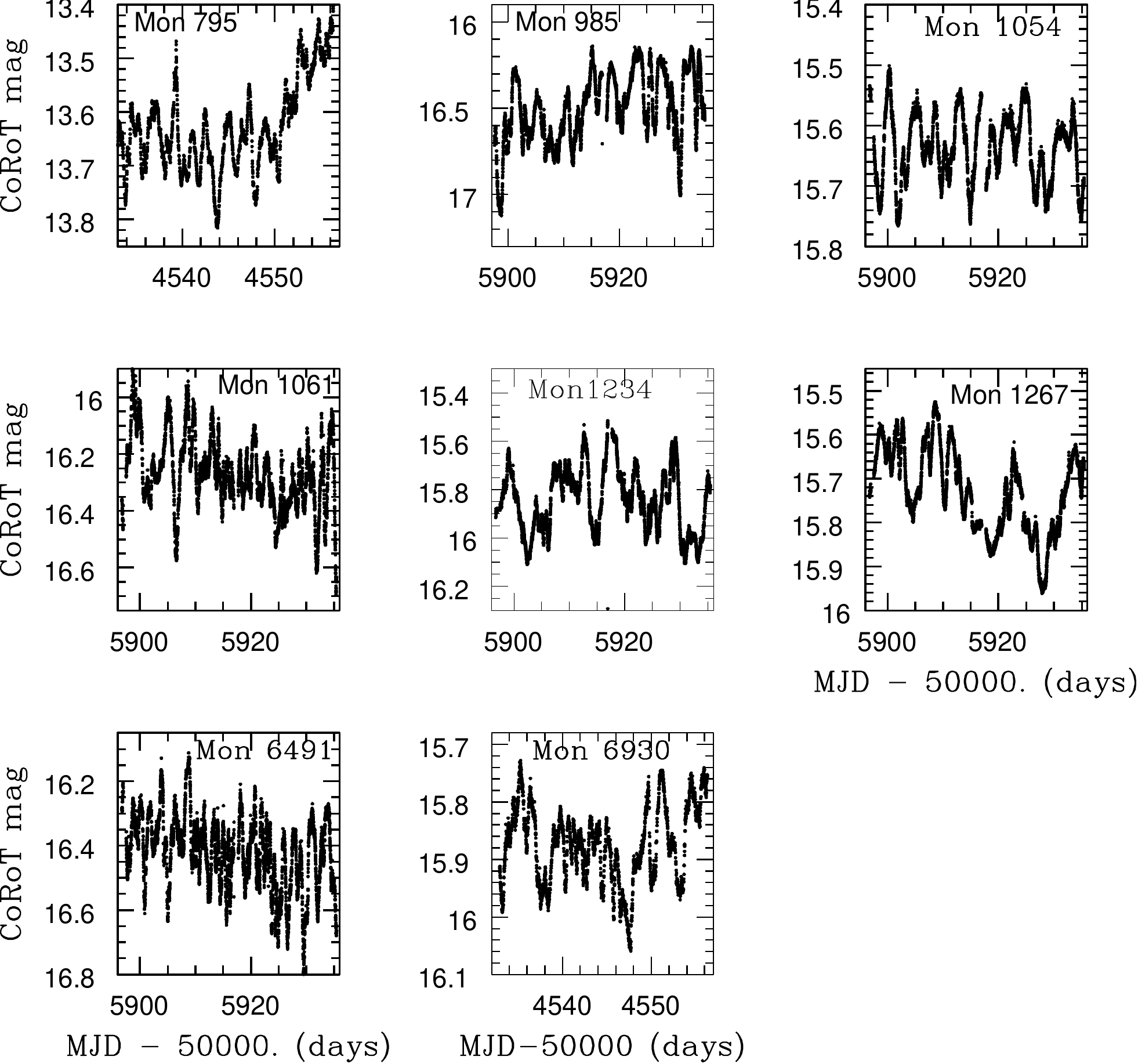}
\caption{{\em CoRoT} light curves for the other seven NGC~2264 CTTS which
we consider as having stochastic light curves.
\label{fig:stochastic_stars.set2}}
\end{figure*}

Because Cody et al.\ (2014) considered only the 2011
{\em CoRoT} data (and not the 2008 campaign) and because
our membership list has evolved slightly since that paper
was written, we have conducted a thorough review of all of
our data in order to construct the list of stochastic stars
to analyse in this paper.  We also have chosen to add two
additional criteria in order to make the sample more characterizable
and more ``pure".  First, we require that the full amplitude of the
{\em CoRoT} light curve be greater than 0.1 mag; this criterion makes
it more likely that we will be able to detect correlations between
the {\em CoRoT} light curve shape and other data.   Second, we
have attempted to exclude stars whose light curve morphology appears
to be better described as a mix of similar amplitude accretion
bursts and extinction dips.   Light curves with features arising from
a mixture of mechanisms is both possible and expected; however, we
have associated specific physical mechanisms already with accretion
burst signatures and extinction dips.  Separating truly random walk
light curve morphologies from this kind of mixed mechanism morphology
is admittedly subjective, and we probably have not been completely
successful.  However, because our goal is not to identify a complete
set of objects but is instead to identify the physical mechanism
driving this light curve morphology, we believe this process is acceptable. 

Table~\ref{tab:basicinformation}
presents the list of NGC~2264 YSOs whose light curve satisfy our criteria.
The table provides our
internal name for the star, as well as the 2MASS and {\em CoRoT}
designations.  In addition, we provide spectral type, H$\alpha$ 
equivalent width, light curve amplitude in the optical and infrared,
an optical-IR Stetson index (Stetson 1996), a spectral energy
distribution (SED) slope, and
a timescale (see \S 5.4).   The suite of single-epoch broadband photometry 
we have for each of these stars is provided in 
Table~\ref{tab:basicphotometry}; most of these data have appeared previously
in Venuti et al.\ (2014) and Cody et al.\ (2014) -- we repeat their values here
in order to allow readers to easily associate specific stars to specific points in 
the color-color and color-magnitude plots we show.
The {\em CoRoT} light curves for each of these stochastic light curve stars
are provided in 
Figure~\ref{fig:stochastic_stars.set1} and
Figure~\ref{fig:stochastic_stars.set2}.   There are of order 175 CTTS
with good {\em CoRoT} light curves; with seventeen stars in the stochastic
light curve class, the frequency of occurrence for light curves of
this type and our selection criteria is of order 10\% (versus 13\% in
Cody et al.\ 2014 using their slightly different selection criteria and
their slightly different membership list).

\begin{figure*}
\begin{center}
\epsfxsize=.99\columnwidth
\epsfbox{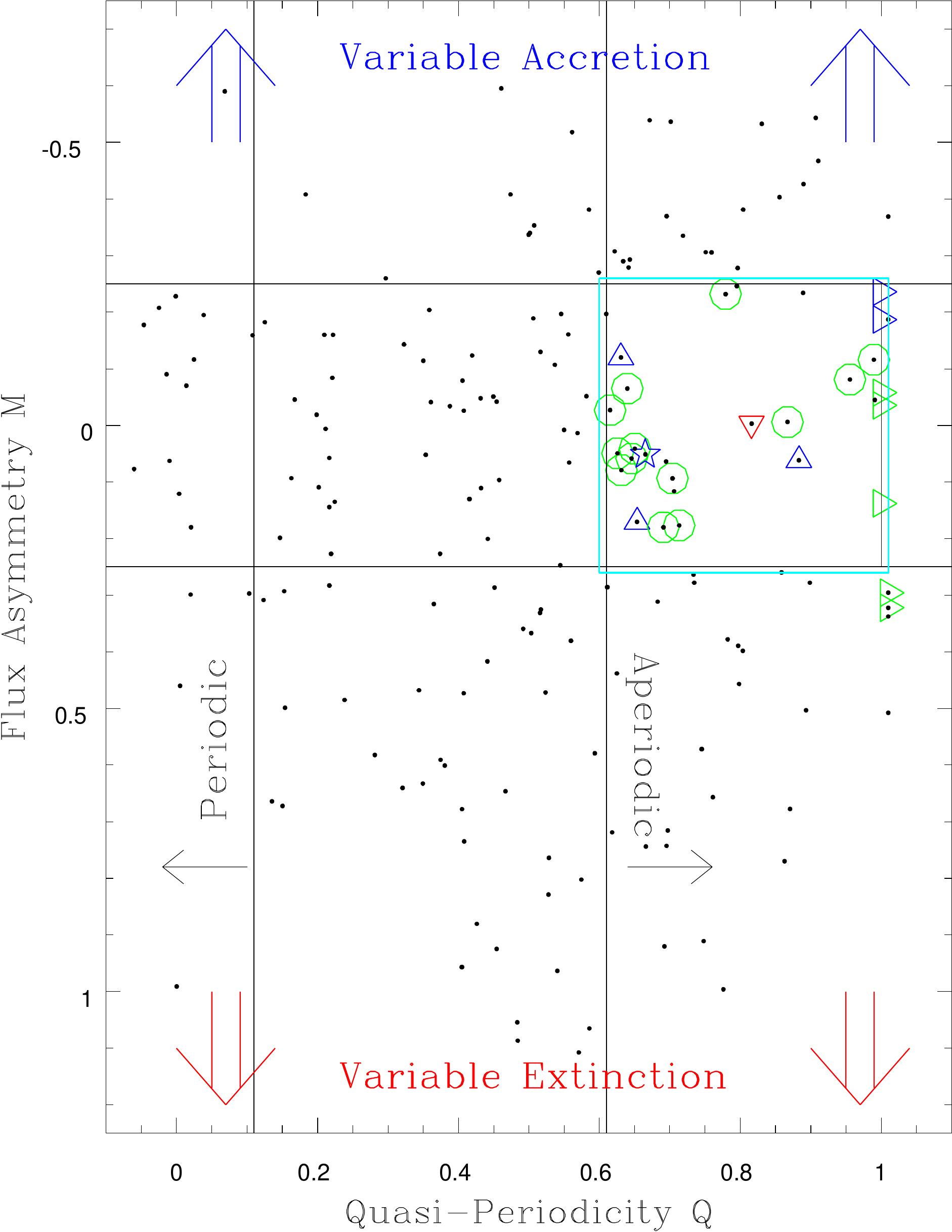}
\end{center}
\caption{Plot derived from the {\em CoRoT} light curves for CTTS in
NGC~2264 illustrating the quantitative light curve morphology
classification technique, and the location of the stars studied in 
this paper within that plot (the region inside the cyan box).  
Q $<$ 0.11 corresponds to periodic light
curves (typically reflective of rotational modulation of a stellar
photosphere mottled by relatively stable, cold spots); 0.11 $<$ Q $<$ 0.61
indicates quasi-periodicity (Cody et al.\ 2014) typically corresponding
to light curves showing periodic structure whose waveform evolves
significantly from period to period as often true for AA~Tau-analogs;
and Q $>$ 0.61 indicates stars with little or no periodic signature.
Blue triangles denote stars with accretion burst light curves
located within the stochastic region (cyan box); blue stars have light
curves with both accretion bursts and narrow flux dips;
the red triangle marks a star with a variable extinction light curve; and
green circles  and triangles are stars we classify as having stochastic
light curves (with the triangles being those with only limits on Q).
\label{fig:stochastic_stars.QM}}
\end{figure*}

Though we have not required the stars in
Table~\ref{tab:basicinformation} to have Q,M metrics that satisfy
the Cody et al.\ (2014) criteria, most of them do so, as is
illustrated in
Figure~\ref{fig:stochastic_stars.QM}.
Our sample includes two stars whose light curve 
metrics fall slightly
outside the nominal stochastic-class QM box boundary.
One of them (Mon-457) has {\em CoRoT} light curves
in both 2008 and 2011; the 2008 light curve falls within the
QM box, whereas the 2011 light curve falls slightly below
the QM lower boundary.  The other star whose light curve is
slightly below the lower QM boundary is Mon-242; we include it
in Table~\ref{tab:basicinformation}
simply because its light curve morphology seems best
matched to this class.

Seven additional stars also satisfy the QM criteria for having stochastic
light curves, but we believe they are better classified as belonging
to other groups.  The light curves for these seven stars are
shown in  the Appendix in
Figures~\ref{fig:otherCorot}.
The stars in 
Figure~\ref{fig:otherCorot} include three stars (Mon-11, Mon-510
and Mon-996) which we 
classified as having accretion-burst dominated light curves and
included in Stauffer et al.\ (2014);
two additional accretion burst dominated stars (Mon-766
and Mon-1048), one star which we believe is best interpreted
as having a light curve dominated by a combination of accretion
bursts and flux dips due to variable extinction 
(Mon-1294), and one star (Mon-774) whose 2011 light curve we
attribute to aperiodic extinction dips.  Other interpretations
could be made for the 2011 Mon-774 light curve; our choice was
strongly influenced by the fact that its 2008 light curve was
clearly that of an AA~Tau analog (periodic, deep, broad flux
dips due to variable extinction) and that stars with AA~Tau-type
light curves often switch to aperiodic flux dip dominated light
curves at other epochs (McGinnis et al.\ 2015).

\begin{figure*}
\begin{center}
\epsfxsize=.99\columnwidth
\epsfbox{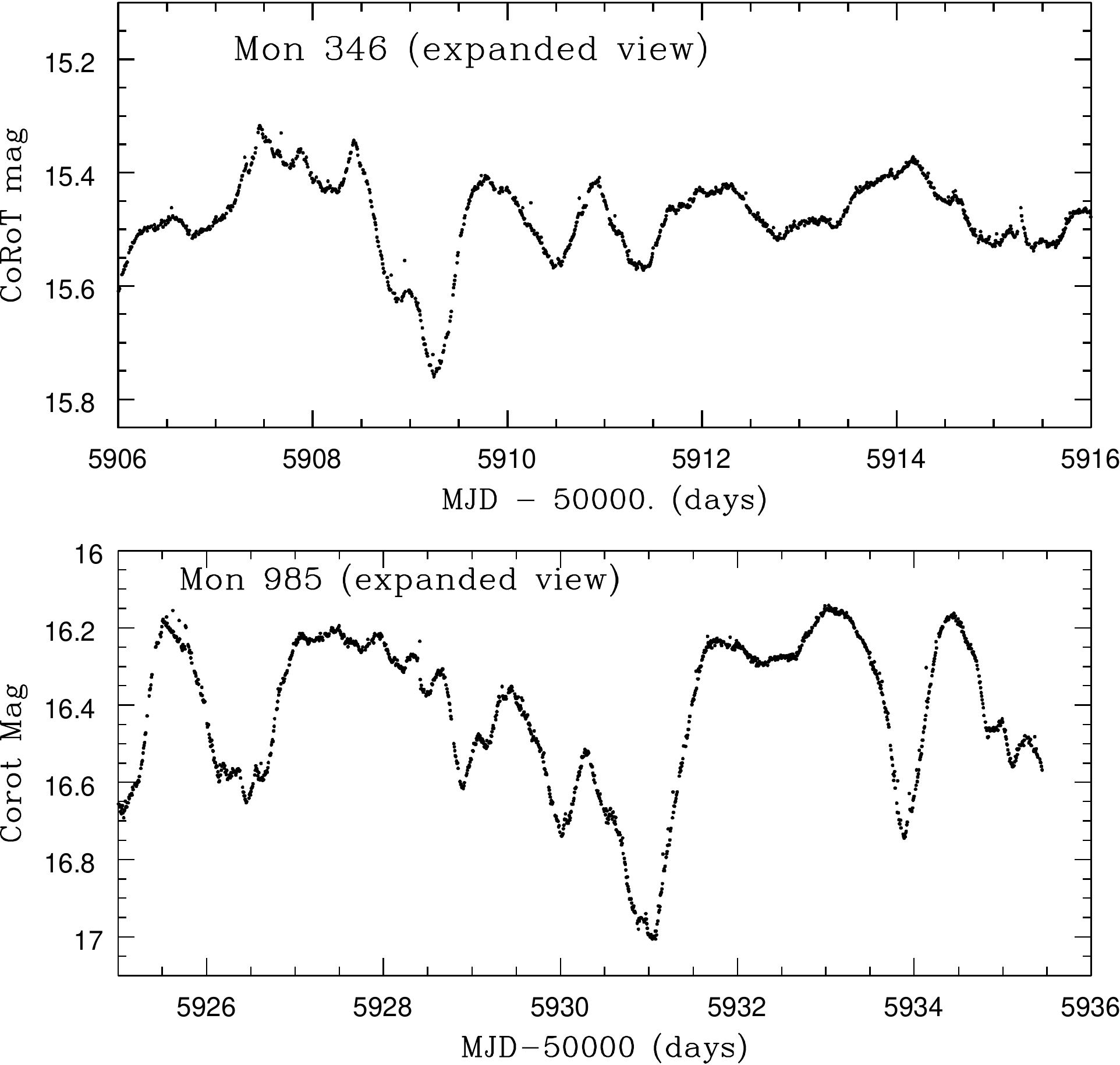}
\end{center}
\caption{{\em CoRoT} light curves for two members of the stochastic
light curve class, zooming in on a 10 day section of their data, in order
to illustrate better the similarity in morphology to other members
of the class with longer timescale variations.
\label{fig:stochastic_stars.expanded}}
\end{figure*}

While the Q and M statistics for the stars in 
Figure~\ref{fig:stochastic_stars.set1} and
Figure~\ref{fig:stochastic_stars.set2} assign
them to the same morphological light curve class, at first glance their light
curves nevertheless appear somewhat heterogeneous.  In significant part, that
is because the characteristic timescales for their variations differ
(see \S 5).  Specifically, some of the stars have quite short
characteristic timescales (e.g., Mon-346, 577, 985 and 6491), whereas
others have relatively long timescales (e.g., Mon-425, 457, 491).  In
order to better illustrate the similarity of their light curves, 
Figure~\ref{fig:stochastic_stars.expanded}
shows the light curves of two of the stars with short timescales
in an expanded view.  With the x-axis expanded roughly five times in scale,
the light curves for these two stars (Mon-346 and Mon-985)
look much more similar to the light curves for the long timescale
members of the class.

\begin{figure*}
\begin{center}
\epsscale{1.0}
\plottwo{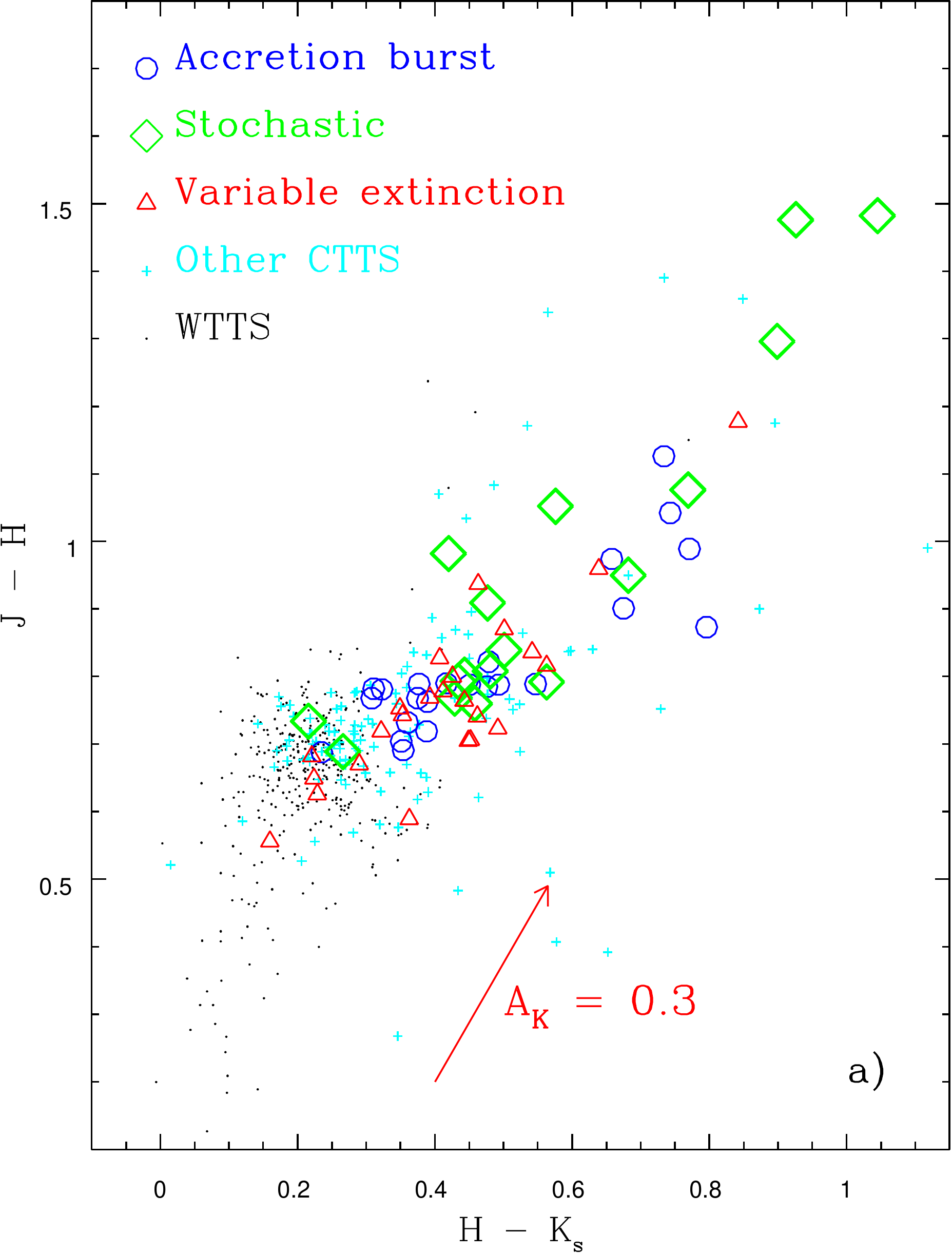}{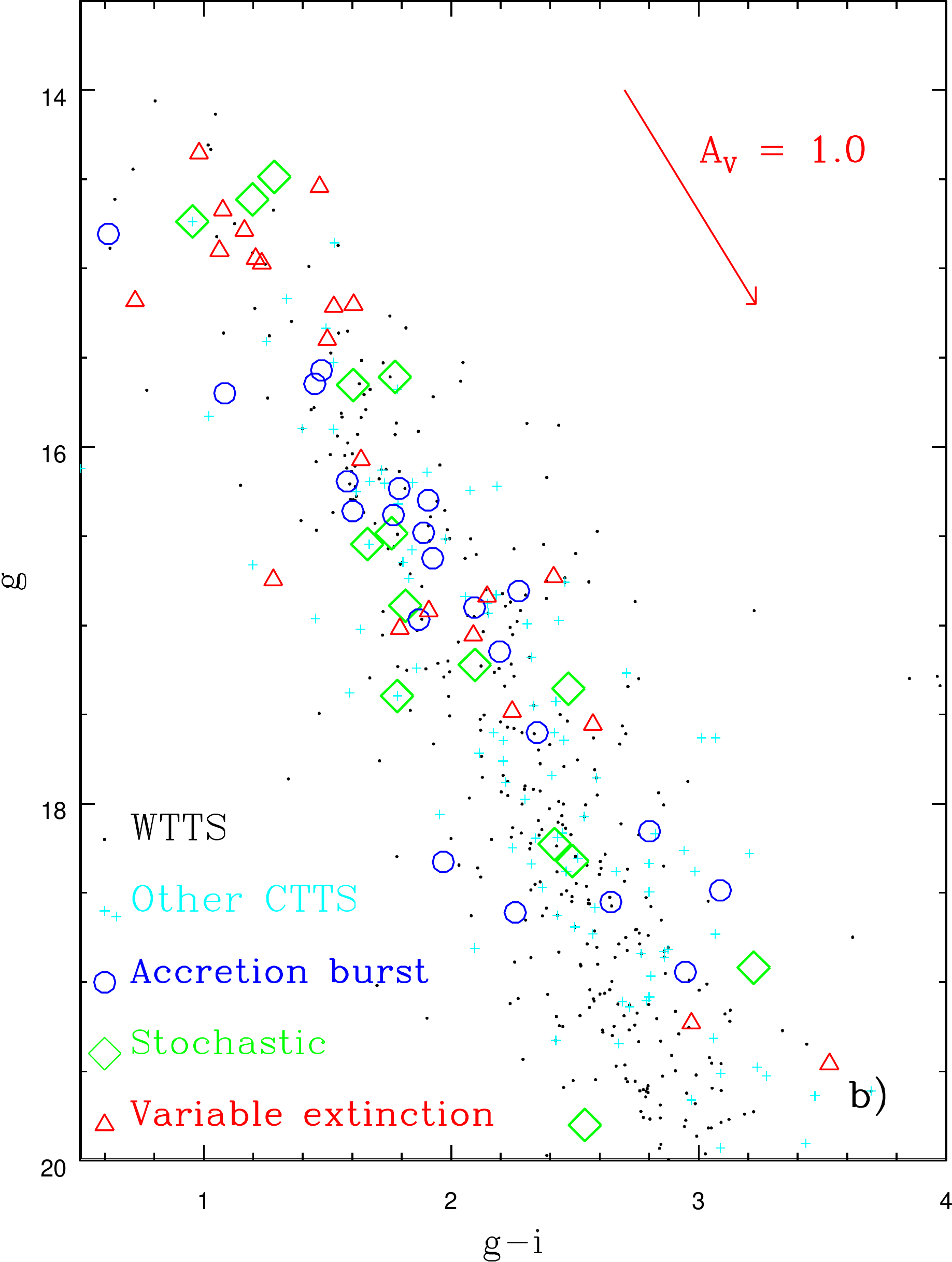}
\end{center}
\caption{a) Near-IR color-color diagram 
for stars with stochastic
{\em CoRoT} light curves, comparing those stars to CTTS whose light curve
morphologies we ascribe to variable extinction (AA~Tau and Ap classes in
McGinnis et al.\ 2015) and to stars whose light
curves show short-duration accretion bursts (Stauffer et al.\ 2014).  
Small black dots are WTTS
NGC~2264 members; small cyan plus signs are CTTS that are not members of the three light
curve classes highlighted in the plot.  
The stochastic light curve stars have locations in
this diagram that are not greatly different, on average, compared to
the other light curve classes, but are somewhat weighted to redder
colors (most extinction or more warm dust); and  
b) an optical color-magnitude diagram for the same set of stars.  This
diagram shows no obvious systematic age differences between
the stochastic light curve stars and the members of the other two
CTTS
light-curve classes or the WTTS.
\label{fig:optCMD_nearIRccd}}
\end{figure*}

\begin{figure*}
\begin{center}
\epsscale{1.0}
\plottwo{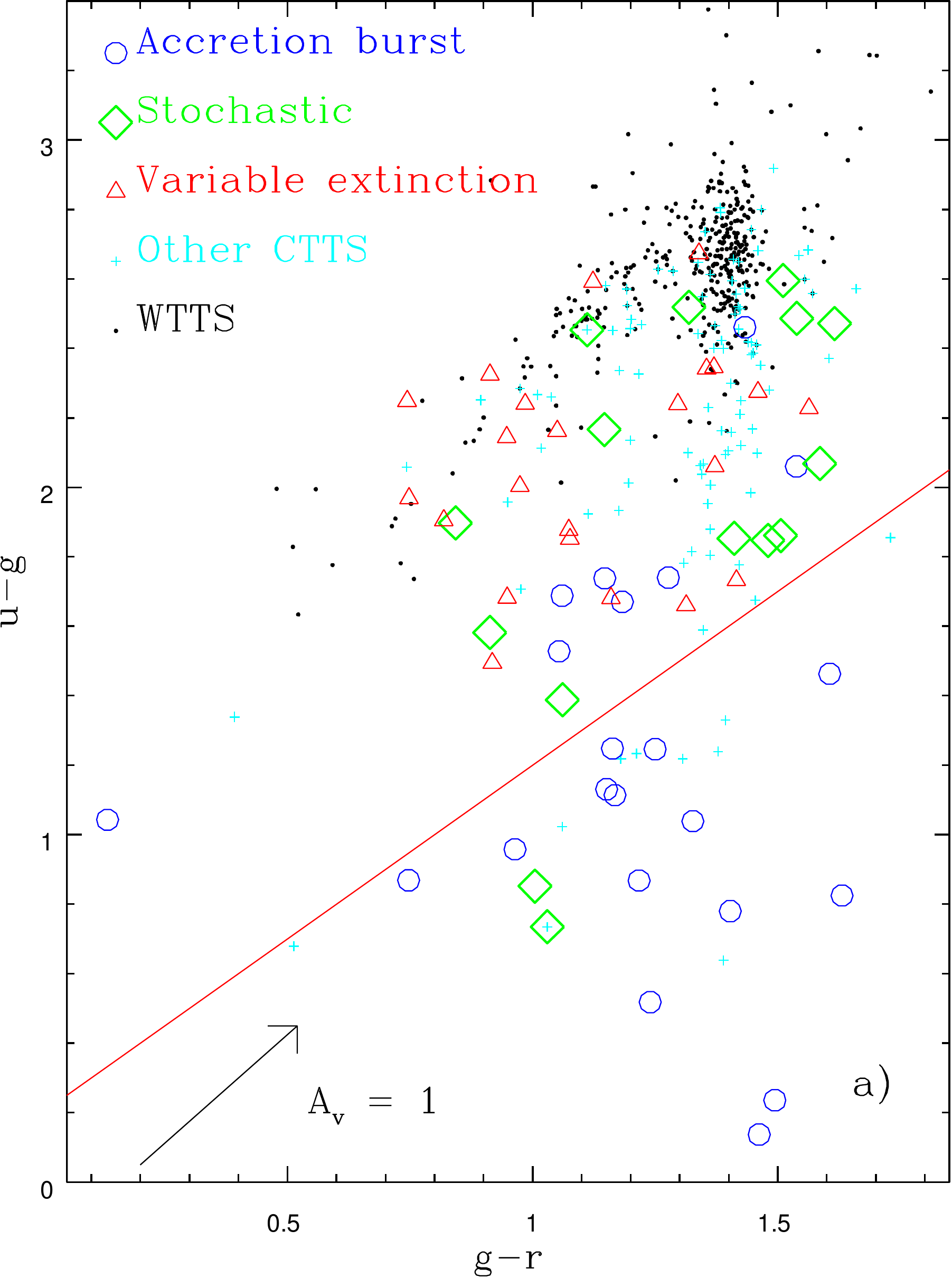}{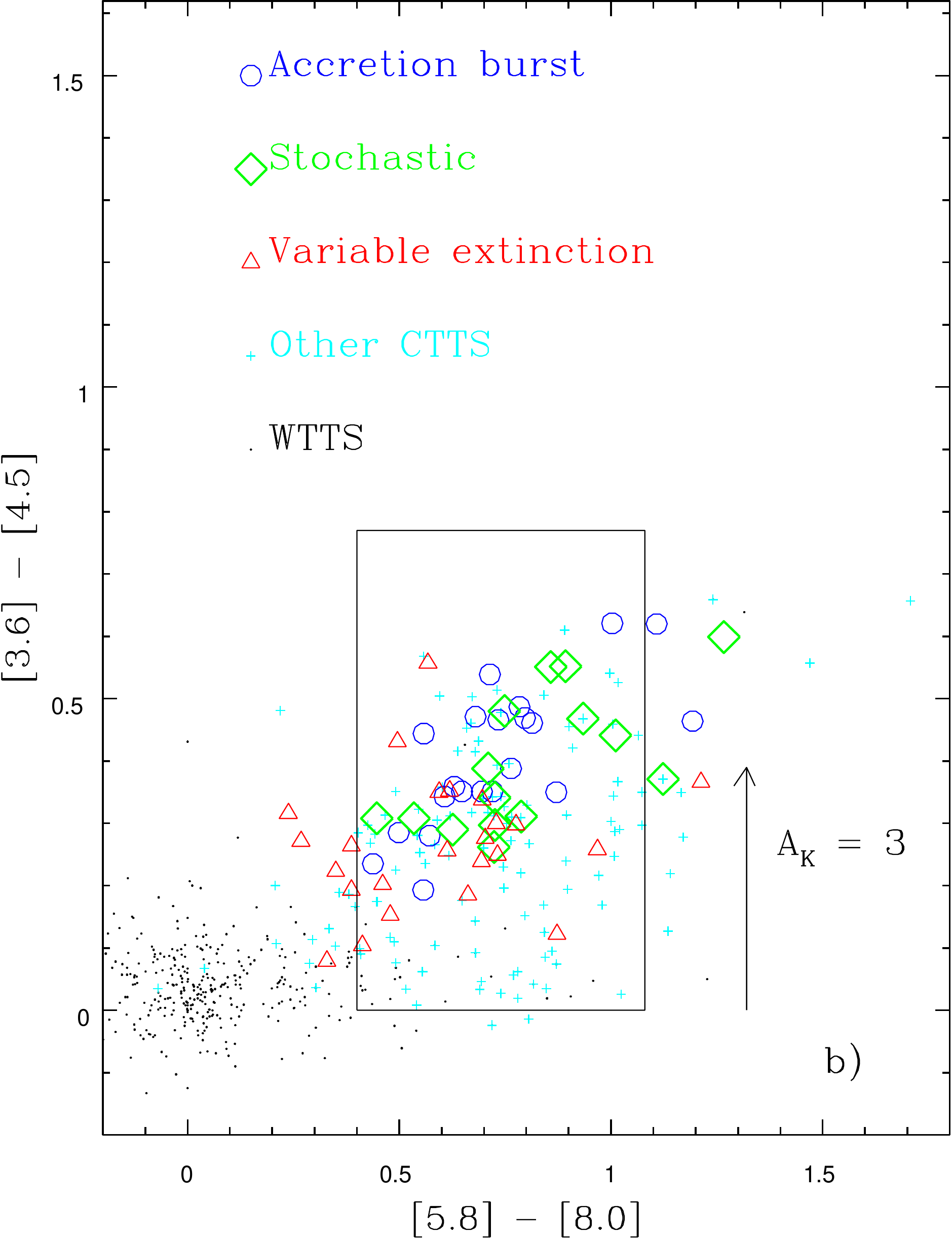}
\end{center}
\caption{a) $u-g$ vs.\ $g-r$ color-color diagram for stars with stochastic
{\em CoRoT} light curves, comparing those stars to CTTS whose light curves
we ascribe to variable extinction
and to stars whose light
curves show short-duration accretion bursts.  Small black dots are WTTS;
small cyan plus signs are CTTS that are not members of the three light curve classes
highlighted in the plot.  The diagonal
red line is a rough boundary between stars with weak and strong UV
excesses (Venuti et al.\ 2014).  Note that one of the stochastic stars - Mon-1267 -
falls outside the limits of hte plot at $g-r$ = 2.06, $u-g$ = 2.94.   The locus of
points for the stochastic light curve stars in this diagram is similar to
that for the variable extinction stars (and has much lower UV excesses on
average than the accretion burst stars);
and  b) mid-IR color-color diagram for the same set of stars.  The rectangular
box encloses the region dominated by Class II sources as defined in
Allen et al. (2003).  In this
diagram, the stochastic light curve stars have IR colors similar to those
for the accretion burst stars, and distinctly redder, on average, compared
to the stars in the variable extinction class.
\label{fig:uvexcess_irac-ccd}}
\end{figure*}

Before proceeding to our discussion of the {\em CoRoT} light curves for
the stochastic stars and the probable physical mechanism causing their
optical variability, it is useful to consider the physical properties of
these stars as derived from their mean observed colors and spectral types.
If these stars differ in some significant aspect from the other
CTTS in NGC~2264, that could help us interpret their photometric
variability.   Towards this end, we provide four diagnostic plots:
a $g$ vs.\ $(g-i)$ color-magnitude diagram; a $(u-g)$ vs.\ $(g-r)$ color-color
plot; a $(J-H)$ vs.\ $(H-K_s)$ near-IR color-color plot, and the IRAC
[3.6] $-$ [4.5] vs.\ [5.8] $-$ [8.0] color-color plot.   Our conclusions from
these plots and the data listed in 
Tables~\ref{tab:basicinformation} and
\ref{tab:basicphotometry} are:
\vskip0.01truein
\noindent
1) From the near-IR color-color diagram (Figure~\ref{fig:optCMD_nearIRccd}a),
we infer that on average the stochastic stars have relatively small
extinctions, comparable to, but probably on average somewhat larger than
those for the accretion burst and variable
extinction classes (a list of the accretion burst and variable extinction
stars plotted in these four figures and used elsewhere in the paper
is provided in Table~\ref{tab:sptypes}).
This agrees with extinction estimates derived from
combining the observed $(V-I)$ colors  (Sung et al.\ 2008),
published spectral types (Table~\ref{tab:basicinformation} and
Cody et al.\ 2014), and standard colors for young stars from
Pecaut \& Mamajek (2013), where 
the median $E(V-I)$ colors are 0.42, 0.05 and 0.15 for the stochastic,
accretion burst and variable-extinction classes, respectively.  The
derived $E(V-I)$ for the accretion burst stars is probably an underestimate
because their $(V-I)$ colors are significantly influenced by hot spots, while
the $E(V-I)$ colors of the variable extinction stars may be underestimated if
scattered light from the disk contributes significantly to their optical
flux.  The
three stochastic stars with the largest $(H-K_s)$ colors (Mon-457, 491,  and 985) 
also have the
largest disk to photosphere contrast ratios at 4.5 $\mu$m\ (see 
Table~\ref{tab:basicinformation}, column 7), demonstrating that their light
is disk-dominated in the IR.
\vskip0.01truein
\noindent
2) From the optical color-magnitude diagram (CMD) 
(Figure~\ref{fig:optCMD_nearIRccd}b), we conclude
that the stochastic stars are not, on average, displaced ``vertically" relative to the
other CTTS groups nor relative to the weak-lined T~Tauri (WTTS) sample.   
If the CMD can be used to
infer relative ages, this indicates no significant age difference between
these stars despite their quite different light curve morphologies.
As shown in
Table~\ref{tab:basicinformation}, the stochastic class includes YSOs with
spectral types spanning the entire G to mid-M range.  The stochastic and variable
extinction classes dominate the upper-left portion of the CMD, indicating they
probably include a larger fraction of high mass CTTS compared to the accretion
burst class.  Comparison of the spectral type distributions for the three classes
confirms this difference (see \S 5.2).
\vskip0.01truein
\noindent
3) Figure~\ref{fig:uvexcess_irac-ccd}a, the $(u-g)$ vs.\ $(g-r)$ diagram, has
been used as an accretion diagnostic in Venuti et al.\ (2014).  Young stars
without on-going accretion form an inverted V in this diagram, with early
G stars located near $g-r$ = 0.5, $u-g$ = 2.0, and
with both colors becoming redder to later spectral types until about M0 
(at $g-r \sim$ 1.4, $u-g \sim$ 2.8).  Later M spectral types retain 
$g-r \sim$ 1.4 but become bluer in $u-g$, presumably due to enhanced
chromospheric emission.   Reddening shifts stars towards the upper right
in the diagram; accretion primarily shifts stars down but also slightly
to the left.  The diagram shows that, on average, the stochastic stars have
smaller UV excesses than the stars with accretion-burst dominated light
curves, but comparable UV excesses as for the variable extinction sample.
A Student-T test yields a less than 10$^{-4}$ probability that the 
accretion burst stars are drawn from the same parent population as the other
two classes, but finds no significant difference between the stochastic
and variable extinction classes.
Also in support of these conclusions, the median H$\alpha$ equivalent widths for
the stochastic, accretion burst and variable extinction classes are 
-18 \AA, -68 \AA, and -16 \AA, respectively.
\vskip0.01truein
\noindent
4) Figure~\ref{fig:uvexcess_irac-ccd}b shows the IRAC two-color diagram
for the same set of stars.  All of the stochastic stars fall within the Class II
box defined in Allen et al.\ (2003) or slightly redward of that box.  This
suggests that their inner disks are similarly dusty, on average, as the
accretion-burst dominated class (which have a very similar distribution
of points
in this diagram).   The stochastic stars have redder IRAC colors, on
average, compared to the stars in the variable extinction light curve class (a
Student-T test yields a 1\%\ probability that the two classes have the
same parent population colors for each axis).
Comparison of the mean SED slopes from 2 to 8 $\mu$m yields a similar result, with
mean slopes of $-$1.08, $-$1.21, and $-$1.54 (and statistical uncertainties
of about 0.10 for each of these values) for the stochastic, accretion burst
and variable extinction classes, respectively.

\begin{deluxetable*}{lccccccccc}
\tabletypesize{\scriptsize}
\tablecolumns{12}
\tablewidth{0pt}
\tablecaption{Basic Information for YSOs with Stochastic Optical Light Curve Shapes\label{tab:basicinformation}}
\tablehead{
\colhead{Mon ID\tablenotemark{a}}  & \colhead{2MASS ID} &
\colhead{{\em CoRoT}\tablenotemark{c}} &
\colhead{SpT\tablenotemark{d}} & \colhead{H$\alpha$ EW
\tablenotemark{d}} & \colhead{FR4.5\tablenotemark{e}} &
 \colhead{{\em CoRoT} Ampl.\tablenotemark{f}} & \colhead{J$_{Stetson}$\tablenotemark{g}} &
\colhead{SED Slope\tablenotemark{h}} & \colhead{$\tau$$_{opt}$\tablenotemark{i}}  \\
\colhead{} & \colhead{} & \colhead{} &  
\colhead{} & \colhead{(\AA)} & \colhead{} & \colhead{(mag)} & \colhead{opt-IR} &
\colhead{} & \colhead{(days)} 
 }
\startdata
CSIMon-000119*  & 06412100+0933361 &  223985987 & K6  & -10.6  & 2.9 & 0.27 & 2.30 & -1.42 & 1.6 \\
CSIMon-000242   & 06411185+0926314 &  602079796 & K7  &  -5.4  & 0.4 & 0.14 & 1.16 & -1.39 & 5.3 \\
CSIMon-000346   & 06410908+0930090 &  603402478 & K7  & -27.   & 4.1 & 0.28 & 0.78 & -0.63 & 1.5 \\
CSIMon-000425   & 06411668+0929522 &  616943882 & K5  &  -6.3  & 2.4 & 0.30 & 1.31: & -1.02 & 1.1 \\
CSIMon-000457*+ & 06410673+0934459 &  616919789 & G6  & -49.4  & $>$10. & 0.18 & 0.97: & -0.18 & 3.7 \\
CSIMon-000491+  & 06405616+0936309 &  616895873 & K3  & -67.2  & 5.2 & 0.50 & 0.41: & -0.63 & 4.6 \\
CSIMon-000577   & 06414382+0940500 &  616895846 & K1  &  -5.9  & 2.9 & 0.10 & 0.72 & -1.55 & 1.3 \\
CSIMon-000613+  & 06410577+0948174 &  616849463 & K6.5& -30.1  & 2.1 & 0.14 & 0.92: & -1.51 & 1.4 \\
CSIMon-000636+  & 06404884+0943256 &  616872578 & M0  & -15.5  & 5.4 & 0.49 & 1.63 & -0.57 & 1.2 \\
CSIMon-000795*  & 06411257+0952311 &  500007115 & GVe & -35.3  & 2.5 & 0.22 & \nodata & -1.55  & 0.5 \\
CSIMon-000985   & 06401515+1001578 &  223969098 & G8* &  -9.1  & $>$10. & 0.51 & 1.99 & -1.12 & 0.7 \\
CSIMon-001054*+ & 06403652+0950456 &  400007538 & M2  & -21.1  & 2.0 & 0.15 & 0.53 & -1.29 & 0.8 \\
CSIMon-001061   & 06402416+0934124 & 616919835  & M0  & -33.0  & 3.5 & 0.33 & 1.99 & -1.06 & 0.8 \\
CSIMon-001234+  & 06401113+0938059 & 223968039  & K6  & -52.9  & 1.4 & 0.60 & 1.55 & -1.39 & 1.2 \\
CSIMon-001267   & 06403431+0925171 & 616944066  & G5* & -15.0  & 5.4 & 0.25 & \nodata & -0.47  & 1.1 \\
CSIMon-006491   & 06392550+0931394 & 616920065  & K4* &  -6.2  & 1.7 & 0.29 & 1.64 & -1.52 & 0.5 \\
CSIMon-006930   & 06390409+0916139 & 223948224  & \nodata  & \nodata  & \nodata & 0.18 & \nodata & \nodata & 0.4 \\
\enddata
\tablenotetext{a}{The CSIMon IDs are our internal naming scheme for stars in
the field of NGC~2264 -- see Cody et al.\ (2014) and 
\url{http://irsa.ipac.caltech.edu/data/SPITZER/CSI2264}.  
In the main text of this paper, we omit the ``"CSI" and the
leading zeros in the object names for brevity.  * = synoptic VLT spectra from 2011 available;
+ = synoptic VLT spectra from 2013 available.}
\tablenotetext{c}{{\em CoRoT} identification number.  Three stars were observed in both 2008 and 2011, but
had different ID numbers in 2008 - the 2008 {\em CoRoT} ID number for those three stars are
500007221 (Mon-242), 500007369 (Mon-457), and 500007548 (Mon-1267).  }
\tablenotetext{d}{Spectral types marked with an asterisk have their types estimated
from the dereddened CFHT photometry and the Pecaut \& Mamajek (2013) young star relation
between effective temperature and spectral type; see Venuti et al.\ (2014).  The 
H$\alpha$ equivalent widths for these stars come from our spectra.  All the other spectral types and
H$\alpha$ equivalent widths are from Dahm \& Simon (2005), except for Mon-636 where the
spectral type is from Makidon et al.\ (2004).}
\tablenotetext{e}{FR4.5 is an estimate of the ratio of the flux from the disk to the
flux from the stellar photosphere at 4.5 $\mu$m, based on all the available broad-band
photometry and the published spectral types.  Because the SED models do not include
veiling, these values are likely lower limits.}
\tablenotetext{f}{The 10\% to 90\% amplitude of the CoRoT light curves.  When we
have data for both 2008 and 2011, the CoRoT amplitude corresponds to the epoch 
plotted in Figure~\ref{fig:stochastic_stars.set1} and
Figure~\ref{fig:stochastic_stars.set2}.}
\tablenotetext{g}{Stetson index between the CoRoT 2011 light curve and the IRAC [4.5] light
curve.  A colon following the value indicates we only have a small number of IRAC
photometry points, making the value uncertain.}
\tablenotetext{h}{Spectral energy distribution (SED) slope from 2 to 8 microns, derived
as in Rebull et al.\ (2014).}
\tablenotetext{i}{Optical light curve timescale, as discussed in \S5.4.}

\end{deluxetable*}

\begin{deluxetable*}{ccccccccccccc}
\tabletypesize{\scriptsize}
\tablecolumns{10}
\tablewidth{0pt}
\tablecaption{Photometric Data for the Stars of
Table~\ref{tab:basicinformation}\label{tab:basicphotometry}}
\tablehead{
\colhead{Mon ID}  & \colhead{$u$} & \colhead{$g$} & \colhead{$r$} &
\colhead{$i$} & \colhead{$J$} & \colhead{$H$} & \colhead{$K_s$} &
\colhead{[3.6]\tablenotemark{a}} &
\colhead{[4.5]\tablenotemark{a}} & \colhead{[5.8]} & \colhead{[8.0]} & \colhead{[24]}
}
\startdata
CSIMon-000119 & 17.819 & 15.653 & 14.507 & 14.049 & 12.433 & 11.631 & 11.188 & 10.451 & 10.110 & 9.885 & 9.159 & 6.308 \\
CSIMon-000242 & \nodata & \nodata & \nodata & \nodata & 12.210 & 11.521 & 11.255 & 10.823 & 10.700 & 9.572 & \nodata & \nodata \\
CSIMon-000346 & 19.207 & 17.354 & 15.943 & 14.880 & 12.945 & 11.963 & 11.543 & 10.266 & 9.786 & 9.198 & 8.449 & \nodata \\
CSIMon-000425 & 17.190 & 14.737 & 13.626 & 13.782 & 11.268 & 10.500 & 10.071 & 9.327 & 8.956 & 8.553 & 7.430 & 4.306 \\
CSIMon-000457 & 19.016 & 16.545 & 14.929 & 14.884 & 11.761 & 10.279 & 9.234 & 7.949 & 7.350 & 6.700 & 5.433 & 1.608 \\
CSIMon-000491 & 18.346 & 16.485 & 14.979 & 14.726 & 11.892 & 10.816 & 10.047 & 8.941 & 8.500 & 7.921 & 6.910 & 3.969 \\
CSIMon-000577 & 16.510 & 14.613 & 13.770 & 13.416 & 12.047 & 11.255 & 10.692 & 10.072 & 9.810 & 9.566 & 8.842 & 5.951 \\
CSIMon-000613 & 16.996 & 15.608 & 14.547 & 13.835 & 12.020 & 11.260 & 10.802 & 10.135 & 9.838 & 9.617 & 8.891 & 6.076 \\
CSIMon-000636 & 20.389 & 18.321 & 16.735 & 15.832 & 13.334 & 12.282 & 11.706 & 10.476 & 9.924 & 9.402 & 8.509 & 6.306 \\
CSIMon-000795 & 16.069 & 14.487 & 13.574 & 13.202 & 11.494 & 10.686 & 10.205 &  9.364 & 9.056 & 8.844 & 8.397 & 5.950 \\
CSIMon-000985 & 21.404 & 18.918 & 17.379 & 15.696 & 14.548 & 13.252 & 12.353 & 10.933 & 10.625 & 10.425 & 9.890 & 7.187 \\
CSIMon-001054 & 17.742 & 16.890 & 15.885 & 15.075 & 12.934 & 12.141 & 11.707 & 11.066 & 10.678 & 10.258 & 9.548 & 5.866 \\
CSIMon-001061 & 18.131 & 17.396 & 16.366 & 15.614 & 13.471 & 12.521 & 11.839 & 11.116 & 10.648 & 10.250 & 9.316 & 6.445 \\
CSIMon-001234 & 19.672 & 17.077 & 15.566 & \nodata & 12.955 & 12.116 & 11.615 & 10.907 & 10.596 & 10.318 & 9.530 & 6.650 \\
CSIMon-001267 & 22.738 & 19.800 & 17.740 & 17.260 & 14.651 & 13.175 & 12.249 & 10.940 & 10.389 &  9.790 & 8.932 & 6.329 \\ 
CSIMon-006491 & 20.073 & 18.226 & 16.746 & 15.808 & 13.770 & 12.861 & 12.384 & 11.731 & 11.441 & 11.151 & 10.525 & 7.862 \\
CSIMon-006930 & 19.739 & 17.221 & 15.902 & 15.125 & 13.627 & 12.893 & 12.677 & [12.36] & [12.16] & \nodata & \nodata & \nodata \\
\enddata 
\tablecomments{Broadband photometry for the stars from
Table~\ref{tab:basicinformation}, in AB magnitudes for $ugri$ but in
Vega magnitudes for longer wavelengths. The $ugri$ data are from CFHT,
as reported in Venuti et al.\ (2014); the $JHK_s$ data are from the
on-line 2MASS all-sky point source catalog; the IRAC data are from
Sung et al.\ (2009), or from our own analysis of archival IRAC
imaging.  Typical photon-noise and calibration uncertainties for these
magnitudes are of order 0.02 mag; however, because all of these stars
are photometric variables with amplitudes up to several tenths of a
magnitude, these single-epoch data could differ from absolute,
time-averaged values by 0.1 mag or more.}
\tablenotetext{a}{Magnitudes shown within brackets are from the AllWISE catalog for channels
W1 and W2, which are close to but not exactly matched in wavelength to the corresponding
IRAC channels.}
\end{deluxetable*}

\section{Variable Accretion as the Physical Mechanism to Explain the Stochastic Light
   Curves} 

In the Herbst et al.\ (1994) taxonomy, stars in their Type II light curve
category would generally map into Cody's stochastic class.   Herbst et al.\
advocated that the most probable physical mechanism to explain their Type II
light curves was variable hot spots possibly accompanied by rotational modulation
bringing those hot spots into or out of the visible hemisphere.  We agree
with that assessment.  As noted by Herbst et al., if this is the case, the
light curve modulation should be accompanied by changes in spectral veiling,
resulting in photospheric absorption lines having smaller equivalent widths when
the hot spot contribution to the optical light is greater.   For the 
stochastic stars for which we have synoptic VLT spectroscopy, we will
show that they do exhibit variable veiling consistent with the amplitude
of the optical variability shown in the CoRoT light curves.

The linkage of changes in the spectral veiling to the optical light curve
morphology is made difficult because most of our spectroscopy was obtained
after the CoRoT photometric campaigns.  Also, in some cases, the only
photospheric absorption line we can measure accurately is the lithium 6708
doublet, whose equivalent width could be affected by other processes 
(Basri, Mart\'in \& Bertout 1991; Barrado et al.\ 2001; Baraffe \& Chabrier 2010).
Our analysis therefore proceeds as follows:
(a) for the brighter stars with synoptic VLT spectra, we show that their spectra do
show significant veiling variability as measured by several atomic absorption
features, and the variability shown by the lithium doublet is well-correlated
with the veiling changes;
(b) for all of the stochastic stars where we can measure variations in 
the lithium absorption strength, that variability correlates well with changes
in the strength of the \ion{He}{1} 6678 \AA\ emission line -- an accretion diagnostic;
(c) the inferred continuum level changes needed to explain the veiling
variability is consistent with the measured CoRoT light curve amplitudes; 
(d) the mean veiling of the stochastic stars is small, consistent with
their UV excesses; and
(e) the UV variability seen in the stochastic stars,
as measured by the slope shown in a delta($r$) vs. delta($u-r$) CMD is also best
explained by variable hot spots (see also Venuti et al.\ 2015b).

\subsection{Veiling and Accretion Variability}

\begin{figure*}
\includegraphics[width=6.0in]{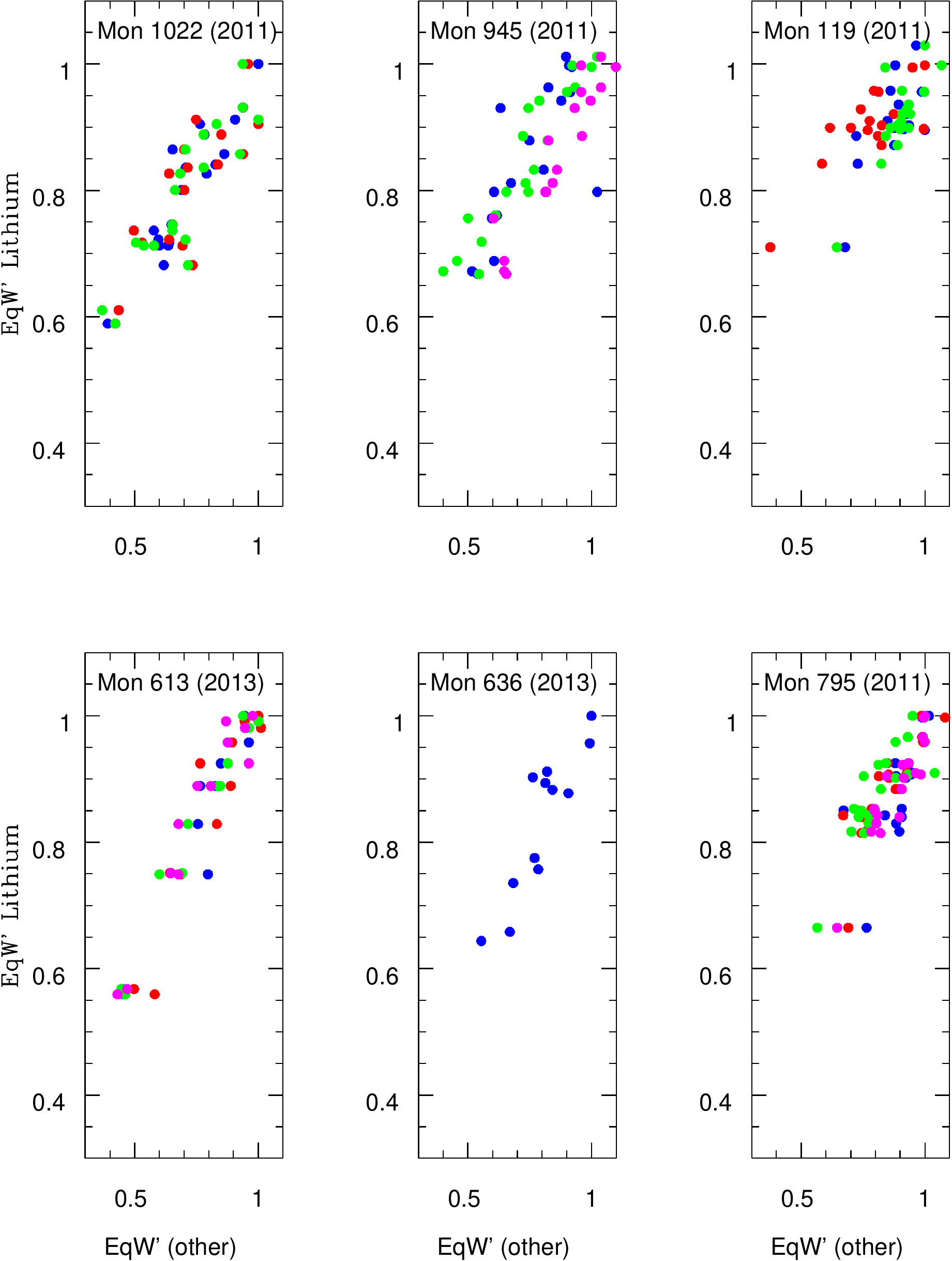}
\caption{Correlation between the equivalent width for the \ion{Li}{1} 6708 \AA\
doublet and that of other absorption lines present in the same FLAMES echelle 
order, for the four stochastic stars with high enough S/N in their spectra
to measure these features accurately.  The equivalent widths at each epoch have
been divided by the maximum equivalent width for that feature, in order to  allow
data for several features to be over-plotted.
Also shown for comparison are
the same plot for Mon-1022 and Mon-945, NGC~2264 members whose light curves are dominated
by accretion bursts (Stauffer et al.\ 2014).  Each color corresponds to
a different absorption feature (blue:  6463$\AA$, CaI+FeI; red: 6492-6500$\AA$, a
blend of CaI, FeI and FeII lines; and green: 6594$\AA$, an FeI doublet).  The good correlation 
and large range in equivalent widths shown for these stars
shows that their spectra have significant contribution from a light
source with relatively smooth continuum and variable strength, 
best explained as due to hot spots.
\label{fig:stochastic_stars.veiling}}
\end{figure*}

For four of the stars with stochastic light curves, we have synoptic VLT
spectra of good enough quality to accurately measure equivalent widths for
\ion{Li}{1} 6708 and at least one other photospheric absorption line.  These observations
show that there are large variations in the absorption line equivalent
widths (veiling variability) over the time period of the VLT campaigns and
that the lithium strength variations track well the variations in the other
absorption features. Figure~\ref{fig:stochastic_stars.veiling} shows the 
equivalent width data for these four stars, along with similar data for two
of the accretion burst stars.  The veiling variability is a natural
consequence of a varying hot spot contribution to the star's continuum
flux and is not expected for any of the other mechanisms that are known to drive
significant photometric variability in low mass YSOs (e.g., cold spots or
variable extinction).

\begin{figure*}
\includegraphics[width=6.0in]{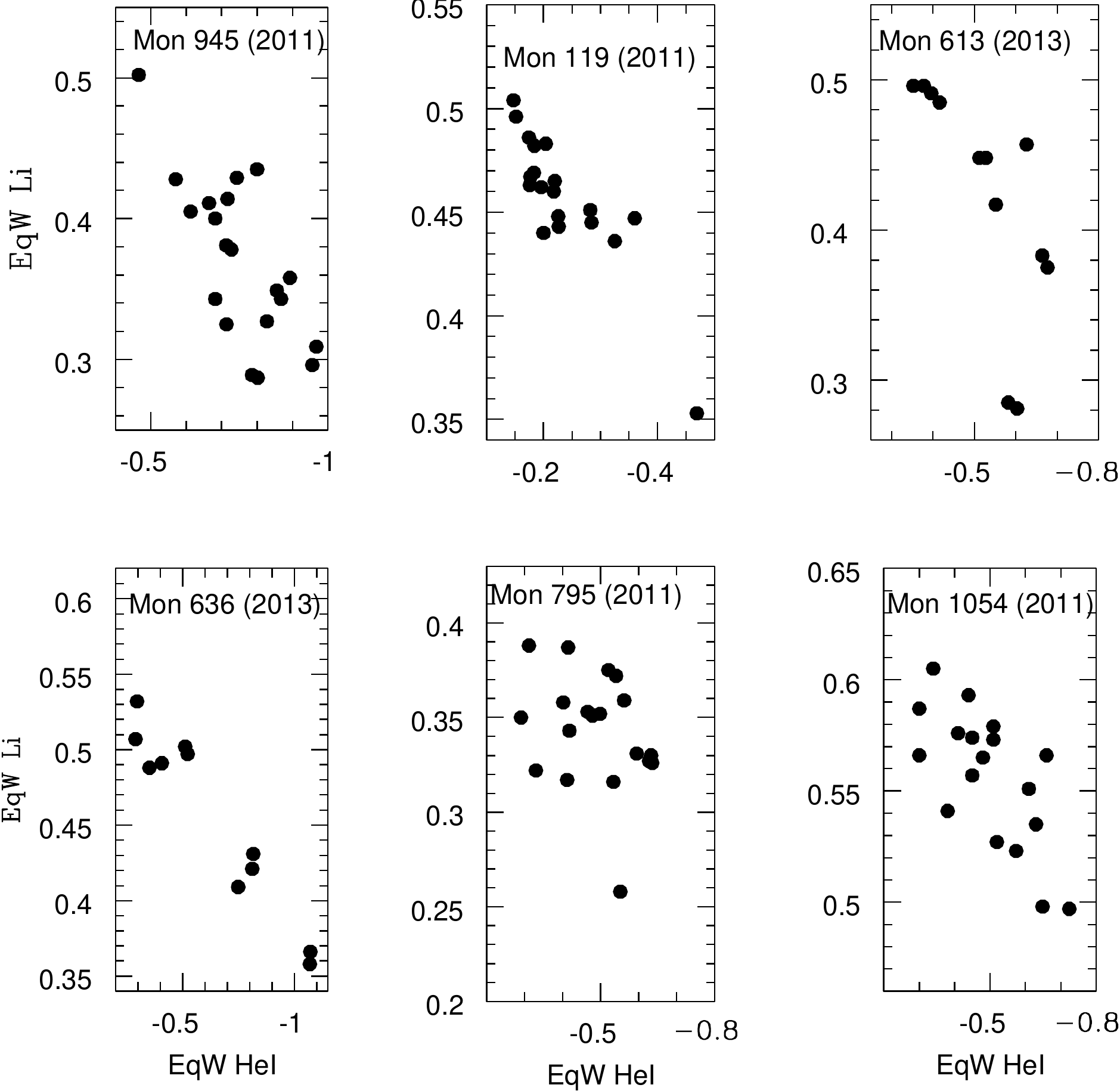}
\caption{Correlation between \ion{Li}{1} 6708 equivalent width and \ion{He}{1} 6678 
equivalent width for one star with an accretion-burst dominated
light curve (Mon-945) and
the five stars with stochastic light curves for
which the \ion{He}{1} line is significantly in emission for at least some
of the spectra.  The generally good correlation implies that the
continuum variations observed for these stars are primarily due to
variations in the accretion rate.  The equivalent width for
\ion{He}{1} 6678 has been corrected for the influence of an FeI absorption
line at almost exactly the same wavelength by subtracting the
expected equivalent width (as a function of spectral type) for the
FeI line from the measured \ion{He}{1} equivalent width -- see 
Appendix \S A2 for more details.
\label{fig:stochastic_stars.liHeI}}
\end{figure*}

For the four stars above, plus one additional stochastic star, 
we have synoptic VLT
spectra from which we can accurately measure the \ion{Li}{1} 6708 equivalent
width and for which the \ion{He}{1} 6678 line is sometimes in emission.  The \ion{He}{1}
6678 line is present in high-resolution spectra for many CTTS, sometimes as
a broad feature possibly associated with a  hot wind (Beristain, Edwards
and Kwan 2001), and sometimes as a narrow (FWHM $\sim$ 36 \kms), 
approximately Gaussian feature,
usually ascribed to hot gas created in the shock where the accretion flow
impacts the stellar photosphere (Johns-Krull et al.\ 2013).  Our
stochastic light curve stars with VLT spectra only show the narrow \ion{He}{1}
emission feature.   Therefore, we would expect stronger \ion{He}{1} emission to
correlate with stronger accretion flux, and the \ion{He}{1} emission and \ion{Li}{1} 6708 absorption
strengths should correlate for stars where accretion
variability is strong.  Figure~\ref{fig:stochastic_stars.liHeI} shows
that this correlation is indeed present for all five stars with stochastic
light curves where we have the data to make this test (the figure also includes
data for Mon-945, a star with an accretion-burst dominated light curve, to illustrate
how this diagram should appear if accretion drives the variability).

\begin{figure*}
\begin{center}
\epsfxsize=.99\columnwidth
\epsfbox{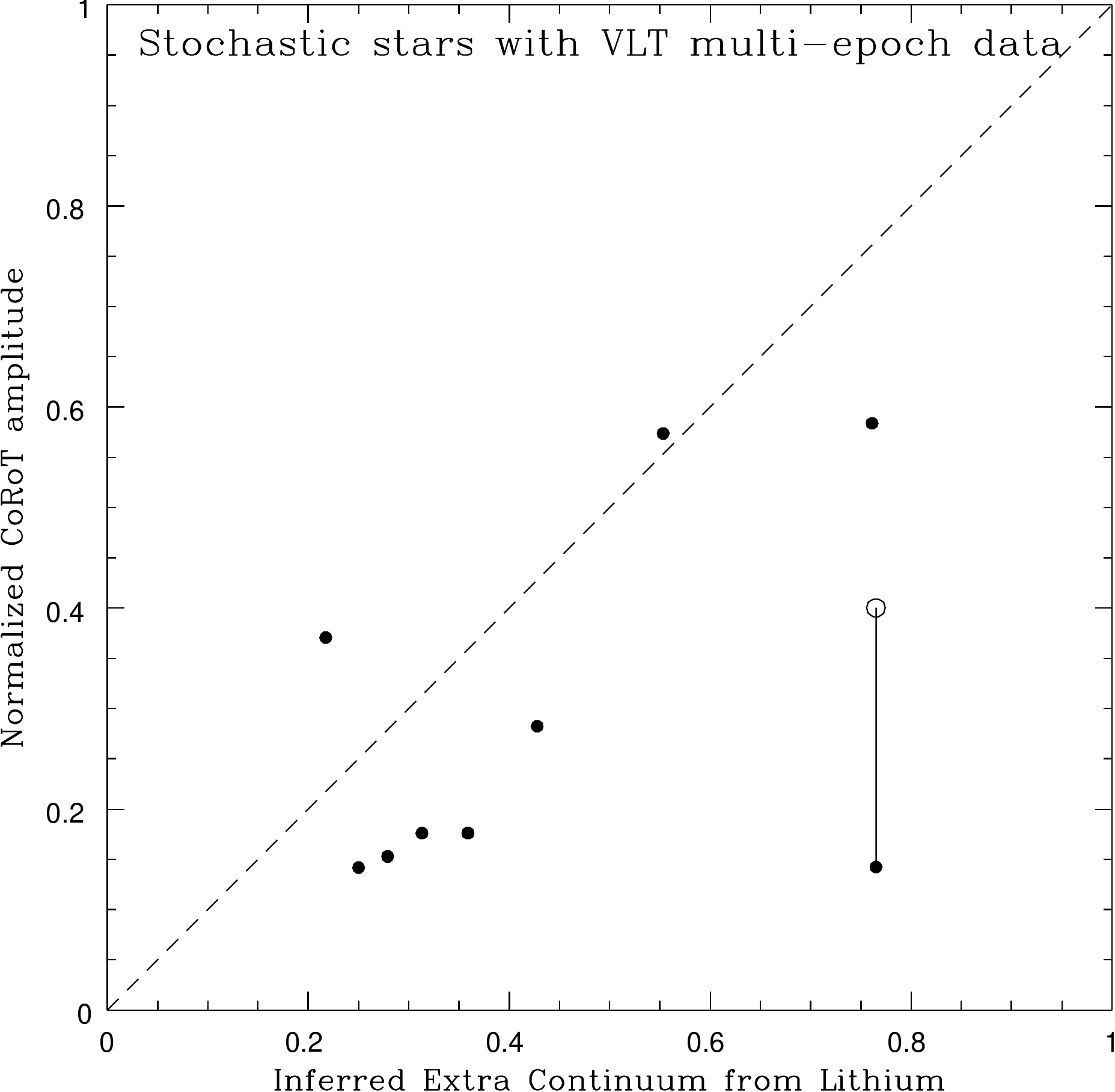}
\end{center}
\caption{Comparison between the inferred range in the continuum flux at
6700 \AA\ derived from our multi-epoch VLT spectra and the observed
light curve amplitudes from our {\em CoRoT} data.  The y-axis is the 
amplitude (in count rate) of the {\em CoRoT} light curve divided by
the mean count rate.  The dashed line indicates
a one-to-one correlation.  The open circle marks the light curve
amplitude derived from our USNO data for a star that we believe had low
variability during the {\em CoRoT} campaign but became significantly more
variable during the following month, as indicated by our USNO and VLT data.  
We interpret the data shown here to be consistent with
most of the optical variability for the stochastic stars as being a
result of variable accretion onto their photospheres.
\label{fig:stochastic_stars.fluxvar}}
\end{figure*}

Figures~\ref{fig:stochastic_stars.veiling} and 
\ref{fig:stochastic_stars.liHeI}
prove that at least these specific members of the stochastic
light curve class have optical brightnesses that vary with time due
to variable accretion.  However, this does not necessarily require that
their light curve shapes and amplitudes are primarily due to variable
accretion luminosity -- other mechanisms (such as variable extinction)
could cause optical brightness variations that in principle could be
larger than the accretion-driven variations.  If our spectroscopy and
photometry were simultaneous, we could directly compare the continuum
flux variations inferred from the spectroscopy with the light curve
amplitudes - however, our {\em CoRoT} data are from 2008 and 2011, whereas
the majority of our spectroscopy is from 2013.   Even for those stars where
we have 2011 multi-epoch spectroscopy, the overlap in time with the
{\em CoRoT} campaign is small, and we therefore cannot make a useful direct
comparison.   However, we can make a statistical test -- is the inferred
range in accretion flux at 6700 \AA\ derived from the veiling
data (mostly from 2013)
consistent with the measured amplitude in the CoRoT light curves (mostly
from 2011)? 
Figure~\ref{fig:stochastic_stars.fluxvar} shows this comparison for the
 stochastic stars for which we have multi-epoch VLT data.   There is
considerable scatter in the plot, but the data are consistent with the
variable continuum inferred from the spectroscopy matching the variable
continuum measured in the light curves - supporting the conclusion that
variable accretion is the dominant cause of the optical 
variability.

Finally, having argued that the lithium equivalent widths for 
CTTS can be used as a relative veiling (and hence accretion) indicator,
we can also use the mean lithium equivalent widths for the stochastic
light curve stars to estimate their mean veiling levels.  In
Table~\ref{tab:other_data}, we provide mean lithium equivalent widths
for the stochastic light curve stars, either measured from our
multi-epoch VLT/FLAMES data, or in six cases, as measured from
single-epoch VLT/FLAMES spectra obtained by the ESO-Gaia project (Randich et al.\ 2013).
We also provide an estimated maximum photospheric lithium equivalent
width for that spectral type, based on a fit to the upper-envelope
of the lithium equivalent widths for all YSOs in NGC~2264 for which we have FLAMES 
spectra.  This allows us to derive an upper limit to the veiling
at 6700 \AA\ for each of these stochastic stars. 
It can be seen that the mean
lithium equivalent widths for the stochastic stars differ only 
slightly from their expected photospheric values, suggesting 
veilings of generally 10\% or less.  The median lithium veiling 
from Table~\ref{tab:other_data} for the stochastic stars is 0.17;
for reference, the median veiling for the accretion burst and
variable extinction classes are 0.36 and 0.17.
The average veiling we derive for the stochastic 
stars therefore agrees with the conclusion drawn from their UV excesses
(Figure~\ref{fig:uvexcess_irac-ccd}a) that stars of this light curve
class have low to moderate accretion rates (significantly lower than
stars with accretion burst dominated light curves), but similar 
to the average accretion rate for stars with variable extinction dominated light curves.

\begin{deluxetable*}{lcccccc}
\tabletypesize{\scriptsize}
\tablecolumns{7}
\tablewidth{0pt}
\tablecaption{Other Quantitative Data Used in This Paper\label{tab:other_data}}
\tablehead{
\colhead{Mon-ID}  & 
\colhead{$<$EqW Li$>$\tablenotemark{a}} & \colhead{Max-EqW-Li\tablenotemark{b}} &
\colhead{Q} & \colhead{M} & \vsini\tablenotemark{c} & \vsini\tablenotemark{d}   \\
\colhead{} & \colhead{} &  \colhead{(\AA)} &
\colhead{}  & \colhead{} & \colhead{\kms} & \colhead{\kms}
 }
\startdata
CSIMon-000119 &  0.48 & 0.52 & 0.62 & 0.05 & 21.0 & 24.8 \\
CSIMon-000242 &  0.59$^*$ & 0.53 & 1.18 & 0.32 & 14.9 & 18.5 \\
CSIMon-000346 &  0.47$^*$ & 0.53 & 0.95 & -0.08 & 21.5 & 18.2 \\
CSIMon-000425 &  0.43$^*$ & 0.50 & $\geq$1 & 0.14 & 16.9 & 16.5 \\
CSIMon-000457 &  0.25 & 0.33 & $\geq$1 & -0.03 & 42.0 & 48.4 \\
CSIMon-000491 &  0.32 & 0.46 & 0.99 & -0.12 & 72.0 & 67.9 \\
CSIMon-000577 &  (0.42) & 0.42 & 0.70 & 0.09 & 35.0 & 34.5 \\
CSIMon-000613 &  0.39 & 0.52 & 0.64 & -0.06 & 22.0 & 15.2 \\
CSIMon-000636 &  0.44 & 0.60 & 0.65 & 0.04 & 25.0 & 24.9 \\
CSIMon-000795 &  0.32 & 0.28 & 0.65 & 0.06 & 28.0 & 35.0 \\
CSIMon-000985 &  0.47$^*$ & 0.37 & 0.69 & 0.18 & 22.6 & 19.5 \\
CSIMon-001054 &  0.53 & 0.62 & 0.71 & 0.18 & 20.0 & 20.1 \\
CSIMon-001061 &  0.29 & 0.60 & 0.78 & -0.23 & ... & 19.9 \\
CSIMon-001234 &  0.43 & 0.52 & 0.61 & -0.03 & 24.0 & 15.9 \\
CSIMon-001267 &  ... & ... & $\geq$1 & -0.06 & ... & ... \\
CSIMon-006491 &  0.50$^*$ & 0.48 & 0.87 & -0.01 & 17.0 & 15.8 \\
CSIMon-006930 &  ... & ... & 0.63 & 0.08 & ... & ... \\
\enddata
\tablenotetext{a}{Lithium equivalent width averaged over all epochs for the
stars for which we have multi-epoch VLT spectra.  Where we do not have our
own spectra, the value is from measurement of the ESO-Gaia spectrum for this star,
and these values are marked with an asterisk.}
\tablenotetext{b}{Inferred photospheric lithium equivalent width
for a YSO of this star's spectral type  from a fit to the upper 
envelope of a plot of lithium equivalent width vs. spectral type for our NGC~2264
stars for which we have VLT spectra.}
\tablenotetext{c}{\vsini\ determined by comparing synthesized, rotationally
spun-up model spectra to VLT/FLAMES observed spectra, as described in
McGinnis et al.\ 2015.}
\tablenotetext{d}{\vsini\ determined by measuring the FWHM for the \ion{Li}{1} 6708\AA\
doublet and converting that to a \vsini\ using a calibration curve derived
from the McGinnis et al.\ values.  Where we have multi-epoch spectra, we
measured FWHM for all the spectra and took the mean FWHM as the input.}
\end{deluxetable*}

\subsection{Variable Accretion Inferred from Broad-Band Colors}

If the primary cause of the photometric variability of the stars
in Table~\ref{tab:basicinformation}
is variable accretion onto the stellar photosphere, it
should also be reflected in the broad-band photometric colors.  When the
accretion rate is high, the contribution to the optical flux from
hot spots should be greatest, and the star should be both brighter
and bluer.  When the accretion rate is lowest, the hot spots should
be at their weakest, and the star should be both fainter and
redder.

\begin{figure*}
\begin{center}
\epsfxsize=.99\columnwidth
\epsfbox{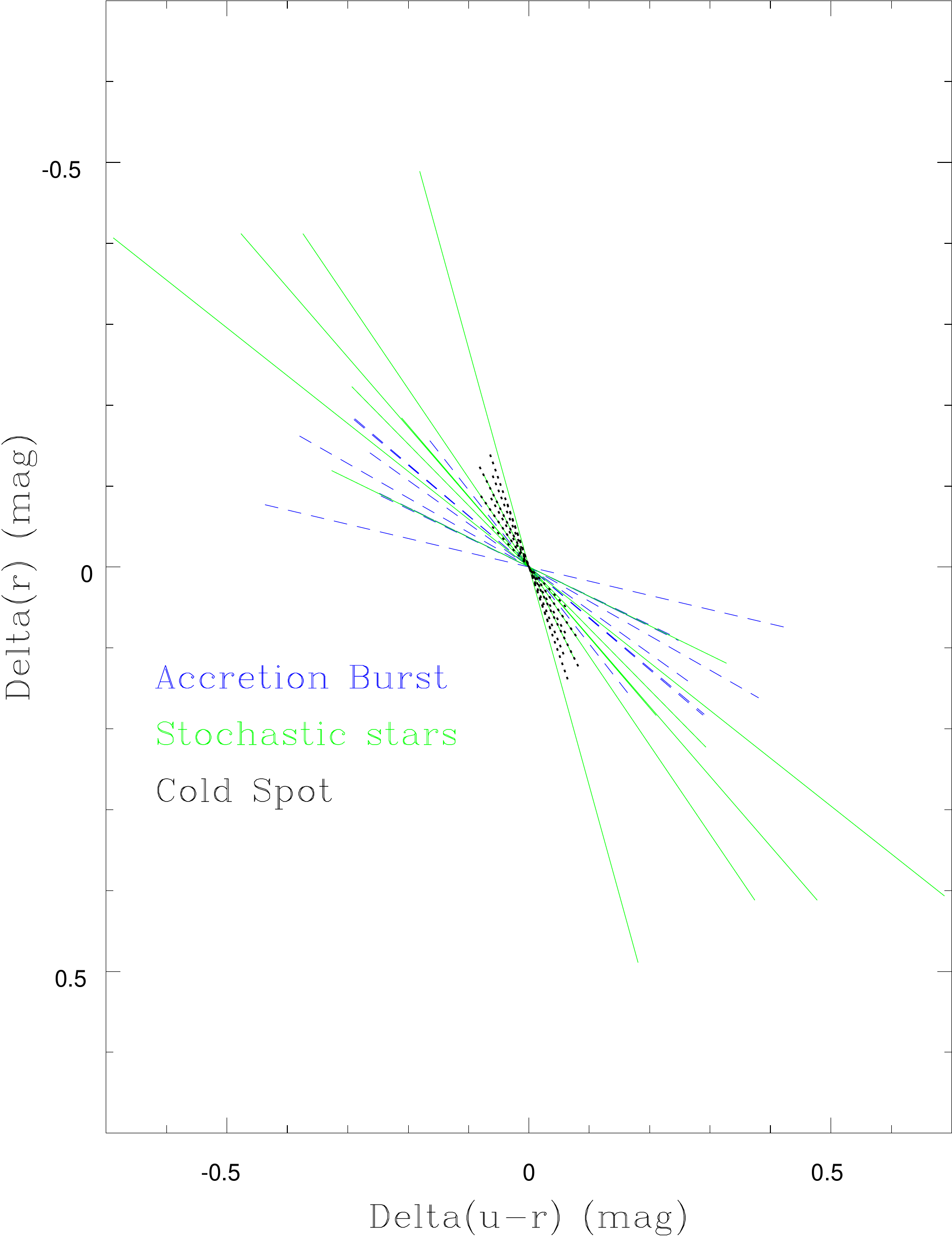}
\end{center}
\caption{
Comparison of the color-slopes ($\Delta r$/$\Delta(u-r)$) derived
from multi-epoch CFHT photometry of NGC~2264 for stars whose {\em CoRoT}
light curves have stochastic light curves to those that are either
dominated by short-duration accretion bursts or by cold photospheric
spots.  The stochastic stars show very similar slopes to the accretion
burst dominated sample, whereas the stars with cold spots have
systematically higher slopes (because the spots are very red, leading
to little change in $(u-r)$ color for a given change in $r$ magnitude).
We have plotted the data for the largest amplitude cold spot light curves 
in order to be able to derive as accurate a slope as possible.  Most stars
with cold-spot light curves have significantly smaller amplitudes.
\label{fig:stochastic_stars.color_var}}
\end{figure*}

As noted in Table~\ref{tab:synoptic_data}, we have multi-epoch CFHT $ugri$ photometry
for NGC~2264 obtained about a month after the 2011 {\em CoRoT}/{\em Spitzer}
campaign.  Venuti et al.\ (2015b) have described these data and used
the data to quantify the level of accretion variability for the CTTS
in NGC~2264.  We now use the same photometry here to test whether
accretion variability is the primary driver of the light curve morphology
we have ascribed to the stochastic variables in the {\em CoRoT} data.
As done by Venuti et al.\ (2015b), for each CTTS, we have plotted the
change in $r$ magnitude vs. the change in $(u-r)$ color; where those data
are well-correlated we have fit a straight-line to the distribution and
derived the slope ($\Delta r$/$\Delta (u-r)$).
Figure~\ref{fig:stochastic_stars.color_var}
shows a plot of these slopes for CTTS whose light curve
morphologies we have ascribed to short-duration accretion bursts, 
to cold spots, and to our stars with
stochastic light curves.  The figure shows that the stochastic light
curve stars have color slopes and amplitudes reasonably well-matched
to the short-duration
accretion burst stars, and qualitatively different from stars whose
light curves are dominated by cold spots.   Stars whose light curves
are dominated by variable extinction can have similar color slopes and
amplitudes to those for the stochastic stars (Venuti et al.\ 2015b), but
the light curve morphologies of the two classes are quite different.

\section{Additional Characterization of the Stochastic Light Curve Class}

\subsection{H$\alpha$ Profiles of Members of the Stochastic Class}

The shapes of emission line profiles for CTTS at least potentially
provide clues to their accretion process (Lima et al.\ 2010; Kurosawa
\& Romanova 2013).   In Stauffer et al.\ (2014), we showed that 
NGC~2264 CTTS with accretion-burst light
curves had H$\alpha$ profiles that were distinctly different, on
average, from those for the variable extinction light curve class.
Specifically, the H$\alpha$ profiles of the  accretion 
burst stars are much more often 
centrally peaked and lacking in distinct absorption features compared
to the variable extinction stars, while the variable extinction stars
much more frequently show H$\alpha$ absorption dips on the red side
of the emission profile.  These characteristics are explicable within
the theoretical models (Kurosawa \& Romanova 2013), 
with the centrally peaked profiles being a
prediction for stars with the highest accretion rates and the redward
displaced absorption dips being a consequence of our line of sight
intersecting warm gas that is accreting onto the stellar photosphere.

We have good multi-epoch VLT spectra with well-defined H$\alpha$ 
profiles for eight of the stars in Table~\ref{tab:basicinformation};
for another five stars,
we have good single epoch H$\alpha$ profiles.   

\begin{figure*}
\begin{center}
\epsscale{1.0}
\plottwo{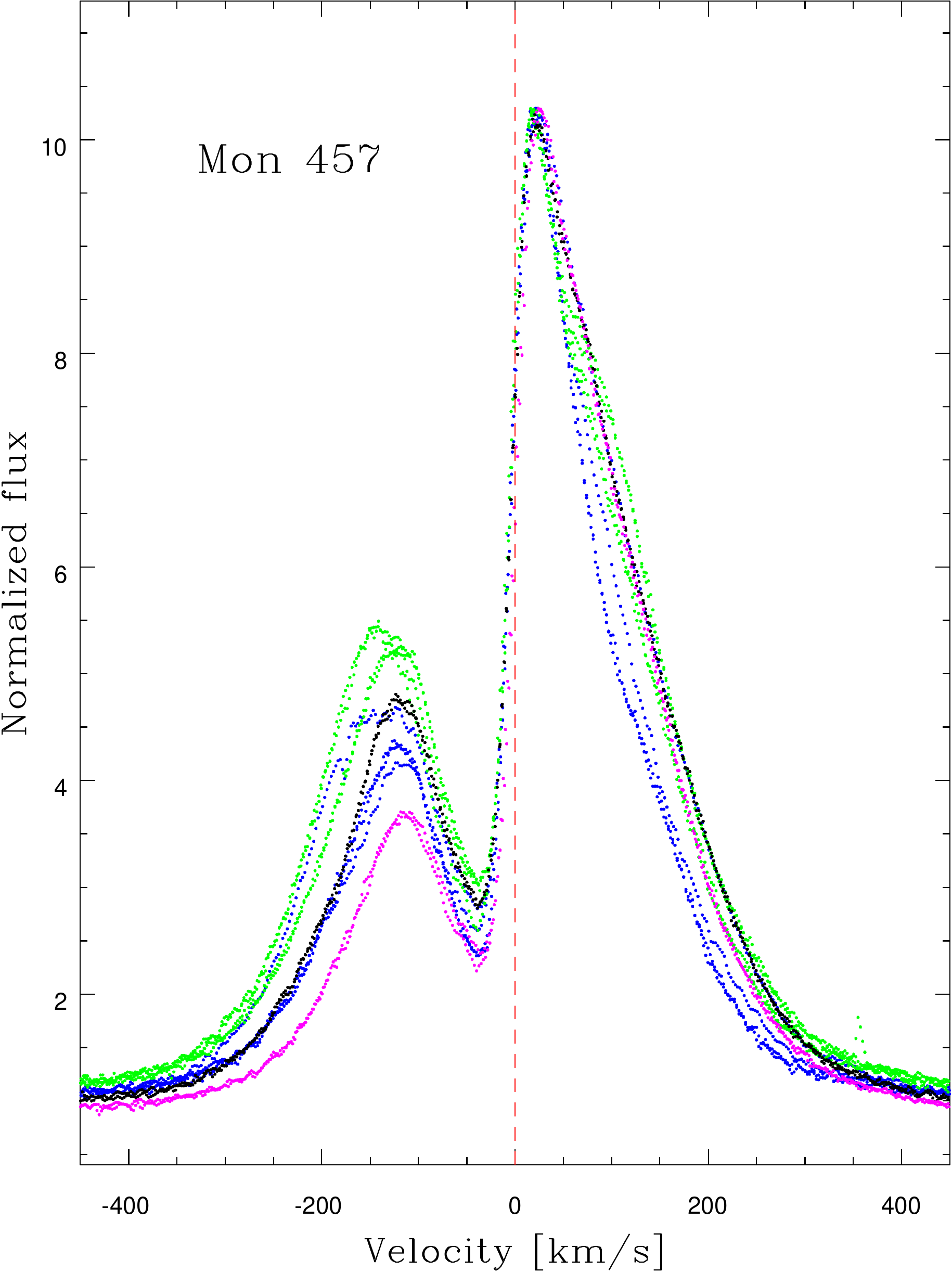}{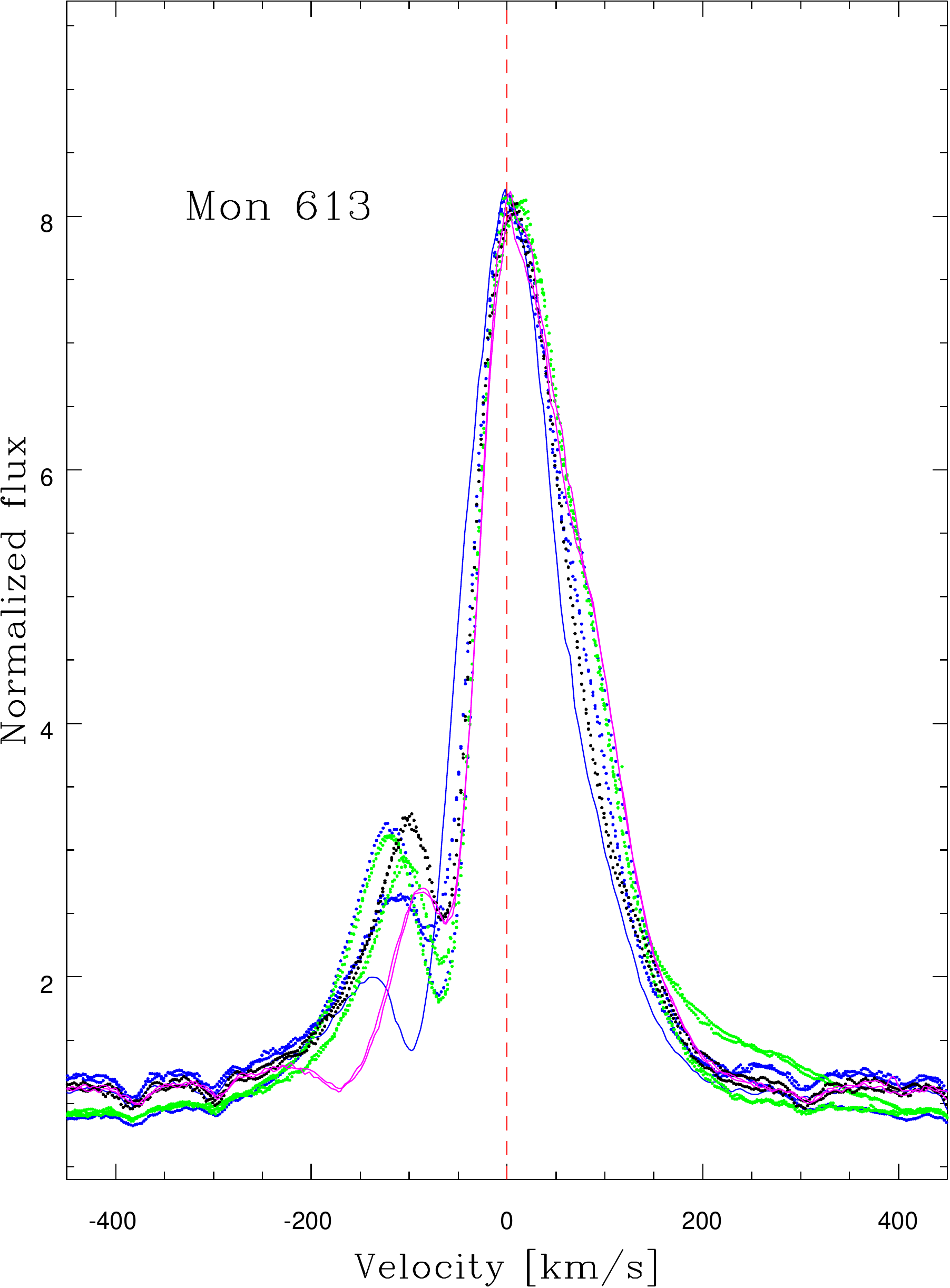}
\end{center}
\caption{a)  H$\alpha$ profiles for all twelve epochs obtained during
the 2013 VLT/FLAMES campaign for Mon-457.   The continuum has been
normalized to 1.0 in all spectra, and the y-axis maximum scale has been
adjusted so that the peak normalized flux at each epoch approximately
coincides.  The dot colors correspond to the observing epochs -- the
spectra from days 1, 4 and 7 are colored blue, spectra from days 8 and
9 are shown in green, day 10 in black, and day 13 in magenta. 
b) Same, except for Mon-613.
\label{fig:multi_halpha.2013}}
\end{figure*}

Figure~\ref{fig:wild_halpha}, in
the Appendix, shows typical H$\alpha$ profiles for twelve
of these stars (the H$\alpha$\ profile of Mon-613 is not shown
in Figure~\ref{fig:wild_halpha}, its H$\alpha$\ profiles are
instead shown in Figure~\ref{fig:multi_halpha.2013}).
H$\alpha$ profiles for stars with accretion-burst
dominated light curves and for stars whose light curves are
dominated by variable extinction events can be found in Figures 6
and 26 of Stauffer et al.\ (2014).

The most common feature of the H$\alpha$ profiles of the stochastic
stars is the presence of a blue-shifted absorption dip.   Six of the
eight stars with multi-epoch data, and four of the five stars with
single-epoch spectra have blue-shifted absorption dips.  For the stars
with multi-epoch data, five of the six stars (Mon-119, Mon-457,
Mon-491, Mon-613 and Mon-795) with blue-shifted absorption dips have
them for every spectral epoch -- as is illustrated in 
Figure~\ref{fig:multi_halpha.2013} for Mon-457 and
Mon-613.  The one star with intermittent blue-shifted absorption
dips is Mon-1234, where only five of twelve spectra show blue-shifted
absorption dips (and eight of twelve epochs show red-shifted
absorption dips).  For the stars in the accretion burst and variable 
extinction light curve classes for which we have multi-epoch VLT
spectra, about a third have persistent blue absorption dips; the number
for the variable extinction stars is difficult to state accurately
because their H$\alpha$ profiles are often very complex and variable.

The spectra used for Figure~\ref{fig:multi_halpha.2013} were obtained
during a thirteen day interval in 2013, with two spectra obtained on
nights 1, 8, 9, 10 and 13, and single spectra on nights 4 and 7.  The
H$\alpha$ profiles for Mon-457 were relatively stable over this
period, with the primary variation being in the height of the blue peak,
which first increased by about 20\% and then decreased by about 50\%.
The velocity offset of the blue-shifted absorption dip was essentially
constant (at about $-$50 km s$^{-1}$) for the entire two week interval.  The
H$\alpha$ profiles of Mon-613 show greater variability, particularly
in the height of the blue peak (which varies by more than a factor of
two) and the velocity offset of the blue-absorption dip (which was
generally about $-$60 km s$^{-1}$, but was $-$100 km s$^{-1}$ in the first spectrum).
The last two spectra of Mon-613 show two blue-shifted absorption dips,
one with $\Delta$v = $-$165 km s$^{-1}$ and the other with $\Delta$v $\sim$ $-$50
km s$^{-1}$.

Similar but independent models of the H$\alpha$ profiles for CTTS
combining both magnetospheric accretion and disk winds have been
created by Kurosawa et al.\ (2006) and Lima et al.\ (2010).
Blue-shifted absorption dips of the type shown in
Figure~\ref{fig:multi_halpha.2013} and Figure~\ref{fig:wild_halpha}
are predicted to be present in the 
H$\alpha$ profiles of CTTS primarily having moderately high mass
accretion rates and generally having intermediate to low inclination
angles (i $\lesssim$ 55$^o$).  Such blue-shifted absorption dips are
believed to arise from gas in a disk wind.    If this is indeed the
case, the fact that (when we have appropriate spectra) the blue
absorption dips are fairly stable on timescales of two weeks or more
indicates that the disk wind structure is also relatively stable on
those timescales (longer than, for example, the expected photospheric
rotation periods of these stars).   A more detailed discussion of
the H$\alpha$ profiles of all the CTTS for which we have FLAMES data
can be found in Sousa et al.\ (2015).

\subsection{Spectral Type and Light Curve Class}

It is reasonable to believe that light curve morphology could depend
on the mass of the YSO.  Higher mass YSOs are likely to have larger
mean rotational velocities based on the Kraft ``law" (J/M $\propto$ M$^{2/3}$; Kraft 1967)
and hence perhaps might have shorter variability timescales.  Higher mass YSOs 
are believed to often have more complex and more non-axisymmetric magnetic fields,
while lower mass (but still $>$ 0.5 M$_{\odot}$) YSOs are expected to have more axisymmetric
magnetic fields with strong dipolar components (Gregory et al.\ 2012). 
These  characteristics are likely to affect the interaction between the star and its circumstellar
disk and the resulting accretion patterns.

\begin{figure*}
\begin{center}
\epsfxsize=.99\columnwidth
\epsfbox{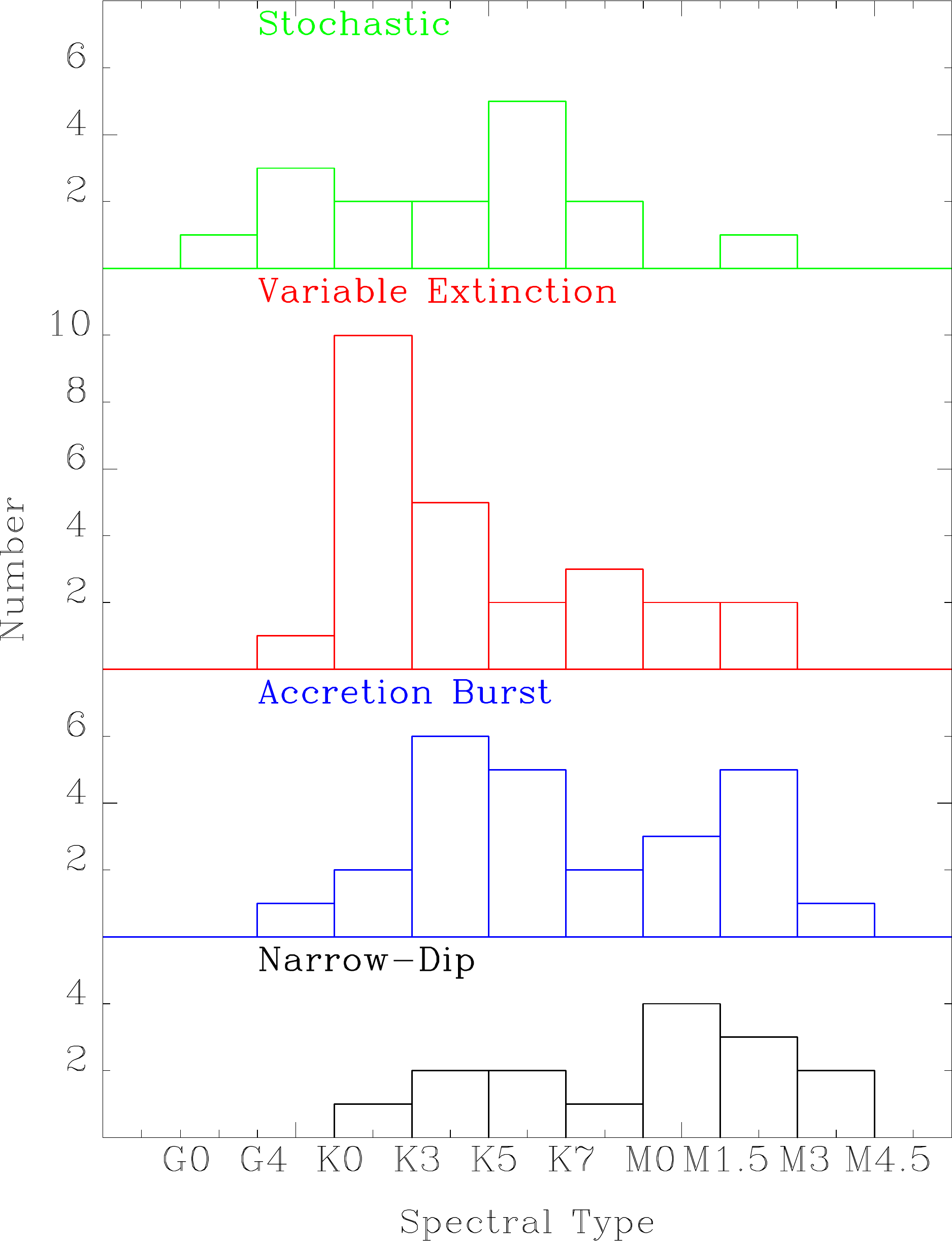}
\end{center}
\caption{Histogram of the spectral type distribution for the stochastic
light curve class members compared to histograms for three other CTTS
light curve classes.  See discussion in \S 5.2.
\label{fig:sptype.histo}  }
\end{figure*}

We do not have any direct mass indicator for the NGC~2264 members, but 
spectral type should provide a reasonable proxy for mass.  Spectral
type estimates are available for most of the NGC~2264 CTTS 
either from the published literature or our
own spectra, or estimates of spectral types derived from multi-band photometry
(Venuti et al.\ 2014).  Figure~\ref{fig:sptype.histo}
compares the spectral type distributions for 
four sets of CTTS in NGC~2264: (a) the stochastic light curve class; (b) stars with
AA~Tau-type light curves or with strong, aperiodic flux dips due to variable 
extinction (McGinnis et al.\ 2015); (c) stars with accretion burst dominated
light curves (Stauffer et al.\ 2014); and (d) stars with short-duration, nearly
Gaussian-shaped flux dips (possibly due to dust entrained in 
accretion columns; Stauffer
et al.\ 2015).    Table~\ref{tab:sptypes}, in the appendix, provides the list of stars we have
assigned to each group and our adopted spectral types.

The stars in all four light curve classes have a fairly wide range of spectral
types.  However, there are significant differences in the spectral type
distributions, with the stochastic class 
having the highest fraction of early spectral type stars (four of 15 stochastic
stars with estimated spectral types
are spectral type G, vs. only two of sixty for the other three classes combined),
while the short-duration flux dip class has a larger fraction of M dwarfs.
Simply assigning numerical equivalents to the spectral types (G5 = 45; K7 = 57;
M2 = 62) and then taking averages, the mean spectral types for the four
groups are K3 (stochastic), K5 (AA~Tau), K7 (burst) and M0 (short-duration
flux dip).

\subsection{Inferences on the Inclination of the Rotation Axis 
   relative to our Line of Sight}

Knowledge of, or at least constraints on, the inclination of a young
star's rotation axis to our line of sight can be very helpful in
attempting to determine a physical mechanism to explain the observed
variability.  For the variable extinction light curve class, if one
assumes that the extinction events occur when structures in the inner disk
pass through our line of sight, then the presence of the flux dips plus a
theoretically modelled disk scale height constrain the inclination of the
disk to be $\gtrsim$\ 60$\arcdeg$.\footnote{Following the simple but
physically well-justified CTTS disk model described in Bertout (2000),
we assume that on average, there is some inclination angle $i$$_{thick}$\ 
greater than which the star is always obscured, there is a range of inclinations
between $i$$_{thick}$\ and $i$$_{thin}$\ where variable extinction events
may occur, and for inclinations less than $i$$_{thin}$\ no disk-related
extinction events occur.  From the census of CTTS light curves in
NGC~2264 (Alencar 2010, Stauffer et al.\ 2015, McGinnis et al.\ 2015) we set
the fraction of optically visible CTTS with variable extinction events
at 25\%.  Because the estimated line-of-sight inclination to the disk for
the prototype CTTS AA Tau is i = 70-75$^o$ (Bouvier et al.\ 2013), we
adopt $i$$_{thick}$\ = 75$^o$.  With those two parameters set, the Bertout (2000)
model then yields $i$$_{thin}$\ $\simeq$\  60$^o$.}
For stars with stable, periodic light curves
due to non-axisymmetrically distributed spots, the shape of the light curve
and the presence or absence of eclipses place constraints on the latitude
of the spot(s) and our view angle to the star.   We can draw some 
inferences on the disk inclinations for the stochastic light 
curve class members simply from their light curve morphology; we can
infer a little more from their spectra -- as we describe below.

Members of the stochastic light curve class are in all cases CTTS.  They
have circumstellar disks that extend inward close enough to the star's
photosphere to produce class II SEDs.   By definition of our light curve
classes, a star in the stochastic class does not have deep flux
dips due to variable extinction events, and therefore the inclination of our
line of sight to their disks is almost certainly $\lesssim$ 60$\arcdeg$.  As noted
in Stauffer et al.\ (2015), two of the stochastic light curve stars do have
probable short-duration flux dips in their light curves (one dip each in 
2008 and 2011 for Mon-119, and two dips in 2011 for Mon-577 - 
see Figure~\ref{fig:stochastic_stars.set1}).  These short
duration flux dips may arise from dust in accretion columns -- and hence from
dust above the expected disk upper surface layer.   Even if the occulting dust
is nominally part of the disk material, the short duration and shallowness
of these flux dips suggest that the dust would be located in the upper surface
layer of the disk. Hence, even for these stars, we expect $i \lesssim$ 60$\arcdeg$.

\begin{figure*}
\begin{center}
\epsfxsize=.99\columnwidth
\epsfbox{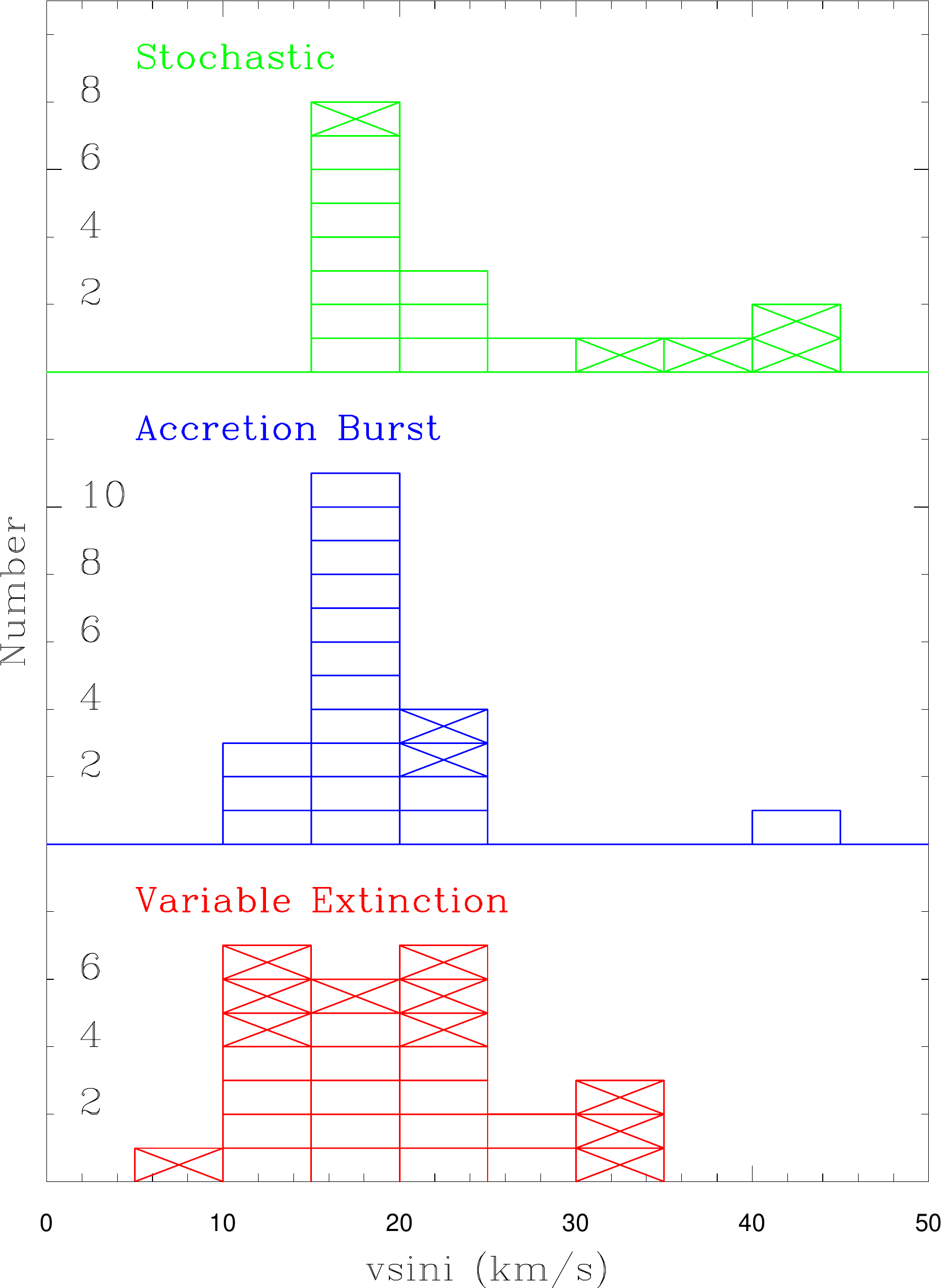}
\end{center}
\caption{Histogram of estimates of the projected rotational
velocity derived from Gaussian fits to the \ion{Li}{1} 6708\AA\ doublet
for the variable extinction, accretion burst, and stochastic
light curve classes.  Stars with spectral type K3 or earlier are marked with
a cross; not unexpectedly, these stars include most of the rapidly
rotating CTTS in the three classes. 
\label{fig:stochastic_stars.Li_vsini_hist}  }
\end{figure*}

If there were well-measured periods, \vsini's, and radii available for the
stochastic stars, it would be possible to directly estimate $\sin i$.   However,
periods are not known (more or less ``by definition") for the stochastic
stars.   Nevertheless, one might at least determine if the stochastic stars
have an unusual \vsini\ distribution compared to other CTTS in NGC~2264, and
from that either be able to draw inferences about their inclinations or
their intrinsic rotational velocities\footnote{A brief description of our
process for estimating \vsini's is provided in the Appendix.}.  Even that exercise is limited
due to the spectral types of the stochastic stars (Figure~\ref{fig:sptype.histo}),
because the stochastic stars include an excess (relative to the other light
curve classes) of relatively early type stars ($\leq$ K3).  These early
type stars will arrive on the ZAMS as A or F stars, and hence even on the
PMS will have systematically higher rotational velocities than lower mass CTTS. 
The actual \vsini's, as illustrated in Figure~\ref{fig:stochastic_stars.Li_vsini_hist}
confirm this bias, with many of the most rapidly rotating stars in all three plotted
CTTS groups having spectral type K3 or earlier.
Excluding the early spectral types, the distribution of \vsini's for the
three CTTS groups looks quite similar.   The average \vsini's confirm this:
the mean \vsini's
for the three groups are 19.0 \kms\ (n=10), 20.8 \kms\ (n=17), and 18.7 \kms\ (n=15) for
the stochastic, accretion burst, and variable extinction classes, respectively.  The accretion
burst mean is significantly influenced by one very discrepant star - Mon-7,
with \vsini\ = 85 \kms; excluding Mon-7, the accretion burst class average
\vsini\ becomes 16.9 \kms.

The essential equality of the mean \vsini's for the stochastic and variable
extinction stars is perhaps unexpected.  Given our conclusions concerning
the line of sight inclination to the rotation axes of the two classes
($i \gtrsim$ 60$\arcdeg$\ for the variable extinction class and $i \lesssim$ 60$\arcdeg$
for the stochastic class), and assuming a random distribution of inclinations
within those ranges, one would predict $<\sin i >_{\rm var-ext}/<\sin i>_{\rm stoch} \sim$ 1.5;
hence for $<$\vsini$>$ = 18 \kms\ for the variable extinction class, one would
have expected $<$\vsini$>$ = 12 \kms\ for the stochastic class.  At face value,
therefore, the equality of their average \vsini's suggests that the stochastic
stars might be comparatively rapid rotators.   However, because the mean value
for the stochastic class was derived from only ten stars, we believe it
prudent to simply conclude that it is unlikely that the stochastic stars 
are unusually slowly rotating.

\subsection{Variability Timescales}

\begin{figure*}
\begin{center}
\epsscale{1.0}
\plottwo{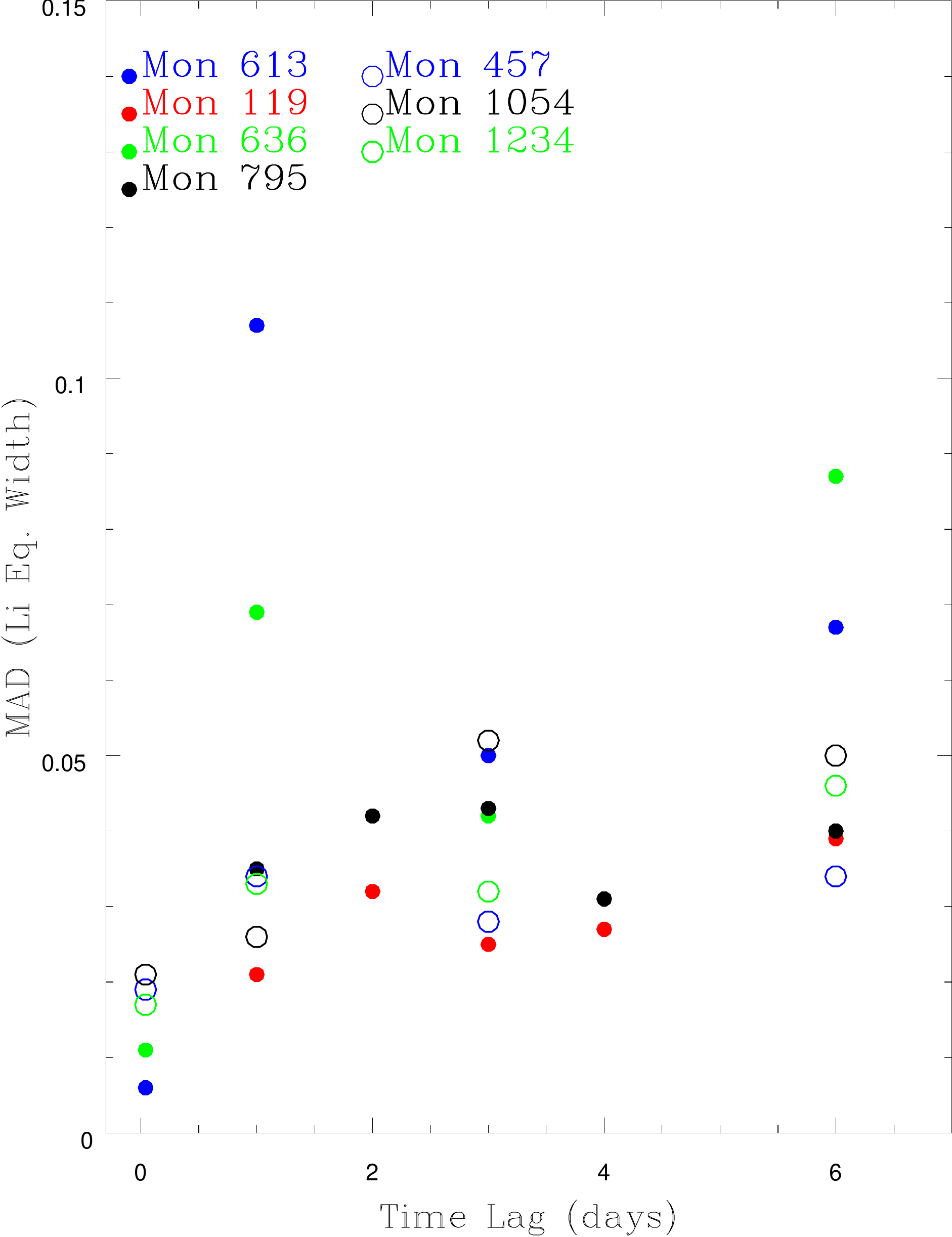}{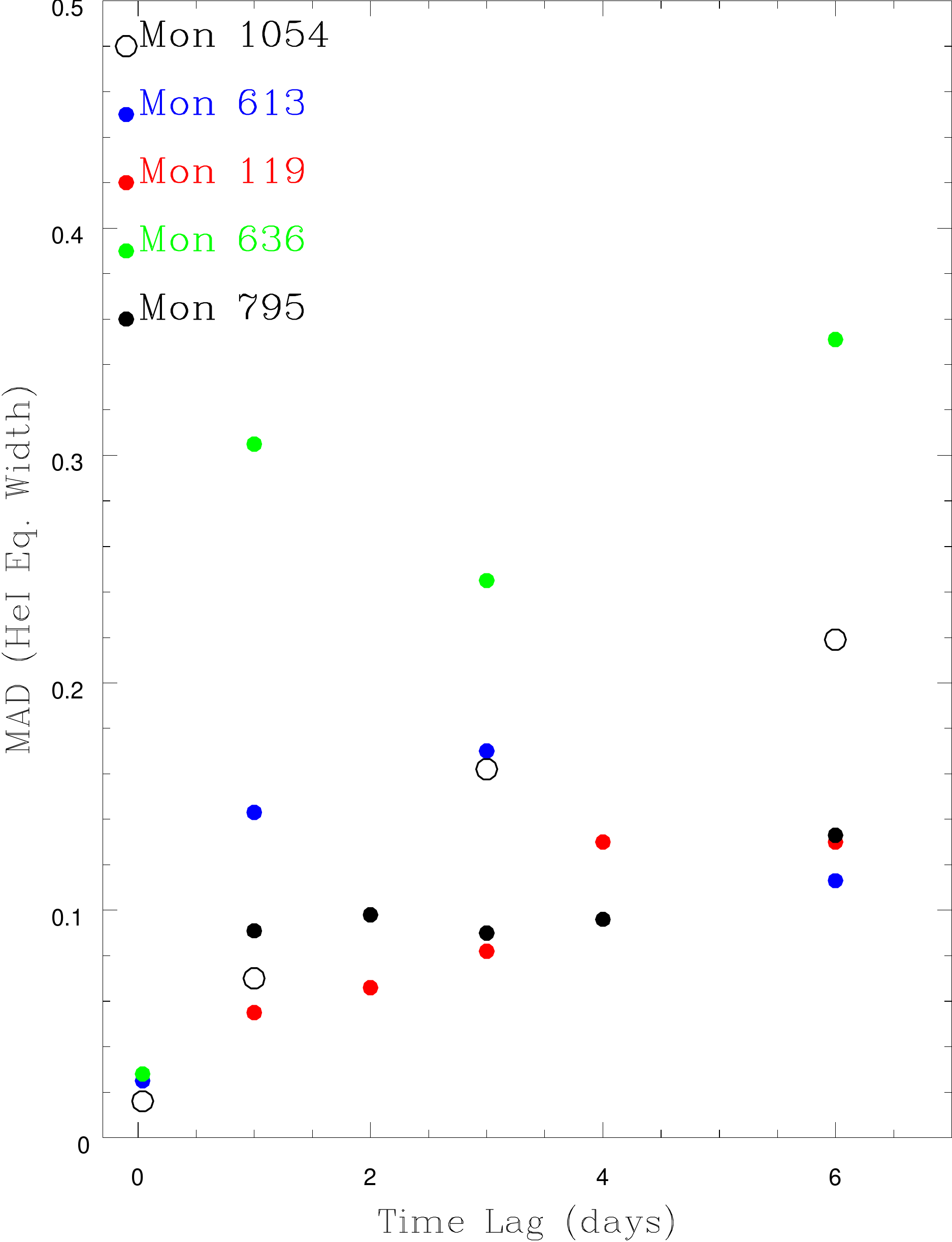}
\end{center}
\caption{a) Mean absolute difference for lithium equivalent widths
measured at two  different epochs as a function of the time
lag between the two epochs.
b) same as for (a), except the feature measured is the narrow
emission component for \ion{He}{1} 6678 \AA.
\label{fig:flames_MAD}}
\end{figure*}

One can obtain useful constraints on the timescale associated with
accretion variability in the stochastic stars from our multi-epoch
VLT/FLAMES spectra.   If the accretion rate changed only slowly
with time (e.g., with a several day time constant), then spectra
taken on the same night, or on successive nights, would have lithium
equivalent widths or \ion{He}{1} 6678 \AA\ fluxes that differ from each
other by significantly less than spectra separated by several days.
We have seven stars with multi-epoch VLT data (from
either 2011 or 2013) where we can make this test using lithium, and
four stars where this test is possible for the helium emission line
(the other stars either never or rarely show helium emission).  For
each star, we identified time lags with at least three pairs of epochs
with that time lag (often there were more than three).  For each time
lag, we calculated the mean absolute deviation between the measured
equivalent widths for each pair of epochs.   Figure~\ref{fig:flames_MAD}
shows the time lag plots for lithium 6708\AA\ and helium 6678\AA.   
As an ensemble, the plotted data indicate that most
of the variability occurs on timescales of order a day or less.

Photometric light curves, of course, provide a much more direct
means with which to derive variability timescales.
For stars with periodic light curves, one can easily determine that
period and often associate that period with a physical mechanism (e.g.
the rotation period of the star).   For stars with aperiodic light
curves, it is much more difficult to derive a timescale which correlates
well with the shape  one ``sees" in the light curve and which can
be convincingly linked to a physical mechanism.   After some experimentation,
we decided to derive timescale estimates based on an algorithm that
measures how long  it takes for the light curve to change by a given
fraction of its full amplitude.  Specifically, for every point in the
light curve, the algorithm looks both forward and backward in time, and
determines the time lag needed for the absolute difference in magnitude
to change by a given amount.  We then average the derived time lag for
every light curve point to determine the mean time lag for that delta
magnitude.  We calculate those timescales for steps
in delta magnitude ranging from 0.005 mag up to one half the full magnitude
range in the light curve.   When portions of the light curve seem to have
linear trends in magnitude (only the first half of the Mon-425 light
curve, and the last third of the Mon-795
light curve -- see Figure~\ref{fig:stochastic_stars.set1} and
Figure~\ref{fig:stochastic_stars.set2}), we first detrend
the light curve to remove that linear trend.   We associate a characteristic
time for the star as the derived mean time lag for a delta magnitude
corresponding to half the full amplitude of the light
curve\footnote{This algorithm returns timescales that are quite similar
to what would be derived from a peak-finding program, for peaks larger than
half the full amplitude, but avoids the difficulty of determining when
a peak occurs.}.  These
characteristic timescales are provided in Table~\ref{tab:basicinformation}.
The timescales range from half a day to of order five days, with most of
the stars having timescales near one day -- in accord with the analysis
of the VLT/FLAMES multi-epoch data.

\begin{figure*}
\begin{center}
\epsfxsize=.99\columnwidth
\epsfbox{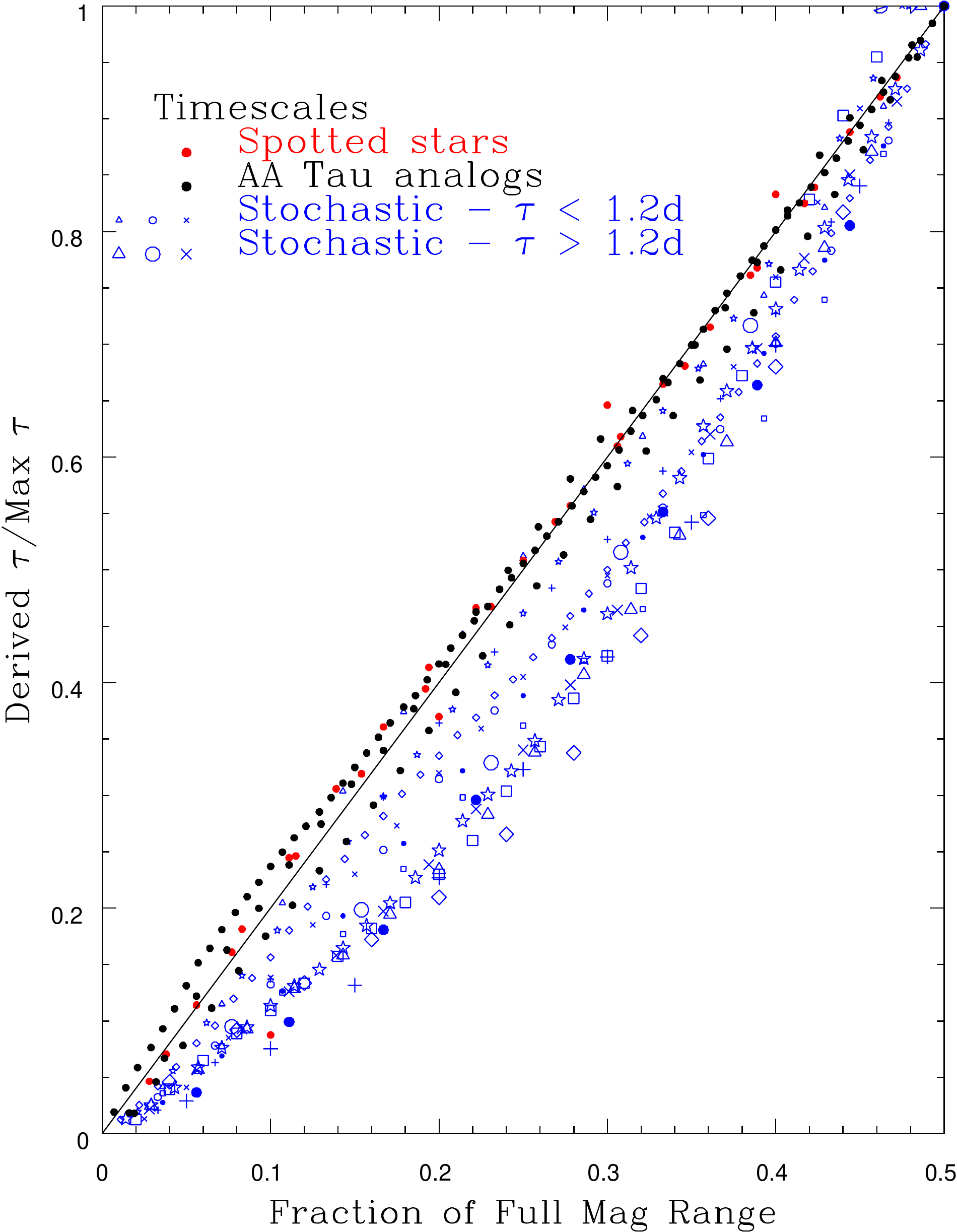}
\end{center}
\caption{Correlation between the mean time lag for the light curve to change by
a given $\Delta$magnitude and $\Delta$magnitude for the stochastic stars and
for a sample of stars with periodic light curves including both AA~Tau
analogs and stars with cold spots.   For each star, the y-axis has been
scaled so that a value of 1.0 corresponds to the time lag
for a $\Delta$mag of half the full amplitude.  Stars with cold-spot dominated
light curves are shown as red dots, AA~Tau analogs as black dots, stochastic
stars with short timescales are shown as small blue symbols (a different symbol
type for each star), and stochastic stars with long timescales are shown as
large blue symbols.
\label{fig:stochastic_stars.deltaTdeltamag}}
\end{figure*}

In principle, the way in which the mean time lag depends on delta magnitude
should encode information on the light curve shape.   Figure~\ref{fig:stochastic_stars.deltaTdeltamag}
plots this relation both for our stochastic stars and for a comparison set
of stars with periodic light curves.  The plot shows nearly linear correlations
for the periodic light curve stars, with all of the stochastic stars falling
below that relationship.  The stochastic stars with short characteristic
timescales ($\tau <$ 1.2 days) in general fall closer to the periodic
relation than the stochastic stars with longer characteristic timescales.

At a macroscopic level, it is reasonable to think of the light curves for
the stochastic variables as reflective simply of a global accretion process
whose $dM/dt$ varies in a seemingly random way.  Such light curves can be
well-matched by simulations with a damped random walk process (e.g., 
Findeisen et al.\ 2015).  However, on a microscopic level, it is possible
that all of the variability arises from individual accretion events creating
transient hot spots of variable size, temperature and lifetime.  If so,
the properties of the observed light curves could be used to place constraints
on the properties of the individual accretion events.  We have created a
simple Monte Carlo process model to explore this idea.

Our model produces light curves of total duration 34 days, from which we
extract the central 30-day portions (to minimize edge effects).  We distribute
N Gaussian flux bursts during the observing window, with each burst having
a height, width and central epoch chosen randomly within specified
bounds from a uniform distribution.  We
also add a continuum level to simulate the contribution to the light from the
stellar photosphere.  We vary the number of bursts and the parameter bounds
(a) to match the observed light curve amplitudes; (b) to match the mean veiling (the
fraction of the light from accretion compared to that from the photosphere);
(c) to yield values of Q and M appropriate for stars with stochastic
light curves;
and (d) to yield timescales as judged from our $\Delta$m vs. $\Delta$t algorithm
that approximately match specific stochastic class stars.

\begin{figure*}
\includegraphics[width=6.0in]{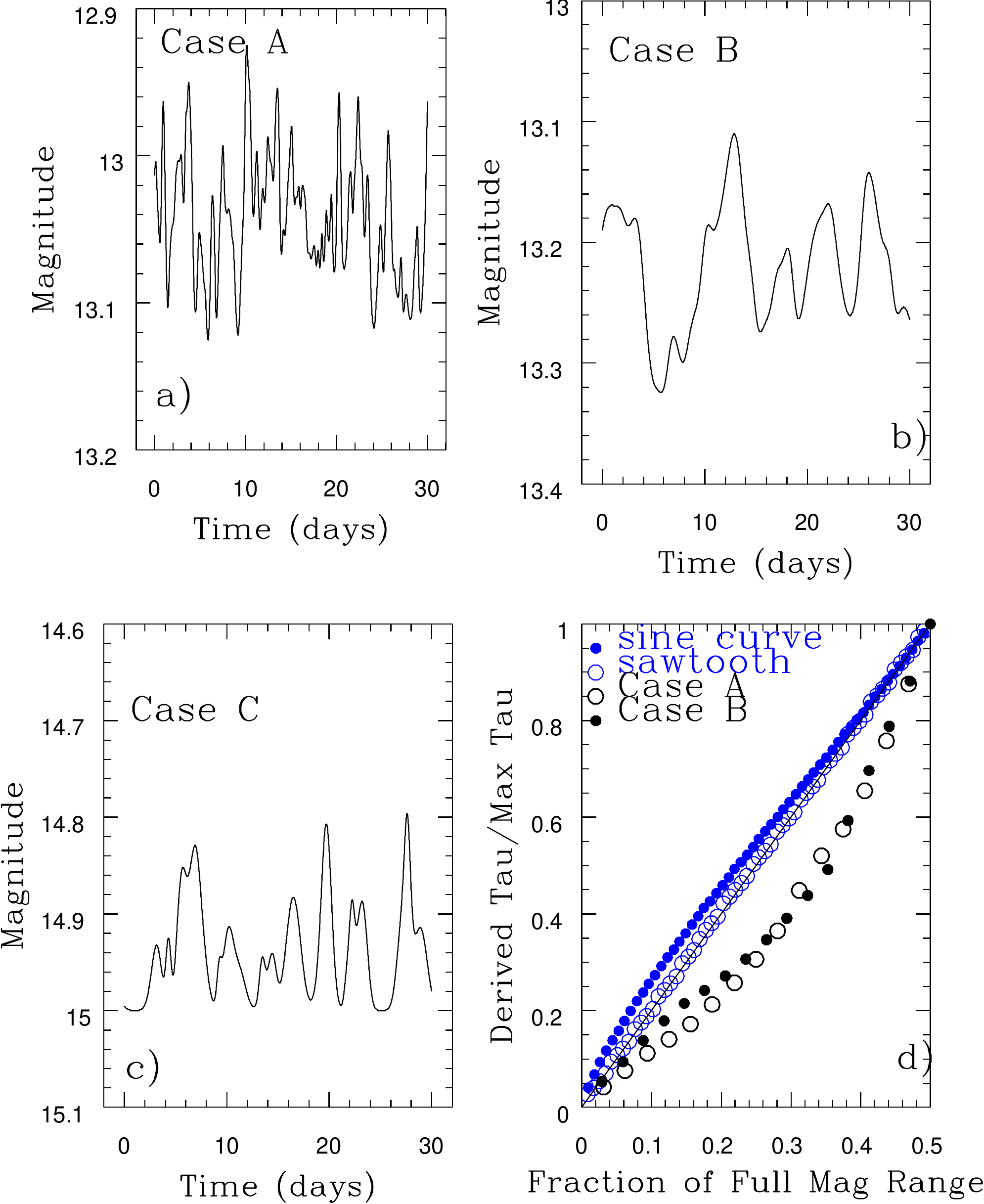}
\caption{
Results from Monte Carlo models aimed at simulating the light curve
morphologies of the stochastic light curve class.
See discussion in \S A3.  The photospheric continuum level in plots a-c
are 13.25, 13.49 and 15.0 magnitude, respectively.
a) Simulated light curve intended to emulate the morphology of a short
timescale stochastic star (e.g., Mon-346 or Mon-577);
b) Simulated light curve intended to emulate the morphology for a long
timescale stochastic star (e.g., Mon-457);
c) Simulated light curve, intended to emulate a light curve
we would classify as accretion-burst dominated (Stauffer et al.\ 2014);
d) $\Delta$t vs. $\Delta$m plot for the light curves in panels (a) and (b),
compared to two idealized light curve types (sine wave and sawtooth).
\label{fig:stochastic_models}}
\end{figure*}

Figure~\ref{fig:stochastic_models}a shows a simulated light curve (``Case A")
whose morphology, amplitude, mean veiling and timescale ($\tau$ = 1.4 days) was designed as
a reasonable match to the {\em CoRoT} light curve of either Mon-346 or Mon-577
(see Figure~\ref{fig:stochastic_stars.set1}).  The parameters for this model
were N (number of bursts in 30 days) = 300, FWHM = 0.2 to 0.7 days; height =
5.0 to 20.0 (in ``counts"), and photospheric continuum level = 500 (in ``counts").
M is 0.14 for this light curve, and the Q statistic indicates the light
curve is random -- hence Q is set arbitrarily to 1.

Figure~\ref{fig:stochastic_models}b shows a simulated light curve (``Case B")
whose morphology, amplitude, veiling and timescale ($\tau$ = 3.3 days) was designed to reasonably
match the {\em CoRoT} light curve for Mon-457 (see Figure~\ref{fig:stochastic_stars.set1}).
The parameters for this model were N = 300, FWHM = 0.9 to 2.7 days,
height = 2.0 to 10.0, and photospheric continuum level = 400.
M is $-$0.19 for this light curve, and the Q statistic indicates the light
curve is random -- hence Q is set arbitrarily to 1.

Figure~\ref{fig:stochastic_models}c shows that as the frequency of bursts
decreases, individual bursts become more visible and the light curve evolves
to better resemble our accretion burst light curve class (Stauffer et al.\ 2014).
The parameters for this light curve were
N = 30, FWHM = 0.5 to 1.5 days, height = 5.0 to 18.0, and photospheric continuum
= 400.

Finally, Figure~\ref{fig:stochastic_models}d provides a $\Delta$t vs. $\Delta$m
plot for the Case A and Case B models.   Also plotted are the $\Delta$t, $\Delta$m
values for two idealized periodic light curves - a sine wave and a sawtooth function.
The models designed to look like our stochastic stars have $\Delta$t, $\Delta$m loci
in the diagram that mimics the loci for the stochastic stars
(Figure~\ref{fig:stochastic_stars.deltaTdeltamag}).

We view this exercise as primarily a proof of concept.  The stochastic light
curves can indeed be simulated with many small, discrete flux bursts.  There must
be many bursts, on average, per day in order to yield light curve shapes that
appear symmetric about their median value.  However, the durations of the events
must be kept short in order that the summed flux from the accretion events
does not become too large and the light curves do not become too smooth (low
amplitude).   The parameters we have chosen for Case A and B are presumably
not unique, but they at least provide possible choices that could yield
light curves similar to those we observe in NGC~2264.

\subsection{Correlation of Optical and IR Light Curves}

The optical photons from CTTS come primarily from the photosphere and
the hot spots produced by accretion shocks.   The IRAC (3.6 and 4.5 $\mu$m)
photons also originate from these regions, as well as 
from warm dust in the inner circumstellar disk.  For stars with only 
small IR excesses from warm dust, and where there is little or no
variable extinction, the optical and IR light curves should be
well-correlated (but possibly with significantly different amplitudes), 
regardless of the physical mechanism driving the variability.
For stars where the disk photons dominate at IRAC wavelengths, 
the optical
and IR light curves could be quite uncorrelated, depending on the physical
mechanisms involved.  The most likely physical processes driving variability
at IRAC wavelengths are: (a) heating of the dust in the inner disk by
reprocessing of the photons emitted from the star's surface; (b) changes in
the inner disk dust surface area exposed to the stellar photon bath (e.g., from
instabilities lifting more dust into the disk photosphere); and (c) viscous,
self-heating of the inner disk.   The optical-IR correlation can also be
affected by details of the geometry -- a transient hot spot on the side of the star
facing us produces a flux burst most evident in the optical light curve, but if
that hot spot is not located in a position where it is seen by the warm dust
that dominates the IRAC light, then it may have only a
minor impact on the IRAC light curve.   Conversely, a transient flux burst on the
back side of the star would have no impact on the optical light curve that we see,
but could have a significant impact on the IR light curve if the IR flux is dominated
by photons from warm dust in the inner disk wall that faces the back side of the star,
as is in fact likely.   As discussed extensively in Whitney et al.\ (2013), the
correlation between the optical and IR light curves of the stochastic stars
therefore can provide important clues to
the physics underlying their variability; we discuss that correlation
for our stochastic stars now.

\begin{figure*}
\includegraphics[width=5.2in]{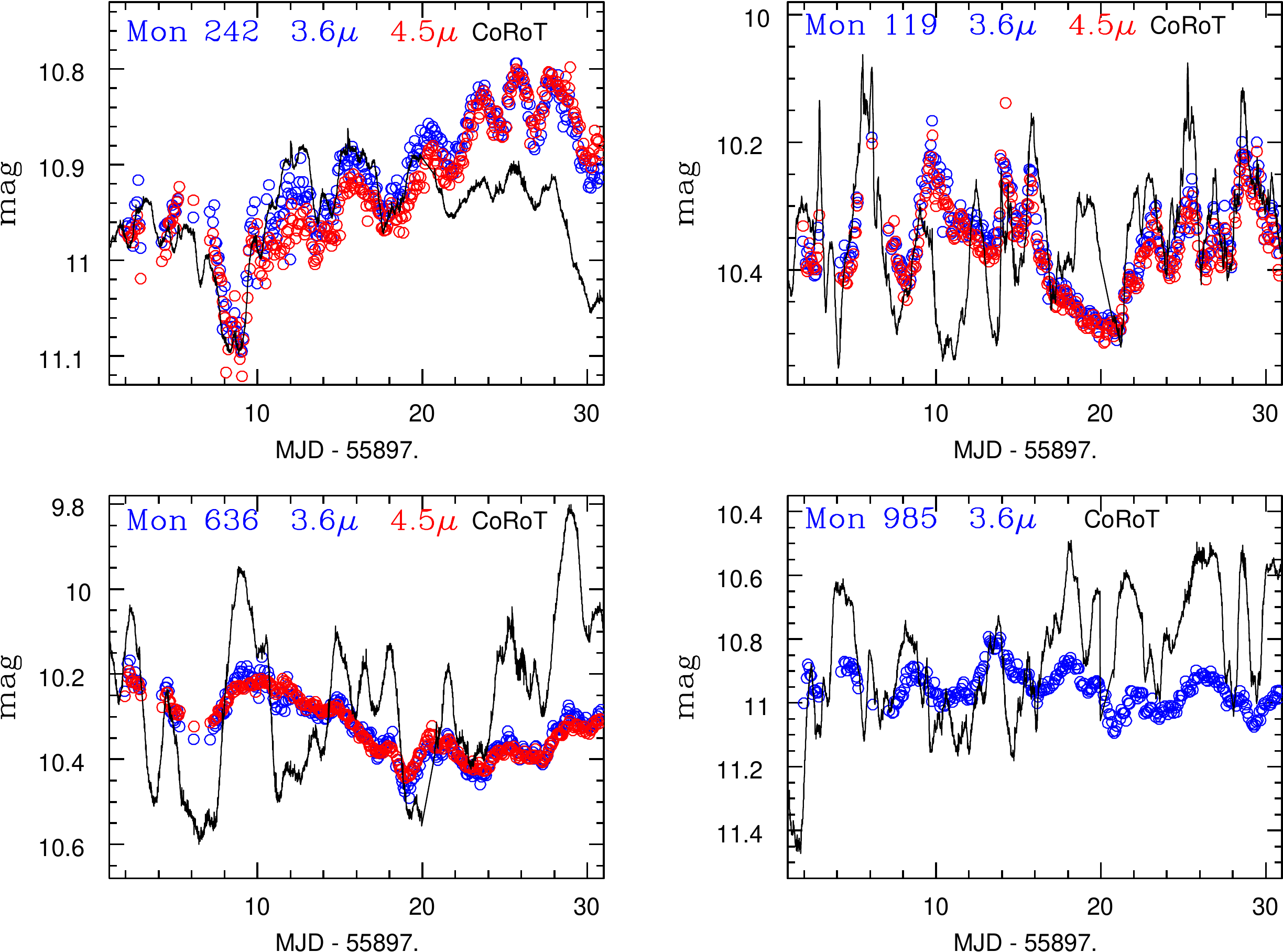}
\caption{Overlay of the {\em CoRoT} and {\em Spitzer} light curves for four
of the stochastic light curve stars.  The plots are arranged according to
their disk-to-photosphere flux ratio at 4.5 $\mu$m, from least disk-dominated
at top-left (Mon-242) to most disk-dominated at bottom-right (Mon-985).  This
results in their also being approximately ordered according to the
degree of similarity between the optical and IR light curve structures.
For each panel, the optical data are shifted in zero-point to align vertically
with the IR data -- the optical and IR delta-mag for full scale is the same.
\label{fig:super.corot_irac}}
\end{figure*}

Table~\ref{tab:basicinformation} provides two quantities 
related to the optical/IR variability issue --
the estimated disk to photosphere flux ratio at 4.5 $\mu$m\ (column 6), derived
from our single-epoch photometry obtained prior to 2010,  and
an optical-to-IR Stetson index (column 8), derived from our 2011 CSI~2264 campaign data.  
The disk to photosphere ratio 
ranges from 0.4 for Mon-242, a star with very little IR excess, to $>$10
for Mon-985, a star with a very large IR excess.  The Stetson index (see
Cody et al.\ 2014 for a more thorough discussion) ranges from 0.41 (largely
uncorrelated) to $>$2 (well-correlated).   
Figure~\ref{fig:super.corot_irac} shows an overlay of 
the {\em CoRoT} and IRAC light curves for four of the
stochastic stars where the Stetson index indicates a good degree of correlation
between the optical and IR light curves.

\begin{figure*}
\includegraphics[width=5.2in]{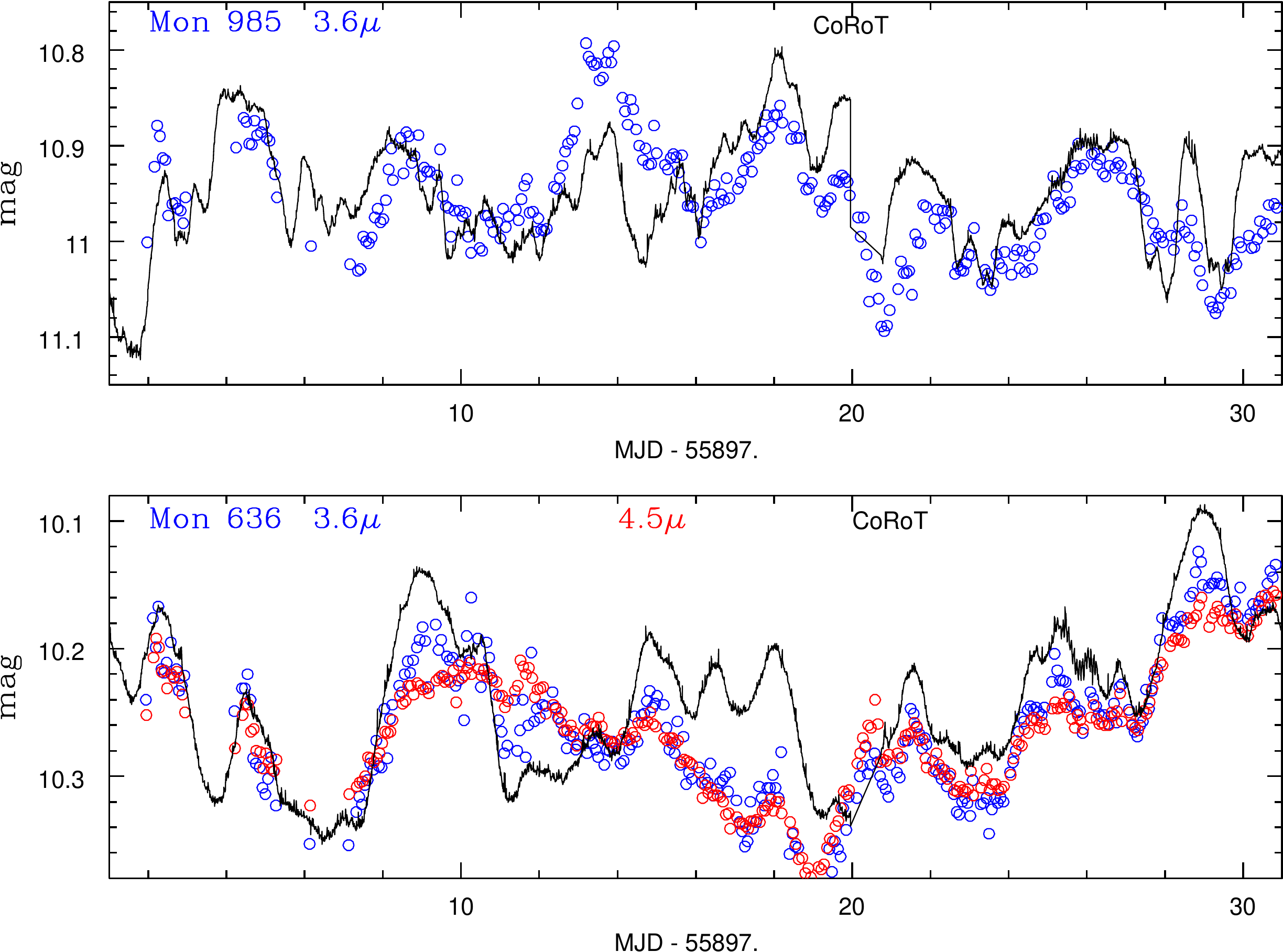}
\caption{Overlay of the {\em CoRoT} and {\em Spitzer} light curves for 
Mon-636 and Mon-985, where in this case the y-axis full-scale for the
IRAC data is one third that for the {\em CoRoT} data (e.g., for
Mon-985, full scale for the IRAC data is 0.4 mag, whereas it is
1.2 mag for the {\em CoRoT} data).  A slowly-varying baseline has
been subtracted from the Mon-636 {\em CoRoT} data in order to better
visualize the short-timescale correlation of
the optical and IR light curves.
\label{fig:super.corot_irac_blow}}
\end{figure*}

The stars in Figure~\ref{fig:super.corot_irac} are arranged in order  (top-left to bottom-right) of 
increasing disk dominance at 4.5 $\mu$m, with disk-to-star flux ratios of 0.4,
2.9, 5.4, and $>$10 for Mon-242, 119, 636, and 985, respectively.  Visual examination
of the four panels indicates that they are also arranged approximately in order
of the degree to which the amplitude and shape of the short-timescale IR fluctuations
match those in the optical.
The optical and IR light curves for Mon-242 track each other
very well on short timescales in both shape and amplitude.  Mon-119
generally shows well-defined IR peaks at location of the optical peaks, with amplitudes
often of order half that as shown in the optical.   Mon-636 and Mon-985, both having
disk-dominated IRAC photometry, show much lower amplitude variability in the IR
compared to the optical, but often appear to have barely discernible features at the location of
structures in the optical light curve.  A better view of the optical-IR correlation
for these two stars is provided in Figure~\ref{fig:super.corot_irac_blow}, where the
IRAC light curves are shown magnified by a factor of 3 in the y-axis.  This plot shows that, in
fact, there is still a good correlation in the optical and IR short-timescale
morphology, but the amplitudes of the IRAC peaks are on average about one third the
amplitude of the optical peaks.

Close examination of the four panels of Figure~\ref{fig:super.corot_irac}, 
shows some optical features which appear
to be completely lacking in IR counterparts.   Good examples are flux excesses in the
optical light curve for Mon-119 at days $\sim$12 and$\sim$20, with no evidence at all
for a feature at those times in the IRAC light curve.  There are no obvious 
inverse situations, where an isolated IRAC flux peak is lacking a simultaneous
{\em CoRoT} peak.

Assume for the moment that the accretion hot spots have effective temperatures
of order 8000 K and that our CTTS are all around spectral type K4.   
If the hot spot emitted as a blackbody, it would have
$R - [3.6] \sim$ 0, and the ``quiescent" photosphere
would have $R - [3.6] \sim$ 2.  In the absence of any warm dust, an accretion
burst causing a 0.2 mag brightening in the optical light curve (comparable to
the light curve amplitude for a typical stochastic star) would cause only a
$\sim$0.03 mag brightening at 4 $\mu$m.  For stars with large IR excess, where the disk
dominates the 4 $\mu$m\ flux by a large factor, the flux variation at 4 $\mu$m
directly from the photosphere would be even lower.  
Because the emitted flux from accretion hot spots has a spectral shape
closer to F$_{\lambda}$\ $\sim$ constant rather than a blackbody spectrum
(Hartigan 1989; Basri \& Batalha 1990), this estimate is somewhat 
inaccurate, but we believe the basic conclusion is still valid.
Therefore, it is 
unlikely that the flux variation at IRAC wavelengths seen in any of the stars 
in Figure~\ref{fig:super.corot_irac} is directly a result of the 
variable photospheric contribution.

Assume now also that radiative heating dominates the energy budget for the warm
dust responsible for the flux at IRAC wavelengths, and also assume that the
system geometry is such that the transient hot spot is well positioned to be
seen by us and by the warm dust in the inner disk.  In that case, a change in the
luminosity of the star of $\Delta$m(bol)$_{star}$ due to appearance of
a transient hot spot should approximately result in a similar
increase in the luminosity of the warm dust.  A 8000 K hot spot should have an SED
peaked near 4000 \AA, close enough to the center of the broad {\em CoRoT} bandpass
such that a 20\% change in the {\em CoRoT} light curve flux should correspond
approximately to a 20\% change in luminosity of the hot spot.  For 1000 K dust,
$\lambda _{max} \sim$ 4 $\mu$m, and a 20\% change in the IRAC light curve
should correspond approximately to a 20\% change in the luminosity of the warm
dust.   This suggests that there could be a good correlation between optical and
IR light curves for stars with disk-dominated fluxes at IRAC
wavelengths, which we in fact do see
(Figure~\ref{fig:super.corot_irac_blow}).  To be explained still are the variable,
and sometimes small IR-to-optical amplitude ratios, and the optical peaks with
no apparent IR counterpart.

It is possible to adjust the geometry to reduce the response
of the disk to the hot spot luminosity variations by placing the hot spots
at higher stellar latitude.  This reduces the projected area of the hot spot
as seen by the inner disk dust roughly as the cosine of the spot latitude,
so that a spot at 70$\arcdeg$ latitude would have about half the disk heating impact of a
spot of the same physical size if at 45$\arcdeg$ latitude.  
For an observer located near the rotation axis of the star, spots at such
high latitudes would always be visible and the portion of the inner disk-rim they
``light up" should have similar view angles to the observer -- which would result in
a good correlation between the optical and IR peaks and a rarity of cases where
there is an optical peak without an IR counterpart.  As either the spot latitude
decreases or the inclination of the observer's line of sight to the system
rotation axis increases, the chance for a transient hot spot
occurring where it is not visible to the observer increases.   Therefore,
the observed correlation between the optical and IR light curves favors both
relatively high spot latitudes and relatively low inclinations of the system
rotation axis to our line of sight.  This geometry also naturally reduces any
periodic signature in our light curve due to the star's rotation.

\section{Summary and Conclusions}

Based on the preceding sections, we can summarize the properties that
characterize stars whose optical light curves fall into our stochastic
class as:

\begin{itemize}
\item By selection, their optical light curves appear symmetric about their
median level, with no preference for discrete downward flux dips (variable
extinction) nor upward flux bursts (short-duration flux bursts).  The lack
of flux dips due to extinction excludes stars where the inclination of our
line of sight to the circumstellar disk is large,
implying $i \lesssim$ 60$\arcdeg$.

\item By selection, their optical light curves show very little periodic
signature.

\item The fraction of the CTTS in NGC~2264 having optical light curves
of this type is about 10\%.

\item As judged by their UV excesses, H$\alpha$ equivalent widths, and veiling
of optical absorption lines, the stars with stochastic light curves have relatively low to moderate
accretion rates.  Their mean veiling level is generally of order 10\% to
20\%.

\item Their {\em CoRoT} light curves show amplitudes generally in the
range of 10\% to 30\%.

\item In all cases, their broad-band IR photometry indicates SED slopes 
consistent with Class II

\item The members of the stochastic light curve class have 
spectral types ranging from early G to early M, with a somewhat earlier mean
spectral type than our other CTTS light curve classes.

\item The stochastic stars have small reddening, 
as judged from the $J-H$ vs.\ $H-K$ color-color diagram and
from comparison of their observed optical colors to colors predicted based
on their spectral types.

\item Their H$\alpha$ profiles are most often characterized by blue-shifted absorption
dips that are usually present with similar shape and amplitude at all epochs, 
best explained as due to gas in a disk wind seen from moderate to low
inclination angles.

\item They have veiling variability and color changes derived from multi-epoch photometry
indicative of a variable blue continuum source, consistent with expectation
for photospheric hot spots whose size or temperature change with time.

\item Variability timescales for the stochastic stars derived either from the light curves or from
multi-epoch spectroscopy are generally of order 1-2 days.

\item There is usually a good correlation between the optical and IR light curve morphologies
on short timescales, but often much lower amplitudes for the IR fluctuations;
short timescale events at one wavelength but not the other are quite rare.
These characteristics are best explained by placing the transient hot spots
at high stellar latitudes and the observer at low inclination angles to the
star system rotation axis.
\end{itemize}

Our data show that the light curve variability of the stochastic class arises
from their having highly variable photospheric mass accretion.
Of course, all CTTS presumably have variable mass accretion
rates.   For some CTTS, most notably the variable extinction light curve class,
other processes produce fluctuations in the optical flux of much larger
amplitude which mask the comparatively small changes in brightness due to
the variable accretion rate.   Where such camouflage is not present, 
there appears to be (at least) three light curve morphologies where accretion
processes drive the variability.
Stars with one or two dominant, stable accretion flows (funnel
flow accretion) where the star's rotation is the dominant timescale can produce
quasi-periodic, symmetric (''hot spot") light curves (Cody et al.\ 2014), like
that shown in Figure~\ref{fig:sixctts}e.
Stars with the highest accretion rates
have light curves dominated by relatively strong, short-duration accretion
bursts that are usually random in time but sometimes occur as periodic burst
groups (Stauffer et al.\ 2014; Blinova et al.\ 2015).   What is the physical
driver for the variability shown by the stochastic light curve class?

We believe a clue to the answer to this question can be found by considering
the members of the variable extinction class, and in particular one of the
prototypes -- AA~Tau.
Based on an extensive Zeeman Doppler imaging study of AA~Tau, Donati et al.\ (2010) 
concluded that (a) its magnetic dipole axis is tilted by $\sim$ 20$\arcdeg$\ relative
to its rotation axis; (b) the primary accretion hot spots are located near
latitude = 70$\arcdeg$; (c) its mean mass accretion rate to the stellar surface is
quite small, with veiling levels ranging from near zero up to $\sim$0.2; (d)
that the accretion rate varies significantly on timescales less than the
stellar rotation period.   Concerning the variability in the inferred accretion
rate, they also concluded that ``... the accretion variability observed for 
AA~Tau mostly relates to the variable efficiency at which the disc material
succeeds at entering the closed magnetosphere in the propeller regime, 
rather than to an intrinsic variability of the accretion rate within the 
inner disc."

This characterization of accretion in AA~Tau very nearly matches our description
of accretion in the stochastic light curve class, and we believe this points
to the most likely physical model for their variability.  Our line of sight
to the variable extinction class members must be $i \gtrsim 60 \arcdeg$.  There
must be at least as many CTTS with the same accretion 
properties but seen at $i \lesssim 60\arcdeg$.  Our data support identifying
at least many of these low-inclination counterparts to the variable extinction
class stars
with members of the stochastic light curve class.  Adopting the Donati 
et al.\ (2010) conclusion for AA~Tau, we suggest that instability processes
in the disk photosphere regulate the flow of material onto the magnetic field
lines in the funnel flow, thereby also leading to the stochastically variable flux
from hot spots on the stellar surface.  Some fraction of the variable extinction
stars may have more stable accretion and hence have hot spots that are
relatively stable in size and location; when seen at moderate inclination, these
stars should have light curves that are more or less periodic.  Cody et al (2014)
labelled such stars as quasi-periodic, symmetric.  Future synoptic studies of
star-forming regions should be conducted to determine whether stars with 
quasi-periodic, hot-spot light curves can transition to the stochastic light-curve
morphology (and, if so, how the statistics for that transition compares to
the statistics for the transition between AA~Tau type light curves and the
aperiodic flux dip light curve class).

\begin{acknowledgements}
This work is based on observations made with the
{\em Spitzer} Space Telescope, which is operated by the Jet Propulsion
Laboratory, California Institute of Technology, under a contract with
NASA. Support for this work was provided by NASA through an award
issued by JPL/Caltech. This research was carried out in part at the
Jet Propulsion Laboratory, California Institute of Technology, under a
contract with the National Aeronautics and Space Administration and
with the support of the NASA Origins of Solar Systems program via
grant 11-OSS11-0074.  RG gratefully acknowledges funding support from
NASA ADAP grants  NNX11AD14G and NNX13AF08G and Caltech/JPL awards
1373081, 1424329, and  1440160 in support of Spitzer Space Telescope
observing programs. SHPA, AS and PTM acknowledge support from CNPq, CAPES
and Fapemig.  
\end{acknowledgements}

{\it{Facility:} \facility{Spitzer (IRAC)}, \facility{CoRoT}, 
\facility{CFHT (MegaCam)}, \facility{VLT (FLAMES)}.}

\newpage
\newpage

\clearpage

\section*{Appendix}

\subsection*{A1: {\em CoRoT} Light Curves for Additional Stars Whose
QM Metrics Match that for Stars with Stochastic Light Curves }

\begin{figure*}
\begin{center}
\includegraphics[scale=0.8]{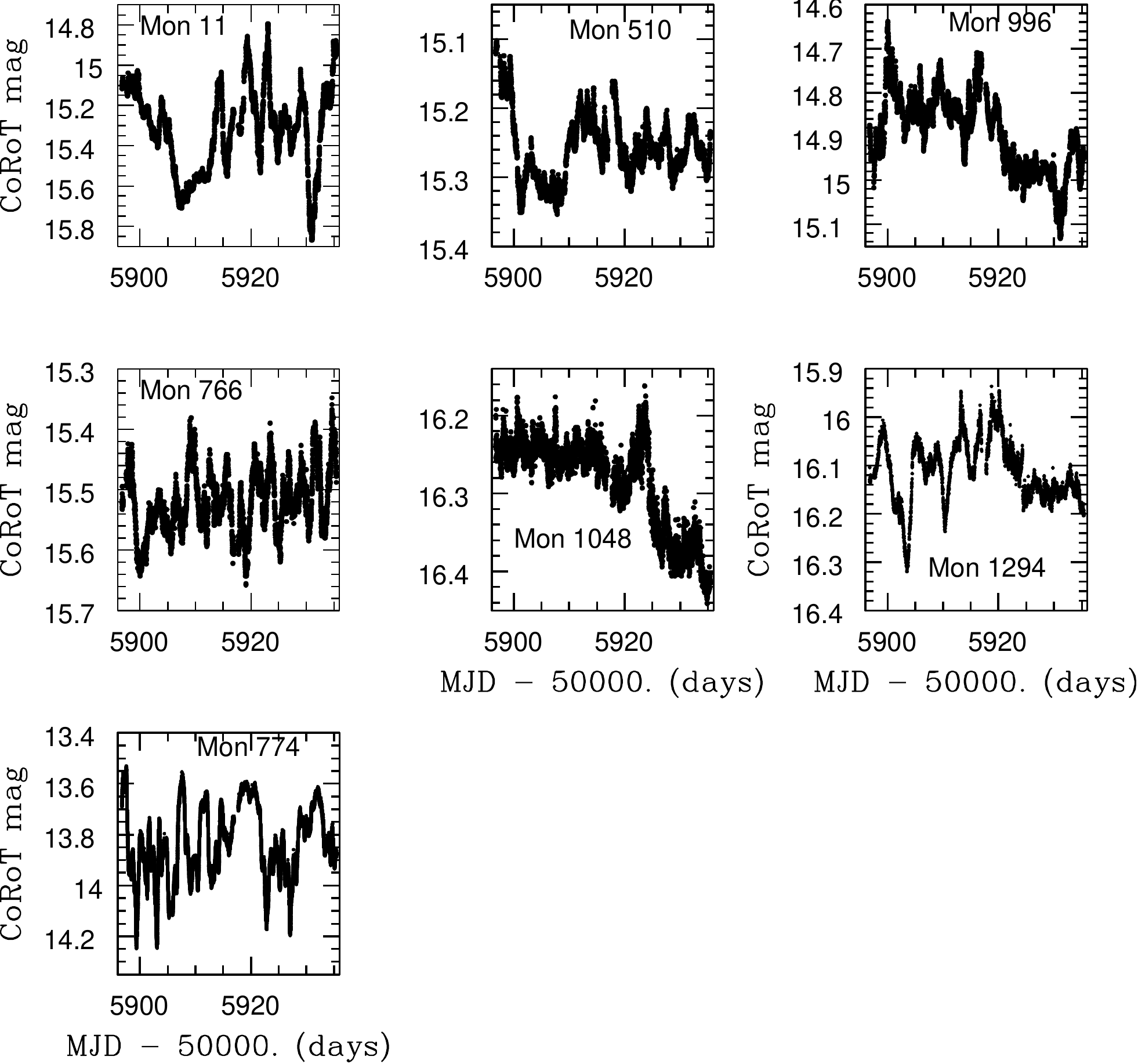}
\end{center}
\caption{{\em CoRoT} data for stars with Q $>$ 0.6 and -0.25 $<$ M $<$ 0.25
(see Figure~\ref{fig:stochastic_stars.QM}) that we believe 
do not belong in the stochastic light curve class.  We instead classify
Mon-11, Mon-510, Mon-996,
Mon-766, and Mon-1048 as accretion-burst dominated; 
Mon-1294 as having light curves exhibiting both accretion bursts and
flux dips due to variable extinction; and the 2011 light curve for Mon-774
as probably due to aperiodic extinction variations (see \S 3).
\label{fig:otherCorot}}
\end{figure*}

We have used two methods to assign stars to the stochastic light curve
class, with each method having its own uncertainties.  One method uses
quantitative formulae to derive metrics for the light curves (Cody et al.\ 2014);
however, the formulae are empirical and the boundaries are to some
extent arbitrary.  The other method is entirely qualitative and is based
on ``by eye" classification of the light curves by sorting them into
groups which seem to have similar properties.   The light curves shown
here in Figure~\ref{fig:otherCorot} 
match the algorithmic definition to be
in the stochastic light curve class, but did not pass our ``by eye"
classification scheme.  They therefore are not included in our final
list of stars with stochastic light curves (Table~\ref{tab:basicinformation}).
See discussion in \S 3.

\subsection*{A2: Spectral Feature Variability and Light Curve Morphology}

In \S 4.1, we showed that when we had appropriate synoptic high-resolution 
spectroscopy, the stars with stochastic light curves have veiling levels which
vary with time and are quantitatively consistent with the observed photometric
variability.  We also showed that the \ion{Li}{1}\ 6708\AA\ doublet - the strongest
absorption feature in our spectra - can be used as a veiling indicator (because
its equivalent width variations correlate well with the equivalent width 
variations of other absorption features in our spectra) and that the variations in
the lithium
equivalent widths also correlate well with variations in the strength of the \ion{He}{1} 6678
emission line.

\begin{figure*}
\includegraphics[width=6.0in]{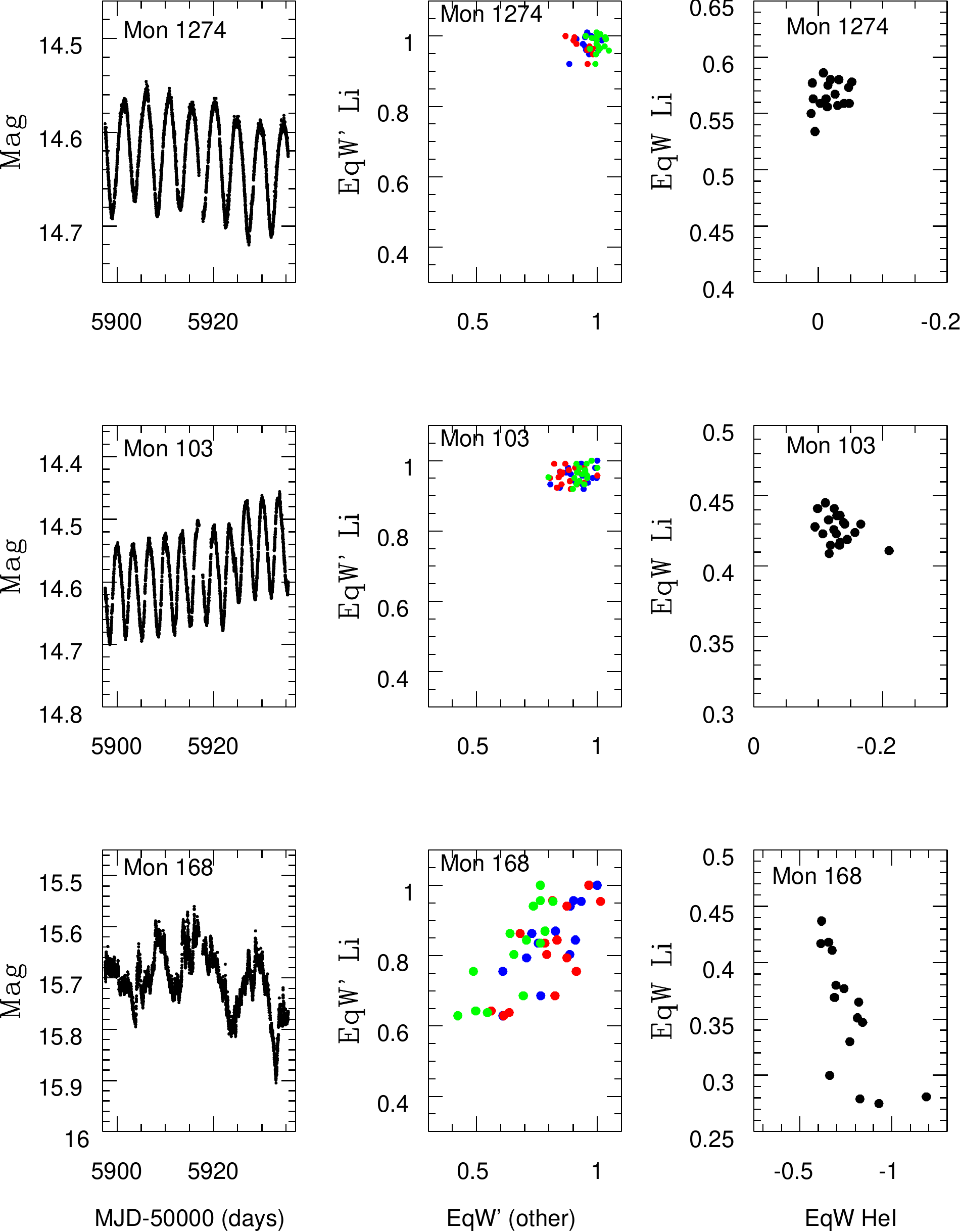}
\caption{{\em CoRoT} light curves and spectral indices plots for three
NGC~2264 members:  (a) Mon-1274, a WTTS with a typical cold-spot light curve;
(b) Mon-103, a CTTS with a cold-spot light curve; and (c) Mon-168, a CTTS
with a hot-spot light curve (see text).   In the middle plot in each row, the different
colors correspond to different absorption features; also, the individual equivalent
widths have been divided by the maximum equivalent width measured for that
feature, in order to allow multiple features to be over-plotted.
\label{fig:morph_vs_spectra.set1}}
\end{figure*}

\begin{figure*}
\includegraphics[width=6.0in]{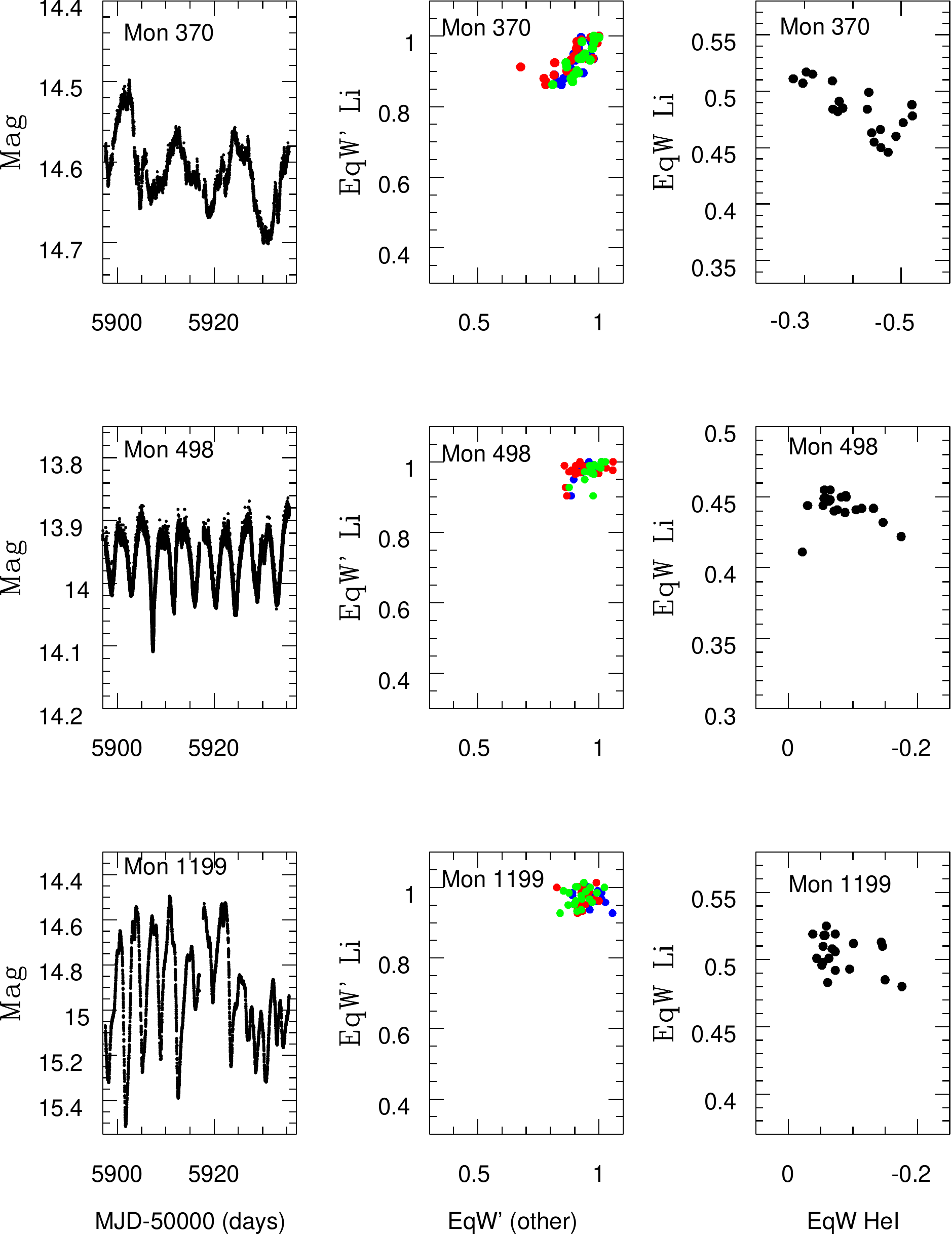}
\caption{{\em CoRoT} light curves and spectral indices plots for three
additional NGC~2264 members: (a) Mon-370, another CTTS with a hot-spot
light curve; (b) Mon-498 and (c) Mon-1199 - two CTTS whose light curves
are unusual and where we are using these diagrams to help determine
the primary physical mechanism driving their optical variability. See
Figure 8 and Figure 21 captions for further details.
\label{fig:morph_vs_spectra.set2}}
\end{figure*}

In this section, we wish to show that these same spectral features can also be
used as diagnostics of the physical processes underlying the variability for other
light curve classes.  For this purpose, we have selected six stars with disparate
light curve morphologies and for which (a) the {\em CoRoT} light curve amplitude
in 2011 is at least 0.10 mag (in order to be comparable to the typical amplitudes
for the stochastic stars and in order to ensure a strong signal in our spectral
indices if variable accretion drives the light curve morphology) and (b) we have
multi-epoch VLT/FLAMES spectra in 2011.  These six stars include (1) a WTTS with a
periodic, stable light curve morphology which we attribute
to cold spots; (b) a
CTTS having a periodic light curve which we would also normally attribute to
cold spots; (c) two CTTS whose {\em CoRoT} light curves show quasi-periodic 
structures which we attribute to hot spots; and (d) two CTTS whose {\em CoRoT}
light curves have unusual morphologies which do not obviously align well with our
standard light curve classes.  
Figure~\ref{fig:morph_vs_spectra.set1} and
Figure~\ref{fig:morph_vs_spectra.set2}
show the {\em CoRoT} light curves,
the correlation between lithium equivalent width and \ion{He}{1} 6678 \AA, 
and the correlation between
lithium equivalent width and the equivalent width of other relatively strong absorption
features in our echelle order.  

For these plots, as well as all the other similar 
plots in this paper, we have utilized sky-subtracted spectra.  That is, for each
object fiber, we have found the nearest sky fiber (which could be up to a few arminutes
away) and subtracted the latter from the former.  Due to the spatially varying
nebulosity, this is a necessarily imperfect process.  For the absorption line and
\ion{He}{1} emission line measures of most interest here, sky subtraction primarily is needed
in order to minimize any contribution from scattered Moon light.  The 2013 spectra
were taken almost entirely during a dark run, and therefore the sky fiber continuum
levels are very low and the uncertainty in sky correction has very little possible
impact.  For 2011, however, spectra were taken at all lunar phases, and the sky
continuum level varies significantly and so could impact equivalent widths for
faint stars.  This is probably the largest source of uncertainty in our plots,
though we believe it should not impact any of our conclusions.  For the \ion{He}{1} 6678 \AA\
equivalent width, an additional source of uncertainty is that there is an \ion{Fe}{1} absorption
feature almost coincident in wavelength with the \ion{He}{1} line, and whose strength is not
negligible with respect to the \ion{He}{1} line we normally see.  We have corrected our measured
\ion{He}{1} equivalent widths for this FeI feature by measuring the strength of the latter line
in WTTS as a function of spectral type, and then subtracting this amount from the
measured \ion{He}{1} equivalent widths for the stochastic stars.  The FeI equivalent widths
range from 0.15 \AA\ at G2, increasing to about 0.26 \AA\ at M0, and then decreasing
to 0.12 \AA\ at M2.

The plots for Mon-1274 (WTTS) and Mon-103 (CTTS) show these stars have little or no veiling
variability despite {\em CoRoT} light curve amplitudes of $\sim$ 0.2 mag.    The
rightmost plot for each star shows that \ion{He}{1} 6678 is not in emission. 
The similar properties for the two stars indicates that cold spots can
dominate the light curve morphology of CTTS, as has been advocated previously 
(e.g., Herbst et al.\ 1994).  The dispersion in the plotted points for the center
and RHS plots for these two stars is primarily due to measurements uncertainties.

The {\em CoRoT} light curves of Mon-168 and Mon-370 (both CTTS) were selected on morphological
grounds as candidates for rotational modulation of hot spots whose size/temperature vary significantly
over the {\em CoRoT} campaign duration.  This results in semi-sinusoidal waveforms that are
more structured than in the case of cold-spots and also are less stable in time.  The spectral
indices plots confirm that these stars have variable veiling and \ion{He}{1} in emission, and that
the absorption line veiling and \ion{He}{1} emission are correlated.  We therefore confirm that hot
spots are the driver of the {\em CoRoT} light curve morphology, and the period derived for
these stars is the photospheric rotational period.

Finally, Mon-498 and Mon-1199 (also both CTTS) have {\em CoRoT} light curves whose
morphology are unusual and not completely consistent with any of our light curve classes.
The center plot for these two stars (lithium versus ``other" absorption line equivalent
widths) shows no significant correlated variability, and hence no firm evidence
for accretion variability.   The RHS plots do show a possible correlation between
the lithium equivalent width and \ion{He}{1} 6678 \AA.
However, the inferred extra continuum flux would only be $\sim$5\% - 10\%\ of the 
stellar photosphere, thus only about half of the {\em CoRoT} amplitude for Mon-498
and only a very small fraction of the {\em CoRoT} amplitude for Mon-1199 could be
attributed to accretion variability.
Instead, we suspect that cold spots plus one or more
short-duration flux dips due to variable extinction
(at day 5906 and possibly day 5930) best explain the light curve for Mon-498.
The lack of veiling variability and the quite large amplitude of the photometric
variations for Mon-1199 argue that its light curve variations are primarily due
to variable extinction arising from MRI or other instabilities in its inner circumstellar
disk temporarily lofting more dust into the disk photosphere.

\begin{figure*}
\includegraphics[width=6.0in]{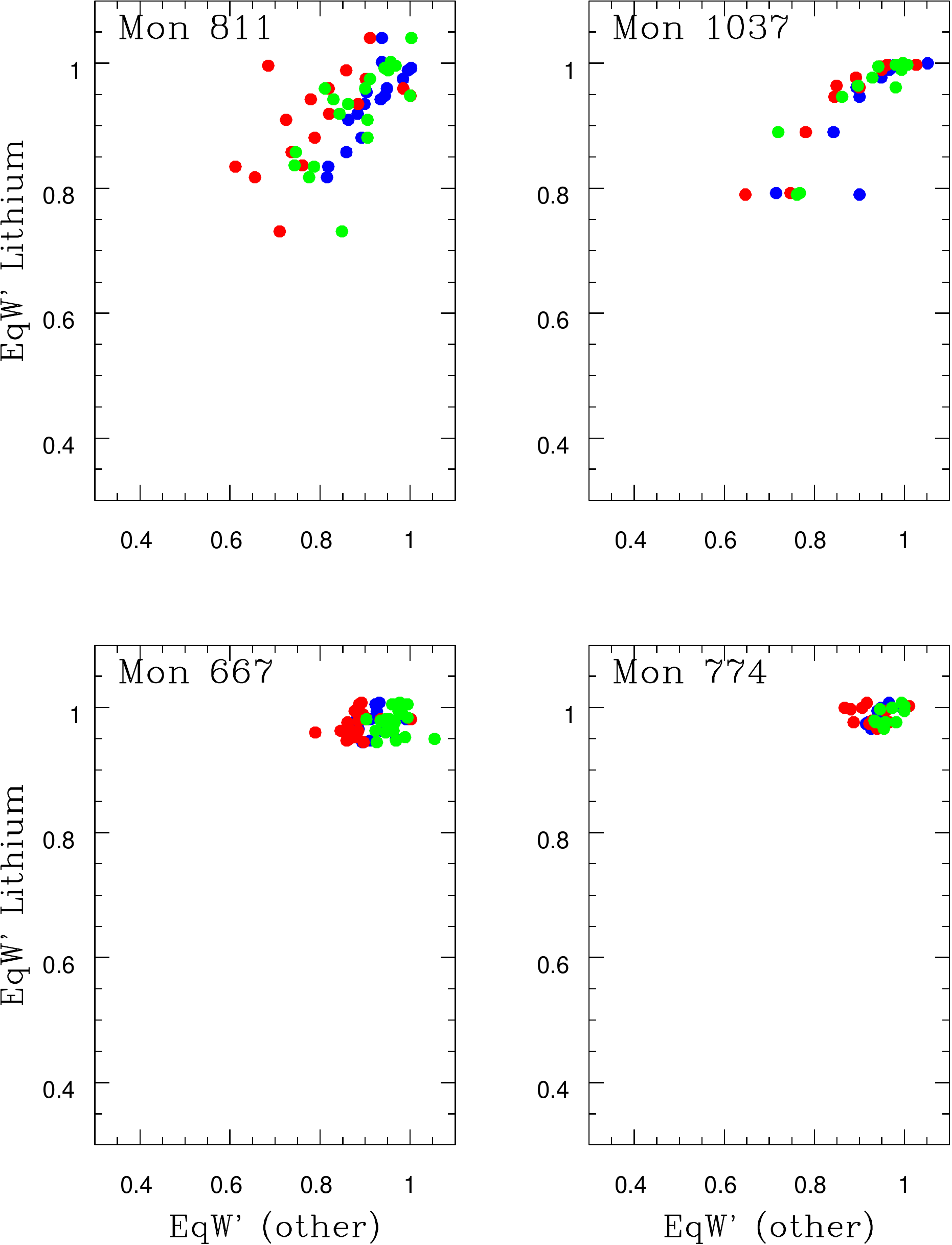}
\caption{Correlation between the equivalent width for the \ion{Li}{1} 6708 \AA\
doublet and other absorption lines present in the same FLAMES echelle
order, for four CTTS whose {\em CoRoT} light curves are dominated by
variable extinction events.  The equivalent widths at each epoch have
been divided by the maximum equivalent width for that feature, in order to  allow
data for several features to be over-plotted.
The top two panels show stars with strong
veiling variability; their \ion{He}{1} 6678 emission line equivalent widths 
also vary and are well-correlated with the variations in the veiling
levels from the absorption features.  The bottom two plots show two
stars with similar {\em CoRoT} light curve morphology, but where there
is no detectable veiling variability.
\label{fig:AATau.veiling}}
\end{figure*}

Stars whose light curves are dominated by deep flux dips produced by variable
extinction events form another common CTTS light curve class.  These flux dips
can either be periodic (like AA~Tau), and likely due to our line of sight
being occulted by a warp in the inner circumstellar disk, or aperiodic, and
possibly due to instabilities temporarily lifting more dust into the upper
layers of the inner disk.  Many stars appear to cycle back and forth between
these two light curve morphologies on long (year or years) timescales
(McGinnis et al.\ 2015).  Based on their UV excesses, mean veiling and H$\alpha$
equivalent widths, these stars have similar mean mass accretion rates as the
stochastic light curve class members, and therefore we might expect similar
veiling variability.  We have investigated this by measuring the absorption
line equivalent widths for eight of the variable extinction stars; 
Figure~\ref{fig:AATau.veiling}
shows plots of the correlation between \ion{Li}{1} 6708 equivalent width and
the other absorption lines (plotted in the same way as was done for the stochastic
stars in Figure~\ref{fig:stochastic_stars.veiling}) for four of these stars.
In summary, we find that the AA~Tau and aperiodic flux dip
stars show a range of accretion variability.  The data for the two AA~Tau stars with the largest
measured veiling variability (Mon-811 and Mon-1037) are shown in the top of 
Figure~\ref{fig:AATau.veiling}.
At the bottom of 
Figure~\ref{fig:AATau.veiling},
we show plots for two of the McGinnis et al.\ (2015) stars
that show essentially undetectable veiling variability ($<$5\% variation in the lithium
equivalent width).  Most of the variable extinction stars in fact show undetected or barely detected
($<$10\%) variations in their veiling levels.    This emphasizes that
their large light curve amplitudes (often more than 0.5 mag, peak to peak) must
be driven by some process other than variable accretion.

\subsection*{A3: H$\alpha$ Profiles for the Stochastic Stars}

\begin{figure*}
\begin{center}
\includegraphics[scale=0.8]{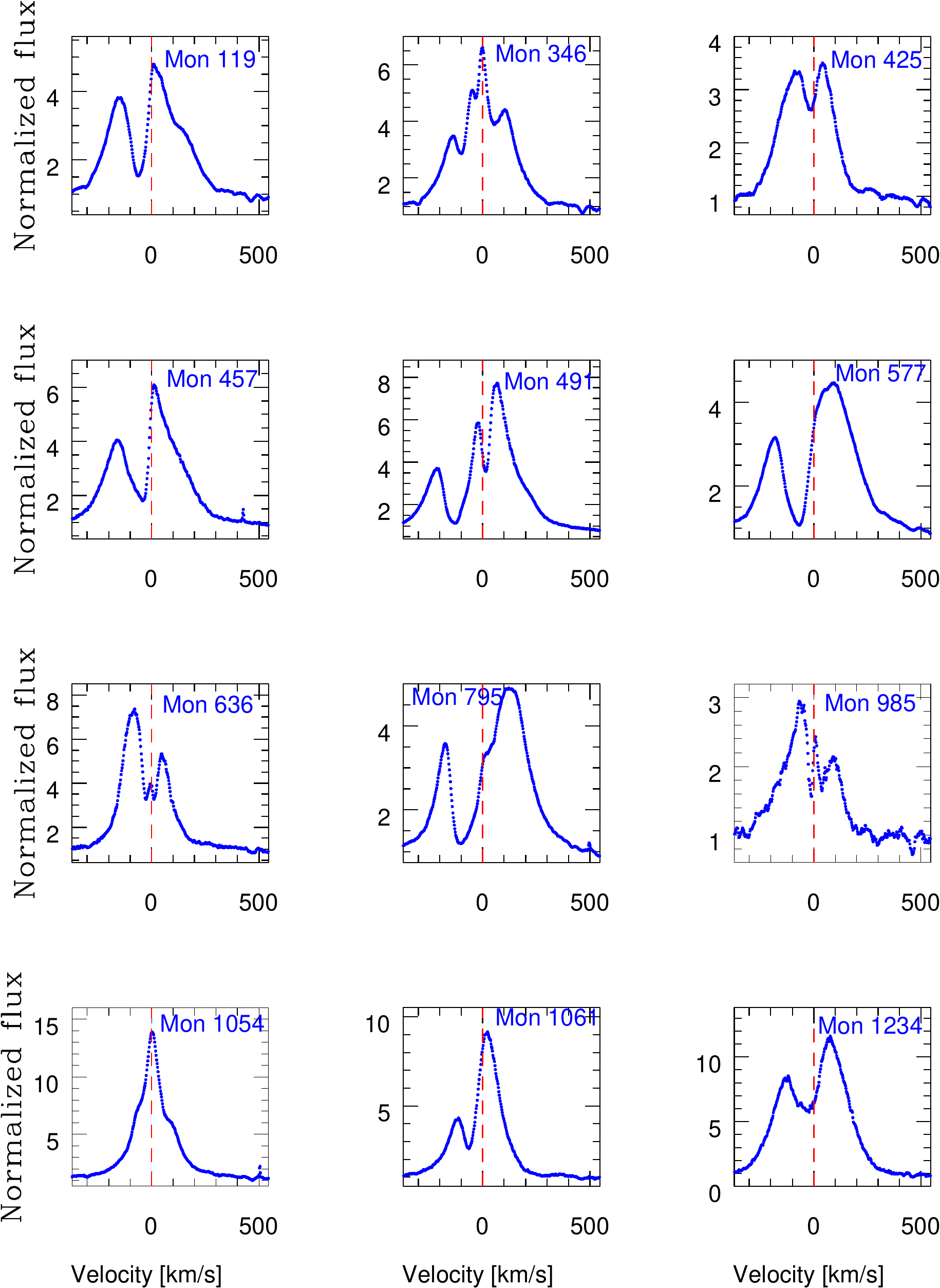}
\end{center}
\caption{H$\alpha$ profiles of the stars whose {\em CoRoT} light
curves are categorized as stochastic, for
those stars for which we have available high resolution spectra. 
The vertical dashed red line marks rest velocity for H$\alpha$.
We omit an H$\alpha$\ profile for Mon-613 here (to make the figure
more compact) - our multi-epoch H$\alpha$\ profiles for this
star are shown in Figure~\ref{fig:multi_halpha.2013},
\label{fig:wild_halpha}}
\end{figure*}

Our VLT/FLAMES spectra only have a wavelength coverage of about 350 \AA.
In the echelle order, we have only two lines normally in emission for CTTS
are present - H$\alpha$ and \ion{He}{1} 6678 \AA.  For eight of the stochastic 
stars, we have multi-epoch data with either 12 or 20 spectra, depending on
the year.  In Figure~\ref{fig:multi_halpha.2013}, we overplot the entire set of
profiles for two of the stars with multi-epoch data.   Here (Figure~\ref{fig:wild_halpha}), 
we provide
single epoch H$\alpha$ profiles for ``representative" profiles for each of
the stars for which we have VLT/FLAMES spectra.  Where we have multi-epoch
data, we have tried to choose an epoch which shows features which are
present in the majority of the epochs.

\subsection*{A4: Spectral Types Used in Figure 13}

In Table~\ref{tab:sptypes} we provide the spectral types used
to construct the spectral type histograms for the accretion burst, variable
extinction and short-duration flux dips classes shown in Figure~\ref{fig:sptype.histo}.
The spectral types for the stochastic class are provided in Table~\ref{tab:basicinformation}.
For all four groups, where possible we use published spectral types, as
described in Cody et al.\ (2014).  For a few of the short-duration flux
dip stars, we have derived spectral type estimates from our own spectra
(Stauffer et al.\ 2015).   Where actual spectral types were not available,
we have estimated spectral type by adopting the estimates for stellar
luminosity and radius described in Venuti et al.\ (2014); combining those two
quantities and the Stefan-Boltzmann law to estimate the effective temperature of
the star; and then using the spectral type to T$_{eff}$\ conversion for young
stars provided in Pecaut \& Mamajek (2013).   We also provide H$\alpha$ 
equivalent widths for the variable extinction and accretion burst members
in the table, because those numbers are used in \S 3.  The H$\alpha$ data
are generally from the literature (Cody et al.\ 2014); when no literature value
was available, we have measured representative H$\alpha$ equivalent widths
from our own VLT spectra, from the ESO-Gaia spectra, or from MMT Hectochelle
spectra described in Furesz et al.\ (2006).

\begin{deluxetable*}{lccccccc}
\tabletypesize{\scriptsize}
\tablecaption{Spectral Type Estimates for Additional Light Curve Classes\label{tab:sptypes}}
\tablecolumns{8}
\tablewidth{0pt}
\tablehead{
\colhead{Mon-ID\tablenotemark{a}}  & \colhead{Spectral Type\tablenotemark{d}} & 
\colhead{H$\alpha$ EW\tablenotemark{e}} &
\colhead{Mon-ID\tablenotemark{b}}  & \colhead{Spectral Type\tablenotemark{d}} & 
\colhead{H$\alpha$ EW\tablenotemark{e}} &
\colhead{Mon-ID\tablenotemark{c}}  & \colhead{Spectral Type\tablenotemark{d}} 
 }
\startdata
CSIMon-000007 & K7 & -12. & CSIMon-000126 & M0 & -26. & CSIMon-000021 & K5 \\
CSIMon-000011 & K7 & -58. & CSIMon-000250 & K3 & -15. & CSIMon-000056 & K5 \\
CSIMon-000117 & M2.5 & -353. & CSIMon-000296 & K2 & -11. & CSIMon-000314 & M3 \\
CSIMon-000185 & K4 & -59. & CSIMon-000297 & K2 & -7. & CSIMon-000378 & K5.5 \\
CSIMon-000260 & K7 & -62. & CSIMon-000325 & G6 & +0.2 & CSIMon-000566 & M3.5 \\
CSIMon-000341 & M0.5 & -161. & CSIMon-000358 & M3* & -6. & CSIMon-0001076 & M1 \\
CSIMon-000406 & K3* & -46. & CSIMon-000379 & K2 & -36. & CSIMon-001131 & M1.5+ \\
CSIMon-000412 & M1 & -31. & CSIMon-000433 & M1* & -7. & CSIMon-001165 & M1.5+ \\
CSIMon-000469 & K7 & -237. & CSIMon-000441 & M2* & -34. & CSIMon-001580 & M1* \\
CSIMon-000474 & G5 & -105. & CSIMon-000456 & K4 & -13. & CSIMon-006975 & M2.5+ \\
CSIMon-000510 & M0 & -102. & CSIMon-000650 & K1 & -6. &  &  \\
CSIMon-000567 & K3 & -84. & CSIMon-000660 & K4 & -17. & CSIMon-000119 & K6 \\
CSIMon-000808 & K4 & -50. & CSIMon-000667 & K3* & -0.1 & CSIMon-000342 & M4 \\
CSIMon-000860 & M2.5 & -261. & CSIMon-000676 & K4* & -1. & CSIMon-000577 & K1 \\
CSIMon-000877 & K4 & -91. & CSIMon-000681 & K3* & -26. & CSIMon-001038 & M0 \\
CSIMon-000919 & M4 & -80. & CSIMon-000717 & M0.5 & -24. & CSIMon-001114 & M1.5 \\
CSIMon-000945 & K4 & -66. & CSIMon-000774 & K2.5 & -14. & & \\
CSIMon-000996 & K7 & -25. & CSIMon-000811 & K6 & -27. & & \\
CSIMon-001022 & K4 & -46. & CSIMon-000824 & K4 & -2. & & \\
CSIMon-001174 & M2 & -130. & CSIMon-000928 & M0 & -16. & & \\
CSIMon-001187 & M2* & -8. & CSIMon-001037 & K1 & -22. & & \\
CSIMon-001217 & K4 & -87. & CSIMon-001140 & K3* & -32. & & \\
CSIMon-001573 & K9* & -26. & CSIMon-001144 & K5 & -50.& & \\
CSIMon-000766 & M0.5 & -70. & CSIMon-001296 & K6* & -16. & & \\
CSIMon-001048 & M3* & ... & CSIMon-001308 & K8* & -41. & & \\
\enddata
\tablenotetext{a}{These stars all have accretion-burst dominated optical light curves
    (Stauffer et al.\ 2014).}
\tablenotetext{b}{These stars all have variable extinction dominated optical light curves
    (McGinnis et al.\ 2015).}
\tablenotetext{c}{These stars all have optical light curves containing short-duration,
    shallow flux dips (Stauffer et al.\ 2015).}
\tablenotetext{d}{Spectral types derived by us in Stauffer et al.\ (2015) are
marked with a ``+" sign.  Spectral type estimates based on the CFHT multi-band
photometry, Venuti et al.\ (2014) and Pecaut \& Mamajek (2013) are marked
with a ``*" sign.  Other spectral types from Dahm \& Simon (2005) or Makidon
et al.\ (2004)}
\tablenotetext{e}{H$\alpha$ equivalent widths primarily from Dahm \& Simon (2005), 
otherwise from our own spectra.}
\end{deluxetable*}

\subsection*{A5: Derivation of Projected Rotational Velocities (\vsinis) from the
 VLT Spectra}

We have derived \vsini\ estimates in two ways for the CTTS in NGC~2264 for which we have
VLT/FLAMES spectra.  Our initial \vsini\ estimation begins by co-adding
all epochs of spectra for each star to produce a high S/N mean spectrum.  For
the adopted spectral type for each star, we produce a synthetic spectrum at
the FLAMES spectral resolution, and then convolve that spectrum with a rotational
broadening function, using software routines available
in SME (Valenti \& Piskunov 1996), Synth3 and Binmag3 (Kochukhov 2007).  We do
this for a range in adopted rotational velocities, and compare the model spectra
to the observed spectrum in a least-squares sense.  The \vsini\ we adopt is that
for the model whose spectrum matches most closely the observed spectrum.  The
formal uncertainties in the derived \vsini's are generally 2 $-$ 3 \kms.  We note that the
model spectra assume solar abundance, hence essentially no feature is present for
the lithium 6708 \AA\ doublet; these \vsini's therefore are based on all the
other absorption features in the FLAMES H$\alpha$ order. 

After deriving \vsini's in this manner for all of our NGC~2264 CTTS, we plotted
those \vsini's versus the mean FWHM of the lithium 6708 \AA\ doublet for each star
(i.e. we measured the FWHM at each epoch, and then took the average of those values;
the FWHM for the stochastic stars ranges from 0.7 \AA\ to 1.8 \AA, with RMS
for the dozen or more epochs for a given star of order 0.03 \AA).
The plot shows the expected relationship, but with a larger than anticipated
scatter for stars with small \vsini.  As one approaches the resolution limit of
the spectrograph, one of course expects an increase in scatter, but it was also
possible that by averaging spectra for the \vsini\ analysis we could artifically
broaden lines (e.g., for an SB1).  Therefore, we adopted a second procedure to
estimate \vsini\ based on combining the lithium FWHM data and the original \vsini\ data to
provide a calibration.  For this purpose, we fit a calibration curve to the
plot of \vsini\ versus FWHM.  Then, for each star, we used that calibration curve
and the measured mean lithium FWHM to provide a new \vsini\ estimate.  Because the
lithium 6708 \AA\ line is always the strongest absorption feature in these spectra,
for faint and/or slowly rotating stars, this latter \vsini\ estimate could be
more reliable than the initial estimate. 
Table~\ref{tab:other_data} provides both \vsini\ estimates - column 7 provides
the initial \vsini\ estimate, and column 8 the \vsini\ estimate derived from
the measured lithium FWHM.

Comparison of the two \vsini\ estimates in Table~\ref{tab:other_data} shows generally
good correspondence.  However, a few stars show significant differences, with
Mon-1234 having the most significant discrepancy (24.0 vs. 15.9 \kms).  The
spectral synthesis method yields essentially the same \vsini\ for Mon-636 and
Mon-1234, while the lithium FWHM method yields 24.9 and 15.9 \kms, respectively.
The mean and RMS FWHM for the two stars are 0.827 $\pm$ 0.036 and 0.725 $\pm$ 0.024,
respectively, with none of the Mon-1234 FWHM values being as large as the
smallest FWHM measurement for Mon-636.  The lithium widths preclude these stars
from having the same \vsini.    Based on this and similar comparisons, we 
adopt the \vsini's derived from the lithium width as our best estimates of \vsini,
and we use those estimates in the plots
in the main body of the paper.

\end{document}